\documentclass[%
 aip,
 amsmath,amssymb,
preprint,
]{revtex4-1}
\usepackage[colorlinks]{hyperref}
\usepackage{CJK}
\usepackage[T1]{fontenc}
\usepackage{amsthm}
\usepackage[utf8]{inputenc} % set input encoding (not needed with XeLaTeX)
\usepackage{setspace}
\usepackage[margin=0.7in]{geometry} % to change the page dimensions
\usepackage{gensymb}
\usepackage{graphicx}% Include figure files
\usepackage{epstopdf}
\usepackage{dcolumn}% Align table columns on decimal point
\usepackage{bm}% bold math
\usepackage{amsmath}
\usepackage{mathtools}
\usepackage{relsize}
\usepackage{multirow}
\usepackage{footnote}
\usepackage[symbol]{footmisc}

\begin{document}
%\begin{CJK*}{GB}{}
%\preprint{APS/123-QED}
\title{Methods to Accelerate High-Throughput Screening of Atomic Qubit Candidates in van der Waals Materials}% Force line breaks with \\

\author{Rodrick Kuate Defo \footnote{Author to whom correspondence should be addressed: rkuatedefo@princeton.edu}}
\affiliation{Department of Electrical Engineering, Princeton University, Princeton, NJ 08540}
\affiliation{John A. Paulson School of Engineering and Applied Sciences, Harvard University, Cambridge, MA 02138}

\author{Haimi Nguyen}
\affiliation{Department of Chemistry, Columbia University, New York, NY 10027}

\author{Mark J. H. Ku}
\affiliation{Department of Physics and Astronomy and Department of Materials Science and Engineering, University of Delaware, Newark, DE 19716}
\author{Trevor David Rhone}
\affiliation{Department of Physics, Applied Physics and Astronomy, Rensselaer Polytechnic Institute, Troy, NY 12180}

\date{\today}
             
\begin{abstract}
The discovery of atom-like spin emitters associated with defects in two-dimensional (2D) wide-bandgap (WBG) semiconductors presents new opportunities for highly tunable and versatile qubits. So far, the study of such spin emitters has focused on defects in hexagonal boron nitride (hBN). However, hBN necessarily contains a high density of nuclear spins, which are expected to create a strong incoherent spin-bath that leads to poor coherence properties of spins hosted in the material. Therefore, identification of new qubit candidates in other 2DWBG materials is necessary. Given time demands of \textit{ab initio} methods, new approaches for rapid screening and calculation of identifying properties of suitable atom-like qubits are required. In this work, we present two new methods for rapid estimation of the zero-phonon line (ZPL), a key property of atomic qubits in WBG materials. First, this ZPL is calculated by exploiting Janak's theorem. For finite changes in occupation, we provide the leading-order estimate of the correction to the ZPL obtained using Janak's theorem, which is more rapid than the standard method ($\Delta$SCF). Next, we also demonstrate an approach to converging excited states that is faster for systems with small strain than the standard approach used in the $\Delta$SCF method. We illustrate these methods using the case of the singly negatively charged calcium vacancy in SiS$_2$, which we are the first to propose as a qubit candidate. This work has the potential to assist in accelerating the high-throughput search for quantum defects in materials, with applications in quantum sensing and quantum computing.

\end{abstract}
\maketitle

\section{INTRODUCTION \label{sec:intro}}
Research in the field of layered materials has grown quickly since the development of the high-quality sample yielding scotch-tape exfoliation technique.~\cite{Geim2013,Toth2019} For instance, the quantum spin Hall effect at up to 100 K was observed in monolayer tungsten ditelluride (WTe$_2$)~\cite{Wu76}. Another important discovery was the observation of a correlated insulator state by tuning the twist degree of freedom in bilayer structures of graphene, due to the presence of flat bands near zero Fermi energy at a twist angle of about $1.1^\circ$, which upon electrostatic doping yields superconducting states with a critical temperature of up to 1.7 K.~\cite{Cao2018corr,Cao2018unc} Research in the field of point defect qubit candidates has also grown rapidly, particularly since the detection of single negatively charged nitrogen vacancy (N$V^-$) color centers in diamond.~\cite{Gruber2012} These qubit candidates consist of defects in the crystal structure involving substitutional or interstitial atoms and/or vacancies, and act as single photon sources as well as sources of electronic spin. Desirable characteristics for such qubits include indistinguishability of emitted photons, negligible spectral diffusion and long spin coherence time. State-of-the-art results include second-long coherence time of N$V$ centers,~\cite{Gill2013} km-scale entanglement of two N$V$s,~\cite{Hensen2015}, and the discovery of spectrally stable germanium vacancy (Ge$V$) and silicon vacancy (Si$V$) point defects.~\cite{Siyushev2017,Becker2018} The properties of spin qubits can be tuned by changing the crystal structure and the nature of the defect. 
Engineering spin qubits with properties that are desirable for applications in sensing, quantum computing and quantum communications are at the forefront of research interests.
Nevertheless, identifying the right combination of crystal structure and defect is enormously challenging due to the combinatorially large number of potential candidates that need to be explored using experiments or first-principles calculations.

The layered material hexagonal boron nitride ($h$-BN) is of particular interest among layered materials as it has a wide bandgap,~\cite{Elias2019} enabling its ability to encapsulate other layered materials and to host a variety of point defects giving rise to optical transitions lying within the range of its bandgap.~\cite{Toth2019,Tran2,IvadyhBN,Mehdi2018,Gottscholl2020} The issue with $h$-BN as a host for a qubit candidate employing electronic spins is that the atomic nuclei of boron and nitrogen have spins as well, which can cause spin-decoherence of the electronic spin state. Such an argument invites consideration of SiS$_2$ as a host, which theory has predicted to exist in layered form between 2.8 GPa and 3.5 GPa for the space group P$2_1$/c.~\cite{Plasienka2016} The SiS$_2$ host would be diamond-like or silicon carbide (SiC)-like in that the atoms that constitute it have a low natural abundance of isotopes with nuclear spin, which can be further suppressed via isotope purification in growth, diamond and SiC being systems that have been extensively studied as hosts for qubit candidates.~\cite{Kuate8,Becker2018,DOHERTY20131,Siyushev2017,Gali2010,SangYun2013,Weber8513,Soltamov2012,Koehl2011,Kraus2014,Wahl,Wang_2019,Dong,Green_2017,Green_2019,Kuate,DEFO2019,sohn2018cont} The point defect we will investigate in SiS$_2$ has the further advantage of exhibiting inversion symmetry, which eliminates the issue of an electric dipole moment making it susceptible to external noise and local fields and thus causing broadening of transitions. 

Given the growing number of experimental investigations, the need to rapidly identify promising point defect qubit candidates is being recognized. Indeed, recent work~\cite{davidsson2021adaq} has looked into automating the process of characterizing point defects through a code named ADAQ, to aid in the more rapid identification of defect signatures from experiments, and a push toward high-throughput point defect calculations was also present in the earlier PyCDT code.~\cite{BROBERG2018165} In identifying point defects in experiments, knowledge of the zero-phonon line (ZPL) transition is essential. We present in this paper two methods that can be incorporated into high-throughput searches for point defect qubit candidates. The first is a quick new method for estimating the error associated with computing the energy of the zero-phonon line (ZPL) transition of a point defect using Janak's theorem and the second is a new method for rapid convergence of excited state self-consistent field (SCF) calculations. These new tools are an important step in accelerating the high-throughput discovery of materials that can support spin qubits for applications in sensing, quantum communication and quantum computing.

Janak's theorem states that,~\cite{Janak1978}
\begin{equation} 
\frac{\partial E}{\partial n_i} = \epsilon_i,
\end{equation} 
where $E$ is the total energy, $n_i$ is the orbital occupation of the $i^{\rm th}$ orbital, and $\epsilon_i$ is the corresponding eigenvalue of the orbital. This theorem is similar to Koopmans' theorem for Hartree-Fock theory~\cite{KOOPMANS1934} in that it relates energy differences under a change in the number of electrons to orbital energies. The theorem is in fact a density-functional theory (DFT) version of a theorem originally proved by Slater for his $X-\alpha$ method, which was introduced in an early attempt to account for both exchange and correlation in electronic structure calculations.~\cite{Slater1970,Kaxiras} As Janak's theorem does not imply that $\Delta E = \epsilon_i \Delta n_i$ for finite $\Delta n_i$, an estimate of the error associated with using the theorem to determine excitation energies where orbital occupations change by integer amounts is necessary.
Applying the theorem and the method of error estimation is inherently faster than computing the energy of the excited state for a given point defect, as the error calculation amounts to terminating the excited state calculation before full convergence. We also show that excited state calculations for systems with small strain converge faster if the charge density is initialized by appropriately mixing the highest occupied molecular orbital (HOMO) and the lowest unoccupied molecular orbital (LUMO) from the ground state calculation, leaving the contribution to the charge density from the remaining orbitals unchanged, as compared to initialization of the charge density from a superposition of atomic charges. 

In this work, we first outline the computational methods in Section \ref{sec:disc}. In Section \ref{sec:res} we include results related to (i) ZPL estimations for silicon monovacancies in $4H$-SiC (see Section \ref{sec:VSi}) and (ii) the singly negatively charged calcium vacancy in SiS$_2$ (see Section \ref{sec:CaV}, which also includes stability calculations) and a discussion of these results, ending with a summary of our conclusions.

\section{THEORETICAL FORMULATION AND APPROACH \label{sec:disc}}

\subsection{Calculation Method\label{sec:calcmethod}}

The purpose of this work is to extend the notion of a mixing parameter~\cite{Johnson1988} to the initialization of the charge density and to bolster the validity of using Janak's theorem~\cite{Janak1978} to calculate the zero-phonon line (ZPL) by demonstrating success for the case of the singly negatively charged calcium vacancy in SiS$_2$, which we investigate as a potential qubit candidate. Our approach builds upon previous work employing Janak's theorem to calculate the electronic properties of excited states of atoms, molecules, and solids.~\cite{PhysRev.184.672,PhysRevA.32.720} In this work, we introduce an innovation by employing Janak's theorem for the calculation of defect properties and additionally providing a lowest order estimate of the error associated with using the theorem for integral change in the occupation of the single-particle states. We argue that the calculations in excess of the ground state calculation in the $\Delta$SCF method are not always necessary, which directly follows from Janak's theorem in the limit where the change in occupation is infinitesimal. To motivate the argument, we compute the error between the $\Delta$SCF method and the approach of using Janak's theorem to lowest order under a change in occupation.

We consider the operator from the single-particle equations in DFT,
\begin{equation}
\mathcal{O}(n(\mathbf{r})) = -\frac{\hbar^2}{2m_e}\nabla^2_{\mathbf{r}}+V(\mathbf{r})+e^2\int\frac{n(\mathbf{r}')}{\left|\mathbf{r}-\mathbf{r}'\right|}{\rm d}\mathbf{r}'+\frac{\delta E_{\rm xc}[n(\mathbf{r})]}{\delta n(\mathbf{r}')},
\label{eq:op}
\end{equation}
where $n(\mathbf{r})$ is the particle number density, the first term represents the kinetic energy of noninteracting quasiparticles with electron mass, the second term represents the external potential, the third term represents the Hartree potential for the Coulomb interaction between the quasiparticles and the last term represents the exchange-correlation potential recapturing the fermionic and many-body nature of electronic interactions.

The single-particle equation for the single-particle state $\phi^{(n)}_i$ with eigenvalue $\epsilon^{(n)}_i$ then reads,
\begin{equation}
\mathcal{O}(n(\mathbf{r}))\phi^{(n)}_i = \epsilon^{(n)}_i\phi^{(n)}_i,
\end{equation}
Under a change in occupation, let $n'(\mathbf{r})$ be the new number density and let $\phi^{(n')}_i$ be the new $i^{\rm th}$ single-particle state such that,
\begin{equation}
\mathcal{O}(n'(\mathbf{r}))\phi^{(n')}_i = \epsilon^{(n')}_i\phi^{(n')}_i,
\end{equation}
is satisfied.

The equation for the total energy from DFT is,~\cite{Kohn1965}
\begin{equation}
E = \sum_i^N\epsilon^{(n)}_i -\frac{1}{2}\int\int\frac{n(\mathbf{r})n(\mathbf{r}')}{\left|\mathbf{r}-\mathbf{r}'\right|}{\rm d}\mathbf{r}{\rm d}\mathbf{r}'+\int n(\mathbf{r})\left[\epsilon_{\rm xc}[n(\mathbf{r})]-\mu_{\rm xc}[n(\mathbf{r})]\right]d\mathbf{r}.
\end{equation}
If we let,
\begin{equation}
 F[n(\mathbf{r})] = -\frac{1}{2}\int\int\frac{n(\mathbf{r})n(\mathbf{r}')}{\left|\mathbf{r}-\mathbf{r}'\right|}{\rm d}\mathbf{r}{\rm d}\mathbf{r}'+\int n(\mathbf{r})\left[\epsilon_{\rm xc}[n(\mathbf{r})]-\mu_{\rm xc}[n(\mathbf{r})]\right]d\mathbf{r},
 \end{equation}
 and 
 \begin{equation}
 \Delta \epsilon = \sum_{i=1}^{N+1}(\epsilon^{(n')}_i-\epsilon^{(n)}_i) -(\epsilon^{(n')}_N-\epsilon^{(n)}_N)
 \end{equation}
 then we can approximate the $\Delta$SCF result as,
 \begin{equation}
E_{\rm \Delta SCF} \approx \epsilon^{(n)}_{N+1}-\epsilon^{(n)}_N+ \Delta\epsilon+F[n'(\mathbf{r})]-F[n(\mathbf{r})]. 
\end{equation}
Therefore, if we only take the difference in ground state eigenvalues, the associated error is,
\begin{equation}
\Delta E_{\rm \Delta SCF} \approx \Delta\epsilon+F[n'(\mathbf{r})]-F[n(\mathbf{r})].   
\end{equation}

We can choose to obtain $n'(\mathbf{r})$ at any arbitrary iteration in the full constrained-occupation calculation for the excited state. Therefore, in the limit where $n'(\mathbf{r})$ is obtained at the final iteration of the full constrained-occupation calculation for the excited state, the error becomes exact. In performing the full constrained-occupation calculation for the excited state, we have also explored initializing the charge density by mixing the HOMO and LUMO by varying amounts. 
 
To obtain the defect levels and total energies, we performed first-principles DFT calculations for the various defects in $4H$-SiC and SiS$_2$ using the VASP code~\cite{Kresse1,Kresse2,Kresse3} and the QUANTUM ESPRESSO code~\cite{Giannozzi_2009,Giannozzi_2017} for $\Delta$SCF calculations. In VASP, atomic structures were first converged using the generalized gradient approximation (GGA) for the exchange-correlation energy of electrons, as parametrized by Perdew, Burke and Erzenhof (PBE)~\cite{Perdew2} and then, for the calculation of defect levels of the uncompressed structure, using the screened hybrid functional of Heyd, Scuseria and Ernzerhof (HSE) with the original parameters (0.2 \AA$^{-1}$ for screening and 25\% for mixing).~\cite{Heyd,Krukau} The terms in $F[n(\mathbf{r})]$ were obtained for the ground state and for the state with the changed occupation, where a non-SCF calculation was performed until convergence keeping the charge density fixed in the latter case. The eigenvalues, $\epsilon^{(n')}_i$, were also provided by the non-SCF calculation. Code from work by Feenstra \textit{et al.}~\cite{Feenstra2013} was used to change orbital occupations. In QUANTUM ESPRESSO, we performed $\Delta$SCF calculations to investigate the SiS$_2$ system using PAW pseudopotentials~\cite{Kresse3} with a 108-atom supercell with gamma-point integration. Modified source code was used to alter the charge density in QUANTUM ESPRESSO. The different values of strain investigated in this work were for biaxial strain. For strain values other than the uncompressed structure, the lattice parameters along the strain directions were rescaled from the lattice parameters of the uncompressed structure. The out-of-plane lattice parameter was held fixed. Constant volume relaxations were then performed for the strained structures as described above. Additional computational details may be found in the Appendix.

For first-principles phonon calculations, performed using Phonopy,~\cite{Togo} the atomic positions in the stoichiometric conventional unit cell were relaxed until the magnitude of the Hellmann-Feynman forces was smaller than $10^{-6}$ eV$\cdot$\AA$^{-1}$ with a Monkhorst-Pack grid of $6\times6\times2$ k-points. Supercells containing 108 atoms ($3\times3\times1$ multiple of the conventional unit cell) with appropriately scaled k-point grids and a cutoff energy of 500 eV where then constructed from the stoichiometric conventional unit cell and used to obtain the force constants to compute the phonon dispersion.

\subsection{Formation Energies}

The formation energies of the calcium vacancy in SiS$_2$ in various charge states were calculated according to the formula,~\cite{zhang1991chemical, RevModPhys.86.253}
\begin{equation}
\label{eq:form_eq}
E_f(q) = E_{\text{def}}(q) - E_0 - \sum_i\mu_in_i + q(E_{\text{VBM}} +E_{\text{F}}) + E_{\text{corr}}(q)
\end{equation}
where $q$ denotes the charge state, $E_{\text{def}}(q)$ is the total energy for the defect supercell with charge state $q$, $E_0$ is the total energy for the stoichiometric neutral supercell, $\mu_i$ is the chemical potential of atom $i$, $n_i$ is a positive (negative) integer representing the number of atoms added (removed) from the system relative to the stoichiometric cell, $E_{\text{VBM}}$ is the absolute position of the valence band maximum, $E_{\text{F}}$ is the position of the Fermi level with respect to the valence band maximum (generally treated as a parameter), and $E_{\text{corr}}(q)$ is a correction term to account for the finite size of the supercell when performing calculations for charged defects.~\cite{Vinichenko} This correction term does not simply treat the charged defect as a point charge, but rather considers the extended charge distribution. The chemical potentials of all the reference elements used in our calculations are listed as follows as a function of their crystal structure and total energy per atom: Si (diamond structure, $-5.42$ eV/atom); S (the total energy of a gas phase S$_8$ molecule was calculated and the sublimation enthalpy was then subtracted,~\cite{Kuate2016,Steudel} $-4.20$ eV/atom); and Ca (face-centered cubic structure, $-2.00$ eV/atom). Consideration of Si-rich or S-rich preparation conditions was made, similar to previous work.~\cite{Kuate}

\subsection{Material Screening Approach}
In order to leverage the vast body of existing knowledge in the field, we utilized electronic databases of two-dimensional materials known as the 2D Materials Encyclopedia~\cite{2Dmatpedia} in a top-down filtering approach and the Computational 2D Materials Database (C2DB)~\cite{Haastrup2018,gjerding2021recent} to find the most promising host materials. The entries from the databases were first selected based on the presence of centrosymmetry in their space groups, yielding 4,822 potential candidates with duplicates discarded. Fulfillment of this criterion allows for the possibility of creating in the host a defect without a dipole moment that could interact with local fields. The following filters were then applied (the number of candidates at each filtration step, discarding duplicates, is listed in brackets),
\begin{enumerate}
\item \textit{PBE band gap above 2 eV (985).} The optical transitions of the qubit candidate implanted in the host must not introduce interference from the electronic states of the host.~\cite{Weber8513} The value is taken to be at least as large as the energy of the optical transitions of well-known point defects such as the N$V^-$ center in diamond and singly negatively charged silicon monovacancies in $4H$-SiC. Given that the PBE~\cite{Perdew2} functional generally underestimates the value of the band gap, such a value allows for the defect levels for transitions similar in energy to those of negatively charged silicon monovacancies in $4H$-SiC and the N$V^-$ center in diamond to be separated from the band edges.
\item \textit{Exfoliation energy below 80 meV/atom (441).} The material must be easy to exfoliate and therefore to fabricate. The value is taken to be commensurate with exfoliation energies of common layered materials such as MoS$_2$ with space group P$\bar{6}$m2.  
\item \textit{Does not contain an atom with a nuclear spin (6)}. Decoherence caused by the interaction of nuclear spins with the electronic ones of the qubit candidate must be minimized. We set the cutoff at $< 5\%$ natural abundance of isotopes containing a nuclear spin using as reference $4H$-SiC, which is the less restrictive among the two commonly studied systems of diamond and $4H$-SiC with defects showing long coherence times at room temperature.
\end{enumerate}
The filtering process yielded SiS$_2$ with space group P$2_1$/c as the host candidate with the lowest decomposition energy. The other host candidates left at the final stage of filtration were CaO with space group P4/nmm, SO$_2$ with space group P$2_1$/c, CO with space group Cmme, Si$_2$O$_3$ with space group P$2_1$/m, and S$_4$O$_9$ with space group P$\bar{3}$. To our knowledge, none of these compounds have been used as qubit hosts. We note that similar criteria have been used in the work of Ferrenti \textit{et al.},~\cite{Ferrenti2020} though in that work the condition regarding the natural abundance of isotopes with a nuclear spin was relaxed to a cutoff of $<50\%$ and bulk materials were considered, yielding more known candidate hosts. Given the possibility of employing isotope purification, we invite researchers wishing to follow up on our work to relax the cutoff if the above candidates do not prove readily amenable to synthesis.

Working from the literature on defects in diamond and $4H$-SiC, we investigated vacancies, germanium-vacancy complexes and lead-vacancy complexes as possible qubit candidates in the SiS$_2$ host. However, these all break inversion symmetry when they are relaxed with spin polarization. On the other hand, the singly negatively charged calcium-vacancy complex is able to preserve inversion symmetry upon relaxation with spin polarization, due in part to the large size of the calcium atom.

\section{RESULTS AND DISCUSSION \label{sec:res}}

\subsection{Silicon Monovacancies in $4H$-SiC\label{sec:VSi}}
We begin by demonstrating the accuracy of Janak's theorem for silicon monovacancies in $4H$-SiC. This system is an example of a success of the theorem though for another prominent system, the N$V^-$ in diamond, Janak's theorem alone is not sufficient to produce good agreement between experiment and theory. The lack of agreement in that case can be explained by the significant difference between the ground and excited state wavefunctions~\cite{Gali} at a location of significant change in the external or ionic potential (at the N atom in that case), which, from the single-particle equations, will in general lead to larger energy changes under a change in occupation compared to differences elsewhere. We can motivate this argument explicitly using what is referred to as ``effective-mass theory'',~\cite{Kaxiras,Lannoo1992} where the potential introduced by the impurity is considered to be a small perturbation to the crystal potential. Let $V^{\rm defect}(\mathbf{r})$ be the change in the external or ionic potential of the crystal due to the presence of the defect and expand the defect wavefunction of interest, $\phi^{\rm defect}(\mathbf{r})$, in a basis of the eigenfunctions, $\psi_{\mathbf{k}}(\mathbf{r})$, of the perfect crystal,
\begin{equation}
\phi^{\rm defect}(\mathbf{r}) = \sum_{\mathbf{k}}\beta_{\mathbf{k}}\psi_{\mathbf{k}}(\mathbf{r}).
\end{equation} 
The wavefunction will obey the single-particle equation,
\begin{equation}
\left[\mathcal{H}^{\rm crystal} + V^{\rm defect}(\mathbf{r})\right] \phi^{\rm defect}(\mathbf{r}) = \epsilon^{\rm defect} \phi^{\rm defect}(\mathbf{r}),
\end{equation} 
where $\mathcal{H}^{\rm crystal}$ is the single-particle hamiltonian for the perfect crystal so that,
\begin{equation}
\mathcal{H}^{\rm crystal} \psi_{\mathbf{k}}(\mathbf{r}) = \epsilon_{\mathbf{k}} \psi_{\mathbf{k}}(\mathbf{r}).
\end{equation} 
Using the orthogonality of the crystal wavefunctions and the expansion for $\phi^{\rm defect}(\mathbf{r})$ we then have,
\begin{equation}
\epsilon_{\mathbf{k}}\beta_{\mathbf{k}} + \sum_{\mathbf{k}'}\left<\psi_{\mathbf{k}}\right|V^{\rm defect}\left|\psi_{\mathbf{k}'}\right>\beta_{\mathbf{k}'} = \epsilon^{\rm defect}\beta_{\mathbf{k}}.
\end{equation}
Thus, for perturbations to the defect wavefunction that equally change the projection of the defect wavefunction onto some perfect crystal eigenfunction (ie. that result in the same $\beta_{\mathbf{k}}$ for some $\mathbf{k}$), the perturbation that will in general cause the greatest change in energy is the one where the other coefficients are such that the defect wavefunction changes most at the location of the defect potential. As we will see in Section \ref{sec:CaV}, the partial charge densities associated with the orbitals changing occupation can be a quick way to identify when Janak's theorem is likely to fail.

\begin{figure}[ht!] 
\centering
\includegraphics[width=0.25\textwidth]{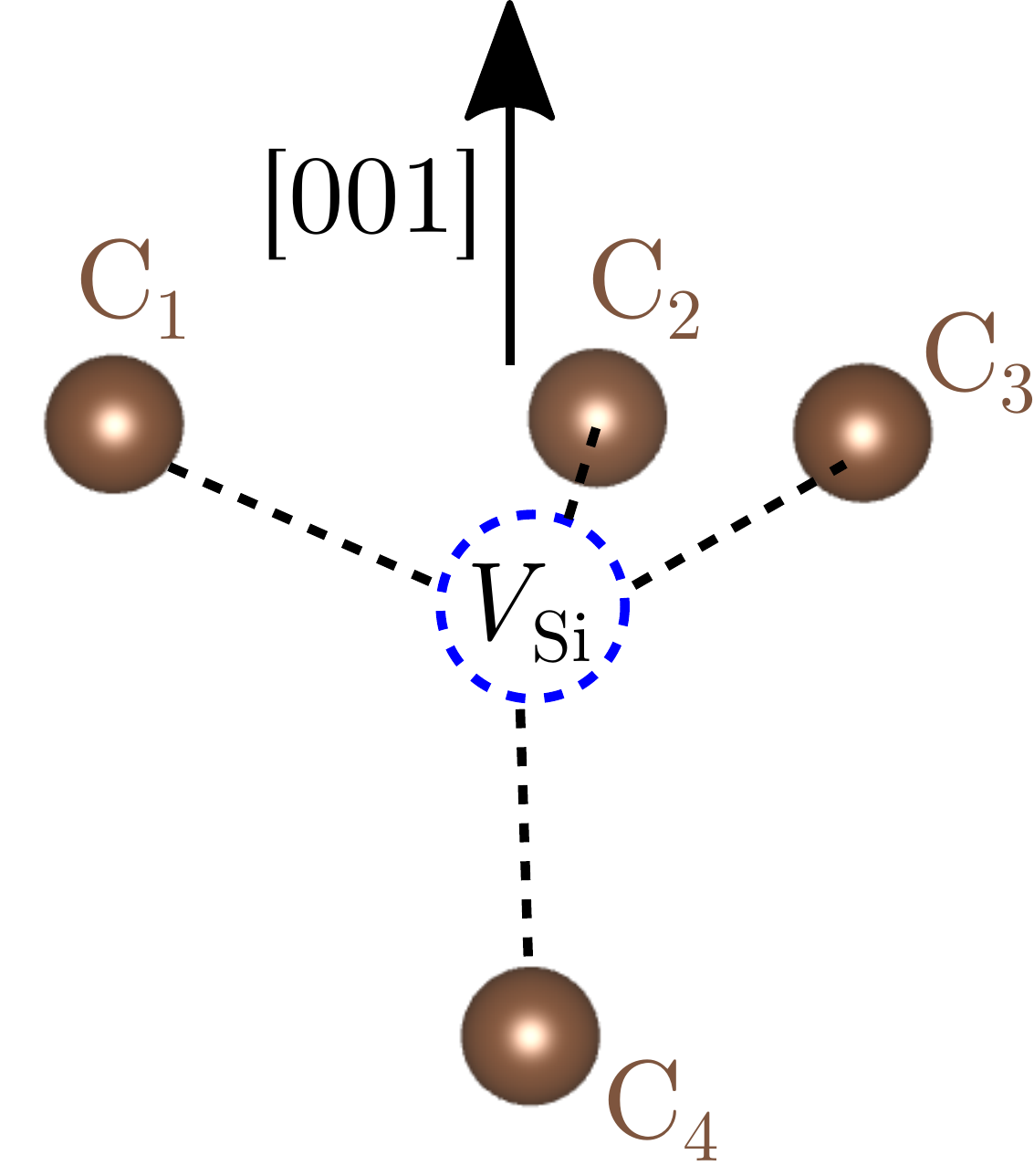}
\caption{ Shown above is the local structure of a $V_{\rm Si}$ in $4H$-SiC, which can be located at either of the two inequivalent $k$ or $h$ sites. Nearest-neighbor carbon atoms to the vacancy are shown in brown and labeled with numbers and the silicon vacancy is shown as a blue dashed circle. The $V_{\rm Si}$ is displayed along the $[001]$ direction.} 
\label{fig:structVSi}
\end{figure}

For the lattice parameters of the stoichiometric hexagonal unit cell of $4H$-SiC using the HSE06 functional, we find $a = 3.08$~\AA~and $c = 10.04$~\AA. These values are in good agreement with experimental values~\cite{Levinshtein} of $a = 3.07$~\AA~and $c = 10.05$~\AA~and other theoretical values~\cite{Yan2020} of $a = 3.07$~\AA~and $c = 10.05$~\AA. The structure of the $V_{\rm Si}$ is shown in Fig. \ref{fig:structVSi}. We observe that the expected $C_{3v}$ symmetry is weakly broken from the distances displayed in Table \ref{tab:VSidists}.

\vspace{100pt}
\begin{table}[ht!]
\caption{Distances in \AA,~$d(X_1-X_2)$, between carbon atoms numbered in Fig. \ref{fig:structVSi} and the silicon vacancy for $V_{\rm Si}$ at the $k$ and $h$ sites.}
\centering
\vspace{1 mm}
\begin{tabular}{|c|c|c|c|c|}
\hline\hline 
site &  $d(V_{\rm Si}-{\rm C}_1)$ & $d(V_{\rm Si}-{\rm C}_2)$ & $d(V_{\rm Si}-{\rm C}_3)$ & $d(V_{\rm Si}-{\rm C}_4)$ \\
\hline
 $k$ &      $2.05$  &   $2.06$ & $2.04$ & $2.03$ \\
 $h$ &      $2.04$  &   $2.04$ & $2.03$ & $2.02$  \\
\hline
\hline
\end{tabular}
\label{tab:VSidists}
\end{table}

\begin{figure}[ht!] 
\centering
\includegraphics[width=0.3\textwidth]{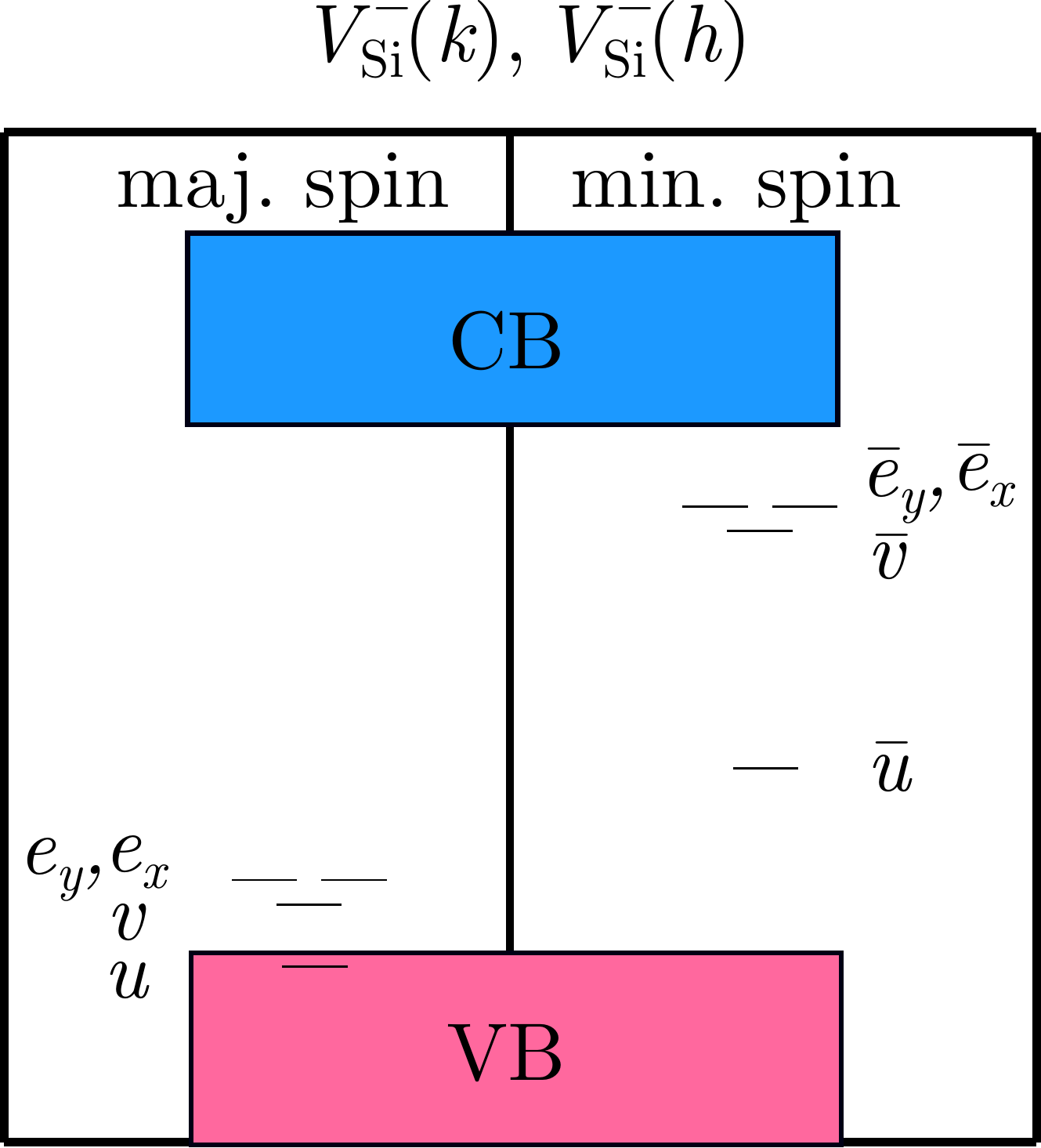}
\caption{Schematic of the majority (without overbar) and minority (with overbar) spin energy levels calculated in the ground state using the HSE06 functional for the $V_{\rm Si}^-(k)$ and $V_{\rm Si}^-(h)$ point defects in $4H$-SiC. The conduction band is indicated in blue and the valence band in red. Single-particle orbitals transforming as the $A_1$ irreducible representation of the $C_{3v}$ point group are represented by $u$ and $v$, while $e_x$ and $e_y$ transform as the $x$ and $y$ components of the two-dimensional $E$ irreducible representation of the $C_{3v}$ point group. The vertical axis of the figure is not drawn to scale.} 
\label{fig:eigenVSi}
\end{figure}

From the work of Soykal \textit{et al.},~\cite{Soykal} we know that for a spin-polarized system the approach of considering holes is equivalent to the approach of considering electrons. Defect levels calculated using the HSE06 functional for the singly negatively charged silicon monovacancy $V_{\rm Si}^-$ in $4H$-SiC with $S = 3/2$ at the two inequivalent $h$ and $k$ sites are illustrated schematically in Fig. \ref{fig:eigenVSi} and are provided in Tables \ref{tab:HSEvalsVSik} and \ref{tab:HSEvalsVSih}. We note that the zero of energy is arbitrary as it depends on the pseudopotential, but, as we consider differences in eigenvalues, the value of the zero is not important. From the work of Soykal \textit{et al.},~\cite{Soykal} in the hole picture the ground state manifold is composed of the states, $\left||ue_xe_y+i\bar{u}\bar{e}_x\bar{e}_y\right>/\sqrt{2}$, $\left||ue_xe_y-i\bar{u}\bar{e}_x\bar{e}_y\right>/\sqrt{2}$, $\left||ue_x\bar{e}_y+u\bar{e}_xe_y+\bar{u}e_xe_y\right>/\sqrt{3}$, $\left||\bar{u}\bar{e}_xe_y+\bar{u}e_x\bar{e}_y+u\bar{e}_x\bar{e}_y\right>/\sqrt{3}$, while the excited state manifold in the hole picture is composed of the states, $\left||ve_xe_y+i\bar{v}\bar{e}_x\bar{e}_y\right>/\sqrt{2}$, $\left||ve_xe_y-i\bar{v}\bar{e}_x\bar{e}_y\right>/\sqrt{2}$, $\left||ve_x\bar{e}_y+v\bar{e}_xe_y+\bar{v}e_xe_y\right>/\sqrt{3}$, \\$\left||\bar{v}\bar{e}_xe_y+\bar{v}e_x\bar{e}_y+v\bar{e}_x\bar{e}_y\right>/\sqrt{3}$. Above, $u$ and $v$ are single-particle orbitals transforming as the $A_1$ irreducible representation of the $C_{3v}$ point group, while $e_x$ and $e_y$ transform as the $x$ and $y$ components of the two-dimensional $E$ irreducible representation of the $C_{3v}$ point group. The overbar denotes the minority spin state. To calculate the ZPL we take the lowest energy hole states, which from Tables \ref{tab:HSEvalsVSik} and \ref{tab:HSEvalsVSih} we see must be the $\left||\bar{u}\bar{e}_xe_y+\bar{u}e_x\bar{e}_y+u\bar{e}_x\bar{e}_y\right>/\sqrt{3}$ (excited) and $\left||\bar{v}\bar{e}_xe_y+\bar{v}e_x\bar{e}_y+v\bar{e}_x\bar{e}_y\right>/\sqrt{3}$ (ground) states. Explicitly, in the following we denote the eigenvalue for the majority spin single-particle orbital $\gamma$ as $\epsilon_\gamma$ and for the minority spin single-particle orbital $\bar{\gamma}$ as $\epsilon_{\bar{\gamma}}$. Then, the ZPL we obtain for the $k$ site $V_{\rm Si}^-$ is $(2(\epsilon_{\bar{v}}-\epsilon_{\bar{u}})/3+(\epsilon_{v}-\epsilon_{u})/3) = 1.34$~eV (the remaining terms cancel), while the ZPL we obtain for the $h$ site $V_{\rm Si}^-$ is $(2(\epsilon_{\bar{v}}-\epsilon_{\bar{u}})/3+(\epsilon_{v}-\epsilon_{u})/3) = 1.43$~eV (where again the remaining terms cancel) in excellent agreement with experimental values of 1.35~eV and 1.44~eV, respectively.~\cite{Bracher} The slight underestimation of the ZPL values may therefore be due to the slightly smaller HSE06 band gap (3.18~eV compared to about 3.2~eV for experiment~\cite{Paufler}). We note, however, that other theoretical calculations yield 1.44~eV for the $k$ site and 1.54~eV for the $h$ site using the HSE06 functional.~\cite{Ivady2017} We believe error compensation in taking the difference of many eigenvalues may be causing the greater accuracy of our approximation to the $\Delta$SCF method using Janak's theorem than the $\Delta$SCF method itself, though later theoretical work shows better agreement with our work and with experiment.~\cite{Udvarhelyi2020} 

We note that, as the many-body states from the work of Soykal \textit{et al.}~\cite{Soykal} have been defined as Slater determinants of products of single-particle orbitals, the only requirement for the correctness of the calculation outlined above is that these single-particle orbitals be orthogonal, which is consistent with the orbitals constructed in the work of Soykal \textit{et al.}~\cite{Soykal} Though after performing structural relaxation with the HSE06 functional the expected $C_{3v}$ symmetry is weakly broken, we have used the assignment and energetic ordering of the orbitals from the work of Soykal \textit{et al.}~\cite{Soykal} as such an ordering and assignment is consistent with obtaining the lowest energy spin-conserving excitation of the $S = 3/2$ states that our calculations produce. Ultimately, these results demonstrate that in cases where the wavefunctions for the states that will change occupation do not appreciably differ at locations of significant change in the ionic potential and where the change in the ionic potential can be viewed as a small perturbation, there is no need to calculate excited state properties to obtain the ZPL as values derived entirely from the ground state calculation can provide excellent agreement with experiment. These conditions are satisfied for the single vacancy as none of the defect wavefunctions are appreciable at the location of the removal of the atom and a single vacancy is a perturbation of less than $1\%$ of the integrated ionic potential for the supercell size.

\begin{table}[ht!]
\caption{DFT eigenvalues in eV calculated in the ground state using the HSE06 functional for the hole single-particle states of the $V_{\rm Si}^-(k)$ point defect in $4H$-SiC.}
\centering
\vspace{1 mm}
\begin{tabular}{|c|cc|}
\hline\hline 
single-particle state &  majority spin & minority spin \\
\hline
 $u$ &      $7.553$  &   $8.116$  \\
 $v$ &      $7.773$  &  $10.011$ \\
 $e_x$ &      $7.828$  &  $10.152$ \\
 $e_y$ &      $7.855$  &  $10.225$ \\
\hline
\hline
\end{tabular}
\label{tab:HSEvalsVSik}
\end{table}

\begin{table}[ht!]
\caption{ DFT eigenvalues in eV calculated in the ground state using the HSE06 functional for the hole single-particle states of the $V_{\rm Si}^-(h)$ point defect in $4H$-SiC.}
\centering
\vspace{1 mm}
\begin{tabular}{|c|cc|}
\hline\hline 
single-particle state &  majority spin & minority spin \\
\hline
 $u$ &    $7.551$  &  $8.119$  \\
 $v$ &    $7.800$  & $10.137$ \\
 $e_x$ &    $7.871$  & $10.227$ \\
 $e_y$ &    $7.906$  & $10.289$ \\
\hline
\hline
\end{tabular}
\label{tab:HSEvalsVSih}
\end{table}

\vspace{110pt}
Based on the results of the calculations for silicon monovacancies in $4H$-SiC outlined above, Janak's theorem shows promise for rapid estimation of ZPL values. Indeed, we will see below that the accuracy of Janak's theorem is maintained for the strain ($\varepsilon$) values with the two lowest total energies, namely the uncompressed and the $\varepsilon = +2\%$ structures, for the singly negatively charged calcium vacancy in SiS$_2$ and that the error calculations we have outlined in Section \ref{sec:calcmethod} provide a clear indication of when Janak's theorem fails.

\subsection{Singly Negatively Charged Calcium Vacancy in SiS$_2$\label{sec:CaV}}
\begin{figure}[ht!] 
\centering
\includegraphics[width=0.9\textwidth]{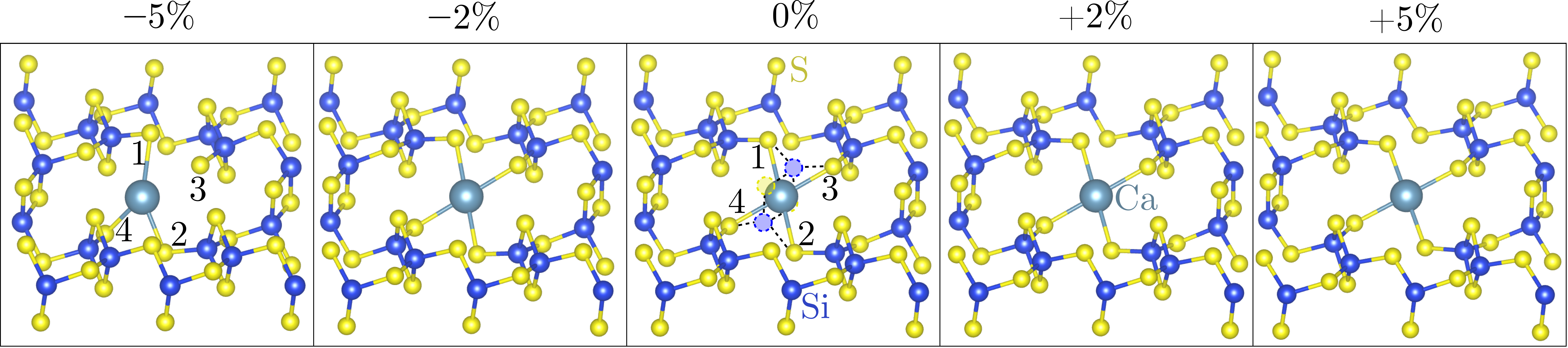}
\caption{Shown above is the structure of a singly negatively charged calcium vacancy using the PBE functional for $\varepsilon = \pm2, \pm5$ and the uncompressed structure. Silicon atoms are in blue, sulfur atoms are in yellow and the calcium atom is in cyan. Silicon and sulfur vacancies are shown as blue and yellow dashed circles, respectively, for the uncompressed structure. The sulfur atoms closest to the calcium atom in the uncompressed structure are numbered as 1-4.} 
\label{fig:structCaV}
\end{figure}

As alluded to above, we now turn to demonstrating the continued accuracy of Janak's theorem for the $\varepsilon$ values with the two lowest total energies for the singly negatively charged calcium vacancy in SiS$_2$. We additionally show that the associated error consistently identifies the larger discrepancies between the results of the theorem and the results of $\Delta$SCF calculations. We also introduce results showing faster convergence of excited state SCF calculations when the charge density is initialized by mixing the HOMO and LUMO compared to when it is initialized from a superposition of atomic charges. For the lattice parameters of the stoichiometric unit cell SiS$_2$ with space group P$2_1$/c using the PBE functional we find $a = 5.93$~\AA, $b = 8.13$~\AA, $\alpha = 90^{\circ}$, $\beta = 102.57^\circ$, $\gamma = 90^\circ$ and a monolayer thickness of $3.65$~\AA~with a vacuum of $16.94$~\AA. Using the HSE06 functional we find $a = 5.88$~\AA, $b = 8.04$~\AA, $\alpha = 90^{\circ}$, $\beta = 103.11^\circ$, $\gamma = 90^\circ$ and a monolayer thickness of $3.59$~\AA~with a vacuum of $16.67$~\AA. The lattice parameters with strain are included in Table \ref{tab:SiS2latparam}.
\begin{table}[ht!]
\caption{PBE lattice constants in~\AA,~and corresponding angles, with strain. The monolayer thickness was not computed as the lattice parameters were directly applied to the defect-containing supercell structures, which were then relaxed at constant volume.}
\centering
\vspace{1 mm}
\begin{tabular}{|c|c|c|c|c|c|c|}
\hline\hline 
 $\varepsilon$ &  $a$ & $b$ & $c$ & $\alpha$ & $\beta$ & $\gamma$ \\
\hline
$+5\%$ & $6.22$ & $8.53$ & $21.09$ & $90^\circ$ & $102.57^\circ$ & $90^\circ$ \\
$+2\%$ & $6.05$ & $8.29$ & $21.09$ & $90^\circ$ & $102.57^\circ$ & $90^\circ$  \\
$-2\%$ & $5.81$ & $7.96$ & $21.09$ & $90^\circ$ & $102.57^\circ$ & $90^\circ$ \\
$-5\%$ & $5.63$ & $7.72$ & $21.09$ & $90^\circ$ & $102.57^\circ$ & $90^\circ$  \\
\hline
\hline
\end{tabular}
\label{tab:SiS2latparam}
\end{table}
Structures for the singly negatively charged calcium vacancy for five strain values are found in Fig. \ref{fig:structCaV}. The values in Table \ref{tab:CaVdists} show that compressive strain tends to destroy the $C_i$ symmetry present in the uncompressed, $\varepsilon = +2\%$ and $\varepsilon =+5\%$ structures, due to buckling of the system.

\begin{table}[ht!]
\caption{PBE distances in \AA,~$d(X_1-X_2)$, between sulfur atoms numbered in Fig. \ref{fig:structCaV} and the calcium atom for $\varepsilon = \pm2, \pm5$ and the uncompressed structure as well as the smallest angles, $\theta(X_1-X_2-X_3)$, between the bonds joining the calcium atom to opposite sulfur atoms. HSE06 values are in brackets.}
\centering
\vspace{1 mm}
\begin{tabular}{|c|c|c|c|c|c|c|}
\hline\hline 
 $\varepsilon$ &  $d({\rm Ca}-{\rm S}_1)$ & $d({\rm Ca}-{\rm S}_2)$ & $d({\rm Ca}-{\rm S}_3)$ & $d({\rm Ca}-{\rm S}_4)$ & $\theta({\rm S}_1-{\rm Ca}-{\rm S}_2)$ & $\theta({\rm S}_3-{\rm Ca}-{\rm S}_4)$ \\
\hline
$+5\%$ & $2.76$ & $2.76$ & $2.83$ & $2.83$ & $179.99^\circ$ & $179.99^\circ$ \\
$+2\%$ & $2.75$ & $2.75$ & $2.80$ & $2.80$ & $179.99^\circ$ & $179.98^\circ$  \\
$0\%$ & $2.76$ ($2.75$) & $2.76$ $(2.75)$ & $2.79$ ($2.79$) & $2.79$ ($2.79$) & $179.98^\circ$ ($179.93^\circ$) & $179.97^\circ$  ($179.89^\circ$)\\
$-2\%$ & $2.76$ & $2.76$ & $2.79$ & $2.79$ & $179.92^\circ$ & $179.88^\circ$  \\
$-5\%$ & $2.97$ & $2.70$ & $3.03$ & $2.72$ & $150.28^\circ$ & $138.39^\circ$  \\

\hline
\hline
\end{tabular}
\label{tab:CaVdists}
\end{table}

\begin{figure}[ht!] 
\centering
\includegraphics[width=0.55\textwidth]{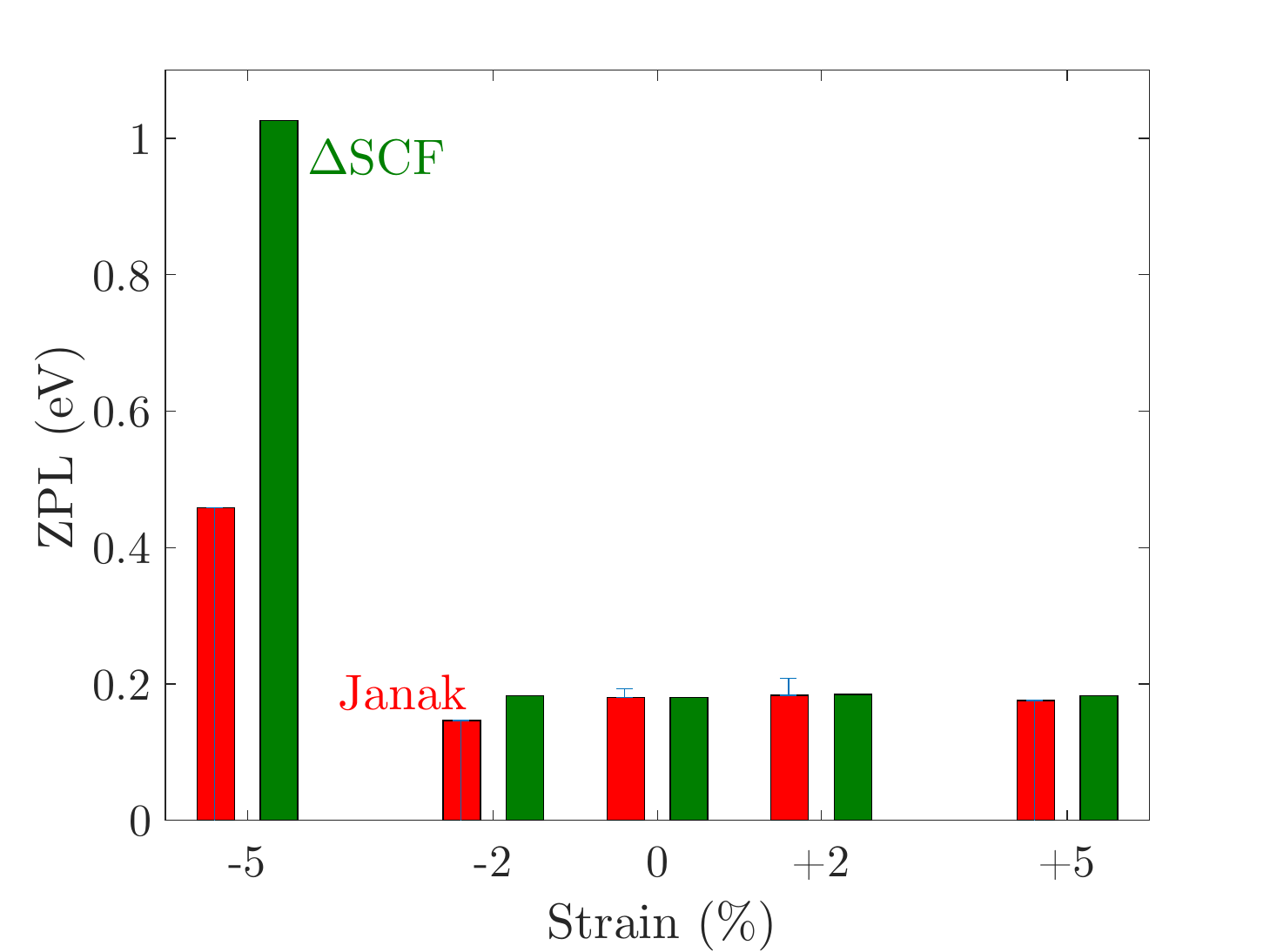}
\caption{ZPL values in eV, calculated using Janak's theorem (red) and using the $\Delta$SCF method (green), for the singly negatively charged calcium vacancy in SiS$_2$ for in-plane $\varepsilon = \pm2\%, \pm5\%$ and for the uncompressed structure. The error associated with the result from Janak's theorem was calculated as described in Section \ref{sec:disc}. The $\varepsilon = -5\%$ case caused the Ca atom to depart from the position that preserved inversion symmetry and was therefore not similar to the other structures.} 
\label{fig:ZPLCavals}
\end{figure}

The approximation to the ZPL using Janak's theorem and using the $\Delta$SCF method for different in-plane $\varepsilon$ can be found in Fig. \ref{fig:ZPLCavals}, where the corresponding eigenvalues can be found in Table \ref{tab:HSEPBEvalsSiS2}. The difference between the LUMO and the HOMO eigenvalues is used in the application of Janak's theorem. We observe little variation of the approximation to the ZPL with tensile $\varepsilon$, but more variation with compressive $\varepsilon$ which can also distort the structure. Since, due to Poisson's ratio, the application of pressure should lead to tensile in-plane $\varepsilon$, we do not expect the ZPL value to change significantly from what we have predicted in experimentally realizable structures. The error calculations perform best for the uncompressed (error of $0.0128$~eV) and $\varepsilon = +2\%$ (error of $0.0249$~eV) structures, which have the smallest energy differences, obtaining the correct sign of the error as well, but fail to capture the correct sign and are much too large for the remaining $\varepsilon$ values (errors of $-17.4096$~eV, $-0.3898$~eV and $-0.6123$~eV for $\varepsilon = -5\%, -2\%$ and $+5\%$, respectively). We note that for $\varepsilon = -5\%$, the Ca atom departed from the position that preserved inversion symmetry and that strain value was therefore not similar to the other structures. The larger errors nonetheless correctly indicate more significant disagreement between the $\Delta$SCF result and the result from using Janak's theorem. We note that based on a statistical learning based prediction of the bulk modulus,~\cite{deJong2016,Evers2015} the structures that should have the two lowest total energies given the requirement of an applied pressure of 2.8 GPa to 3.5 GPa are the uncompressed and $\varepsilon = +2\%$ structures, which is verified by QUANTUM ESPRESSO total energy calculations. Naturally, the uncompressed structure has the lower energy of the two structures.

\begin{figure}[ht!] 
\centering
\includegraphics[width=0.5\textwidth]{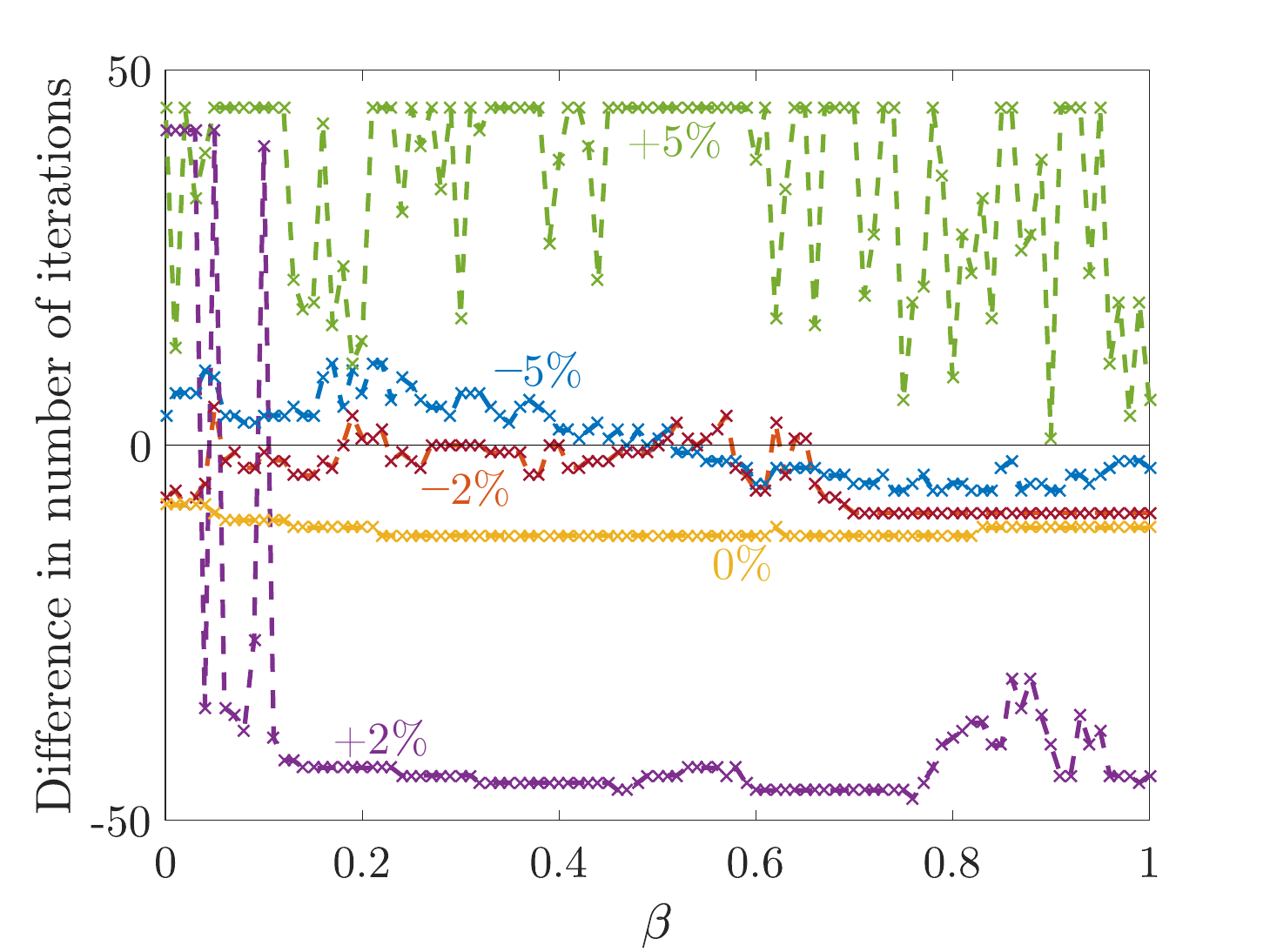}
\caption{Change in the number of iterations to convergence using the charge density initialization presented in Eq. (\ref{eq:betaHL}) and using the default initialization in QUANTUM ESPRESSO. The zero represents the default initialization or standard method so that a data point with a value along the vertical axis of $-7$ indicates that the particular run converged in $7$ fewer iterations than the standard method for the appropriate strain value. In-plane $\varepsilon = \pm2\%, \pm5\%$, where $\varepsilon$ denotes strain, and the uncompressed structure were investigated. The maximum number of iterations was set at 100, which was attained for both $\varepsilon = +2\%$ and $\varepsilon = +5\%$ and explains the plateaus. We took $0 \leq \beta \leq 1$ and used increments of 0.01. Data points are indicated by `$\times$' and the dashed lines are a guide for the eye. The number of iterations required to converge the ground state SCF calculation is included in the text for all $\varepsilon$ values for completeness.} 
\label{fig:beta_conv}
\end{figure}

As shown in Fig. \ref{fig:beta_conv}, these structures with the two lowest total energies also responded best to performing the excited state calculation by replacing the initialization of the charge density from a superposition of atomic charges with the charge density of the ground state calculation, where the HOMO contribution is modified according to, 
\begin{equation}
\label{eq:betaHL}
\left|{\rm HOMO}\right> \rightarrow \beta\left|{\rm HOMO}\right>+(1-\beta)\left|{\rm LUMO}\right>,
\end{equation}
$0\leq \beta \leq 1$. We note that a ground state relaxation is needed for both the standard method or default initialization and the approach of initializing the charge density from a mixture of the HOMO and LUMO orbitals for the excited state SCF calculation. This prerequisite calculation therefore does not change the net increase or decrease in the number of iterations for one method over another. Given that there may be instances where a ground state relaxation is not performed, we report the number of iterations required to converge the ground state SCF calculation for completeness. For $\varepsilon = -5\%, -2\%, +2\%, +5\%$ and the uncompressed structure, the number of iterations required were 29, 35, 34, 31, and 35, respectively.
The value of the band-gap for the uncompressed structure using the PBE functional is 3.55~eV and the defect has total spin $S = 1/2$. The $\Delta$SCF calculations and the differences in ground state eigenvalues were from QUANTUM ESPRESSO.

\begin{table}[ht!]
\caption{DFT eigenvalues in eV calculated in the ground state of the negatively charged calcium vacancy in SiS$_2$ using the PBE and HSE06 functionals for various $\varepsilon$ values.}
\centering
\vspace{1 mm}
\begin{tabular}{|c|ccccc|c|}
\hline\hline 
& \multicolumn{5}{c|}{ PBE}  & HSE06 \\
\hline
  $\varepsilon$ & $-5$\% & $-2$\% & uncompressed & $+2$\% & $+5$\% & uncompressed \\
\hline
 HOMO &  $-2.2137$  & $-2.8524$  & $-3.1014$ & $-3.2690$ & $-3.4880$ & $-3.1995$   \\
 LUMO  &  $-1.7557$  & $-2.7062$  & $-2.9211$ & $-3.0855$ & $-3.3124$ & $-2.9319$   \\
\hline
\hline
\end{tabular}
\label{tab:HSEPBEvalsSiS2}
\end{table}

\begin{figure}[ht!] 
\centering
\includegraphics[width=0.55\textwidth]{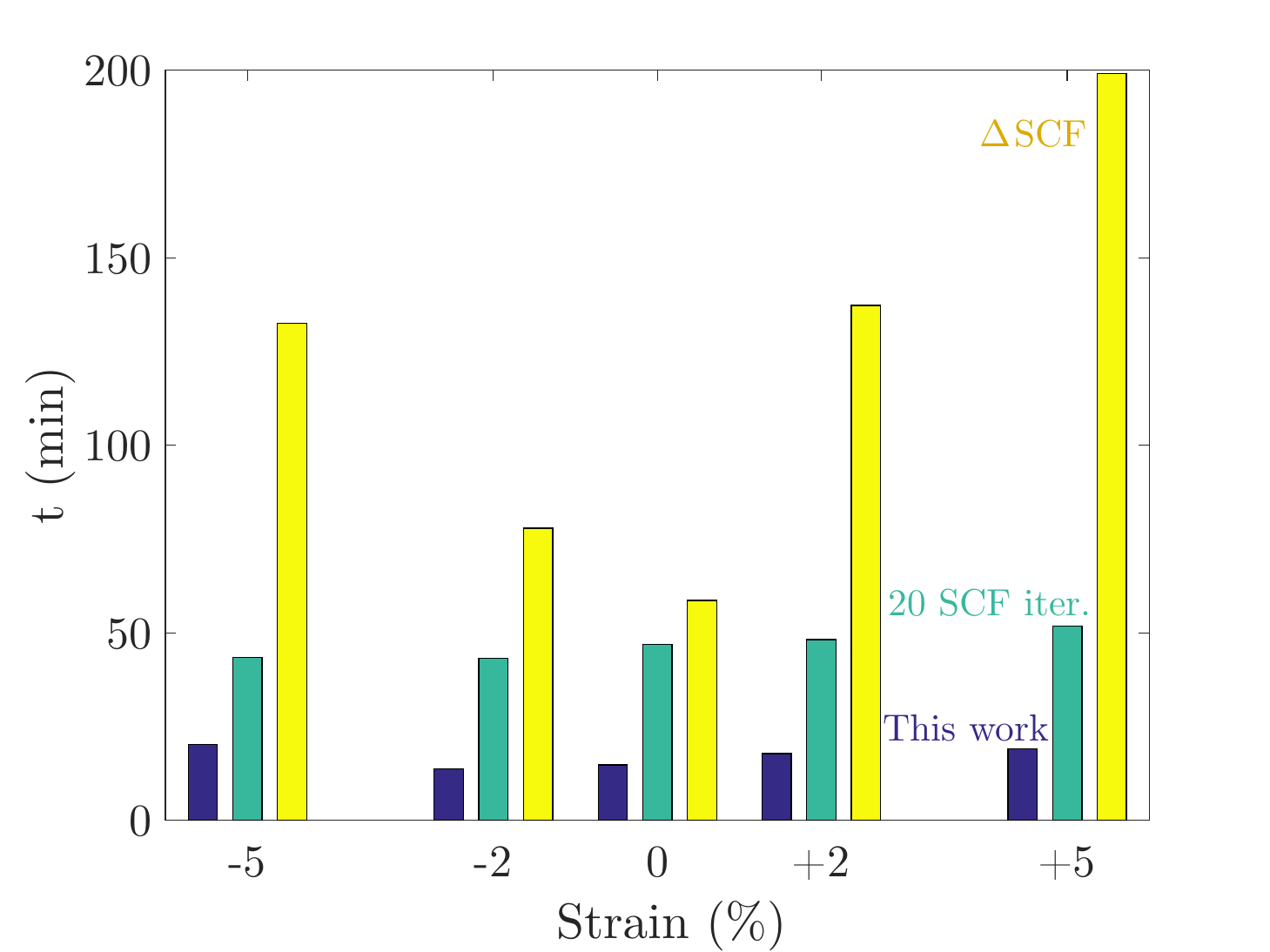}
\caption{Runtimes for convergence of the constrained-occupation calculation for the excited state (yellow), for completion of 20 SCF iterations (turquoise) and for calculation of the error associated wth using Janak's theorem using our method (blue). As our error calculations were done using the VASP code and the constrained-occupation calculations for the excited state were done using the QUANTUM ESPRESSO code, we took the time for a single SCF iteration for a given system from VASP and multiplied by the number of iterations required for the constrained-occupation calculation for the excited state in QUANTUM ESPRESSO to converge.} 
\label{fig:times}
\end{figure}

Based on the computational efficiency of our method for determining the error associated with using Janak's theorem demonstrated in Fig. \ref{fig:times}, we argue that the approach of using Janak's theorem should be acceptable for screening large numbers of potential point defect single-photon emitter candidates for desired ZPL values. Our approach to the initialization of the charge density also shows promise. A key point is that the search would focus on point defect candidates with total spin $S = 1/2$, otherwise more involved group theoretic arguments would be required to determine the correct many-body hole wavefunction as in the work of Soykal \textit{et al.}~\cite{Soykal} Other specific systems, such as triplet systems,~\cite{smart2020intersystem} will also be the focus of future work. The band-gap for the uncompressed structure using the HSE06 functional is 4.75~eV and we again find that the defect has total spin $S = 1/2$. Using QUANTUM ESPRESSO, the HSE06 calculation yields a $\Delta$SCF result of 0.1705~eV and a difference in ground state eigenvalues of 0.2676~eV. The error calculated as described in Section \ref{sec:disc} yields 0.3008~eV. For hosts where the accuracy of hybrid functionals has been demonstrated, clearly their continued use would be recommended. However, one cannot generally state that hybrid functionals will be more accurate than semilocal ones. Indeed, Johari and Shenoy~\cite{Johari2012} find that their PBE band gap for monolayer MoS$_2$ underestimates the experimental band gap of 1.8~eV by just 0.12~eV, while Kuc \textit{et al.}~\cite{Kuc2011} and Ataca and Ciraci~\cite{Ataca2011} found that the PBE0 and HSE06 hybrid functionals overestimate this band gap by approximately 1 and 0.45~eV, respectively. In our work, we have therefore included both the PBE and HSE06 ZPL transitions for the uncompressed structure to reduce the possibility of missing the transition, noting that the transition energy does not vary significantly with strain for low total energy strain configurations. We recommend that any application of our methods using a semilocal functional also be performed with a hybrid functional.

\begin{figure}[ht!] 
\centering
\includegraphics[width=0.5\textwidth]{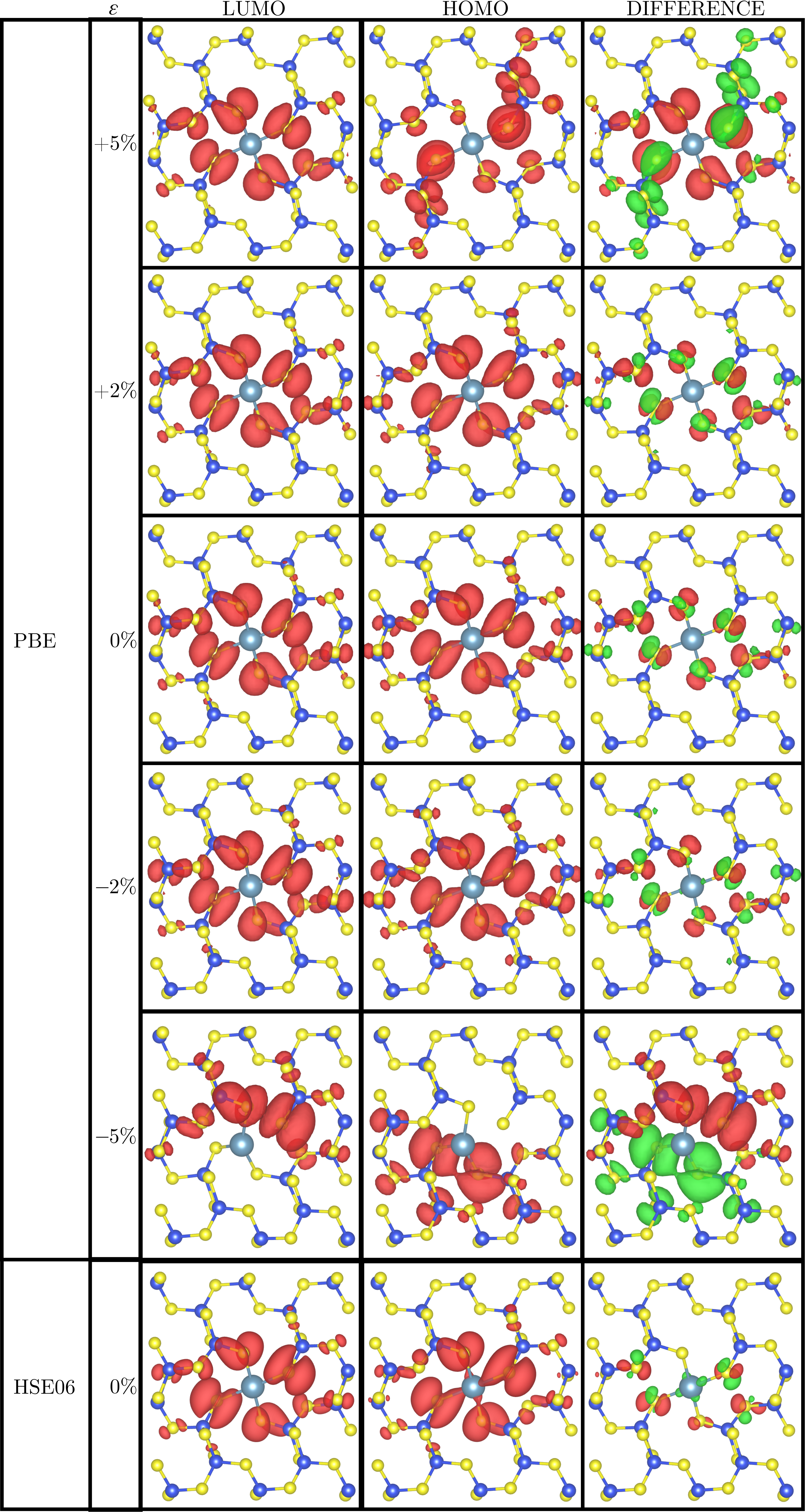}
\caption{DFT calculated partial charge densities for singly negatively charged calcium vacancy in SiS$_2$ using the PBE functional for in-plane $\varepsilon = \pm2\%, \pm5\%$ and for the uncompressed structure and using the HSE06 functional for the uncompressed structure are displayed corresponding to: the LUMO (left), the HOMO (middle) and the difference between the total charge densities with and without exchanging the occupations of the HOMO and LUMO (right). The value of $\varepsilon$ is indicated to the left of the respective panels. Sulfur atoms are in yellow, silicon atoms are in blue and the calcium atom is in cyan. Charge accumulation (depletion) is indicated by translucent red (green). The isosurface of charge density is 0.0005 $e/$\AA$^3$ for all plots.} 
\label{fig:eigenCa}
\end{figure}

\begin{figure}[ht!] 
\centering
\includegraphics[width=0.5\textwidth]{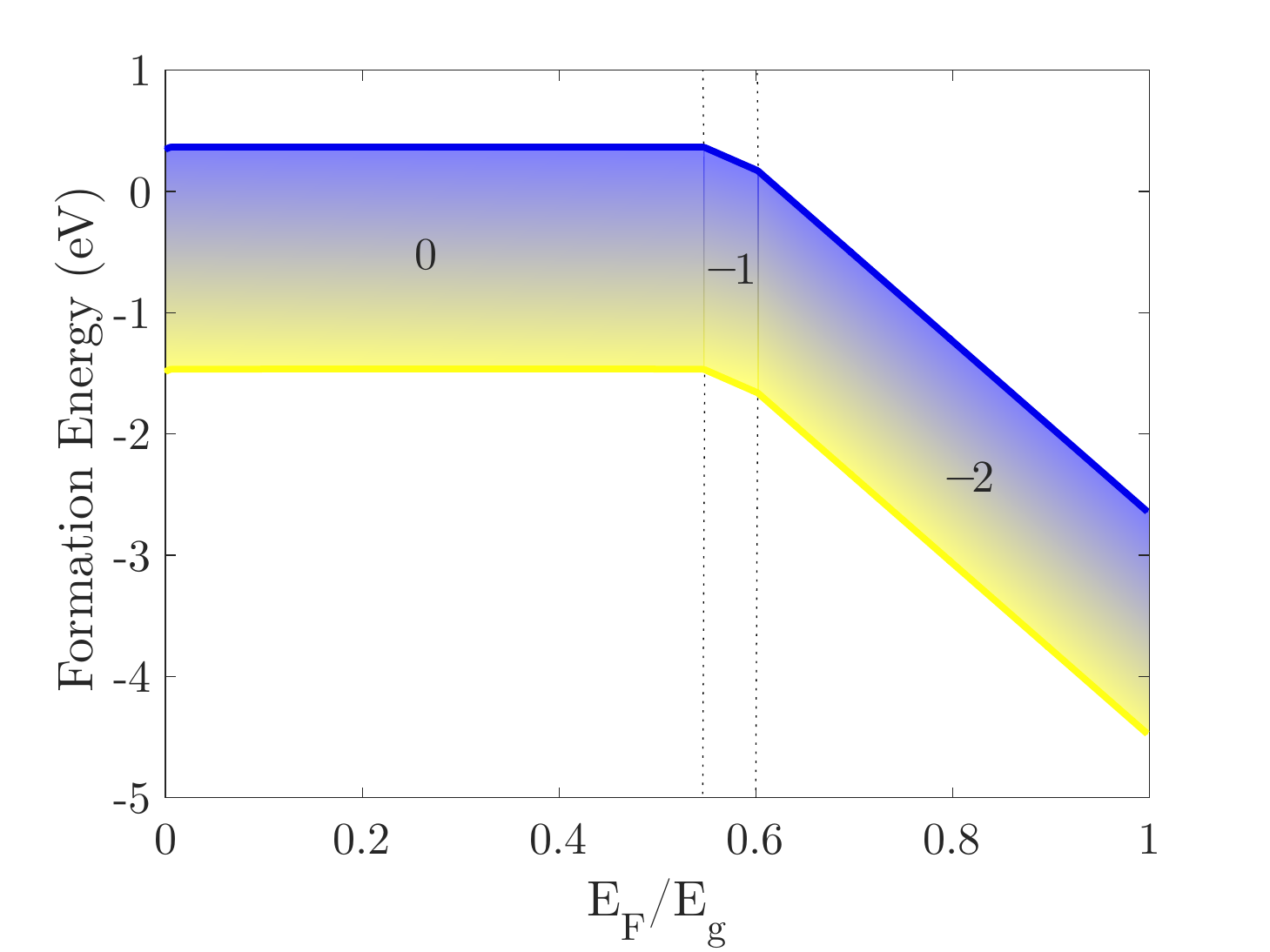}
\caption{Formation energy as a function of Fermi level for the calcium vacancy for a PBE DFT calculated gap $E_g = 3.55$ eV. The lower line (yellow) corresponds to sulfur-rich preparation conditions while the upper line (blue) corresponds to silicon-rich preparation conditions. The integers between the lines indicate the most stable charge state of the defect at the corresponding values of the Fermi level, with each region bounded by vertical dotted lines. Calculations were done for the uncompressed structure.} 
\label{fig:formCa}
\end{figure}

The DFT partial charge densities obtained for ground hole (LUMO) and excited hole (HOMO) states and differences (LUMO $-$ HOMO) are shown in Fig. \ref{fig:eigenCa}. The differences show the closeness of certain LUMO-HOMO pairs. More importantly, Fig. \ref{fig:eigenCa} shows that Janak's theorem succeeds when the differences are not appreciable at the location of the greatest change in the ionic potential of the perfect lattice (at the location of the Ca atom) and does worse otherwise. The structure of the defect is four missing atoms, two silicon and two sulfur, replaced by a single calcium atom such that the resulting point defect is inversion-symmetric, as confirmed by the partial charge densities in Fig. \ref{fig:eigenCa}. This inversion symmetry is broken, however, for in-plane $\varepsilon = -5\%$. The lowest formation energy as a function of Fermi level is displayed in Fig. \ref{fig:formCa} for the uncompressed structure. The plot demonstrates that in sulfur-rich preparation conditions the introduction of a calcium vacancy actually stabilizes the SiS$_2$ structure with space group P$2_1$/c. Given the fact that the singly negatively charged calcium vacancy only exists for a very limited range of Fermi level values, it would be necessary to pin the Fermi level at the appropriate value by doping the system. Using a $^{43}$Ca atom instead of a $^{40}$Ca atom could also lead to coupling between the nuclear spin and the electronic spin to implement a long-lived quantum memory realized with the nuclear spin.~\cite{Pfender2017} 

\begin{figure}[ht!] 
\centering
\includegraphics[width=0.5\textwidth]{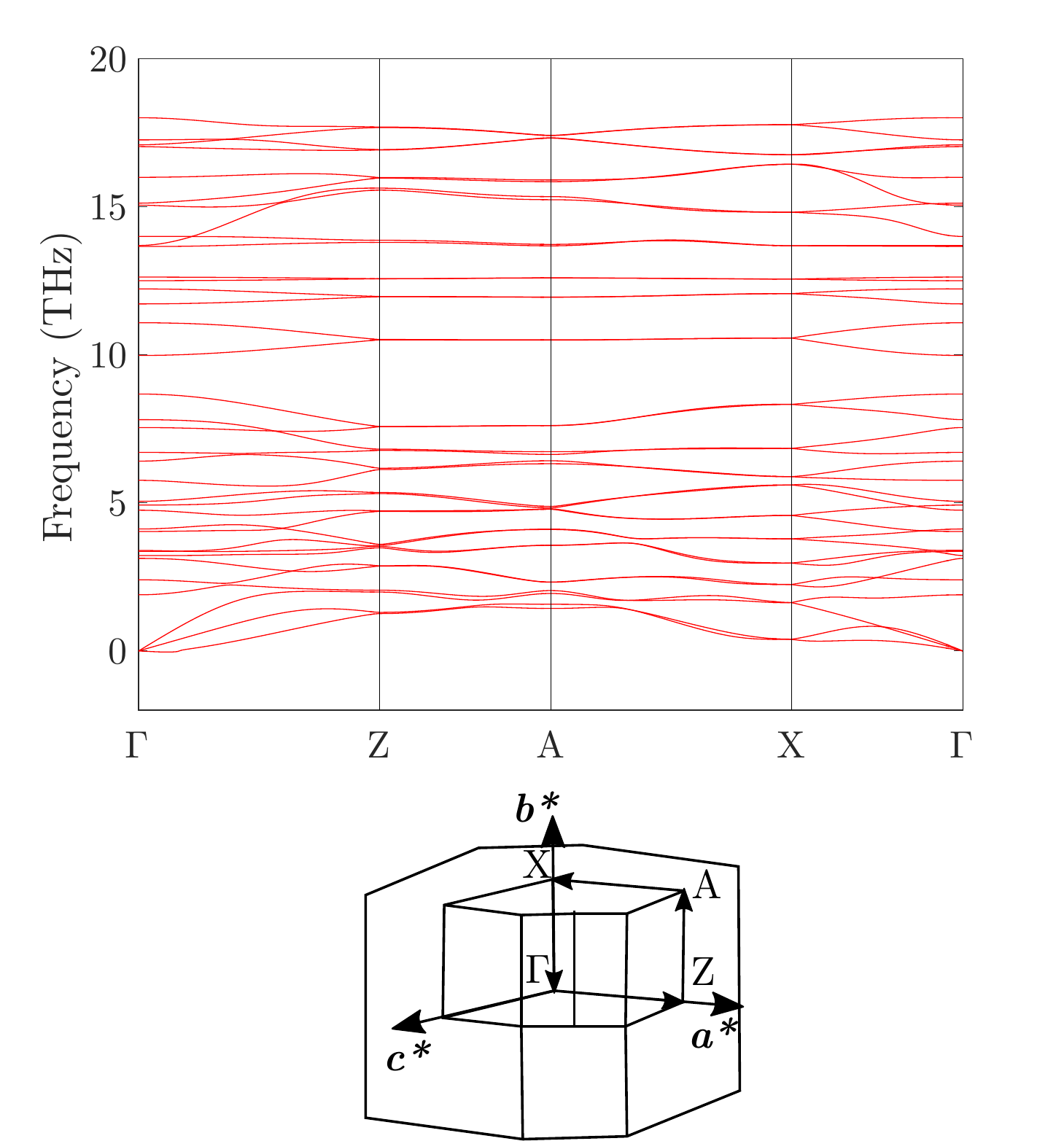}
\caption{Phonon dispersion for the uncompressed P$2_1$/c structure of SiS$_2$ along the indicated high-symmetry path, as calculated with the phonopy code.~\cite{Togo} The structure does show imaginary frequencies near the $\Gamma$ point and would therefore need to be maintained at the 2.8 GPa required for its synthesis to be stable.~\cite{Plasienka2016} The vectors $\mathbf{a^*}$, $\mathbf{b^*}$ and $\mathbf{c^*}$ are reciprocal to the vectors corresponding to the lattice constants $a$, $b$ and the out-of-plane lattice constant, respectively.} 
\label{fig:phondisp}
\end{figure}

We now turn to addressing the possibility of non-radiative decay with phonon emission upon excitation of the singly negatively charged calcium vacancy. Unlike in diamond where optical phonon modes with energies of about 0.17~eV can be found,~\cite{Sato2020} the energies of the phonon modes in SiS$_2$ with space group P$2_1$/c are all below 20 THz as shown in Fig. \ref{fig:phondisp}, corresponding to an energy of about 0.08 eV. This result indicates that non-radiative decay would require multi-phonon processes, which would be rare.

 We note that unlike diamond which is routinely synthesized at high pressure but shows no dynamical instability at ambient condition, our calculations show (see Fig. \ref{fig:phondisp}) that SiS$_2$ with space group P$2_1$/c does exhibit a dynamical instability at ambient condition and would therefore need to be maintained at the 2.8 GPa required for its synthesis to be stable. We would like to make the following points: Experiments at high pressure ($> 10^2$~GPa) are routinely performed. In particular, the discovery of room-temperature superconductors at high pressure (e.g. Snider \textit{et al.}~\cite{Snider2020} at 267 GPa) will undoubtedly propel further technical advances in this area. Experiments of emitters at high pressure (e.g. NV centers in diamond) have also been performed (e.g. Lesik \textit{et al.},~\cite{Lesik2019} Yip \textit{et al.},~\cite{Yip2019} and Hsieh \textit{et al.}~\cite{Hsieh2019}), showing such experiments to be within current feasibility. While it is not ideal that the proposed candidate material requires high pressure, it does not remove one of the key elements of the manuscript $-$ the tools for the rapid screening of defect qubits.

In summary, the case of the singly negatively charged calcium vacancy in SiS$_2$ confirms that using Janak's theorem and the associated error estimation for ZPL calculations is quick and clearly signals when the theorem is applicable. We have also shown that the constrained occupation calculation for the excited state can be completed in some cases with fewer SCF iterations by carefully initializing the charge density.

\section{CONCLUSION \label{sec:conc}}
In conclusion, we have developed a novel computational framework for high-throughput screening of the ZPL of atom-like qubits with applications for materials discovery, quantum sensing and quantum computing. We propose the use of Janak's theorem for ZPL calculations and provide a quick new method for estimating the associated error. We also outline a new method for reducing the number of iterations required to converge excited state SCF calculations, compared to the default procedure for carrying out such calculations. Our ZPL results are consistent with previous experimental work where available and suggest that considering only ground state eigenvalues in the hole approach is a computationally efficient way of calculating optical excitation energies of color centers for screening purposes. Our novel method for initializing the charge density to reduce the number of iterations required for excited state calculations also works well for systems with small strain. Finally, we propose the new singly negatively charged calcium vacancy in SiS$_2$, which has the advantage of being inversion-symmetric and of containing a low density of nuclear spins that would lead to poor coherence properties of spins hosted in the material. However, this new proposed system would need to be maintained at the 2.8 GPa required for its synthesis to ensure stability. We also note that our rapid screening methods can create training data for machine learning and serve in technologies for predicting defect structures with desirable properties. 

\section*{ ACKNOWLEDGMENTS:}

We thank Efthimios Kaxiras of Harvard University and Georgios A. Tritsaris for useful discussions. R.K.D. gratefully acknowledges financial support from the Princeton Presidential Postdoctoral Research Fellowship. We also acknowledge support by the STC Center for Integrated Quantum Materials, NSF Grant No. DMR-1231319. This work used computational resources of the Extreme Science and Engineering Discovery Environment (XSEDE), which is supported by National Science Foundation Grant Number ACI-1548562,~\cite{Towns} on Stampede2 at TACC through allocation TG-DMR120073, and of the National Energy Research Scientific Computing Center (NERSC), a U.S. Department of Energy Office of Science User Facility operated under Contract No. DE-AC02-05CH11231. 

\section*{DATA AVAILABILITY:} 

The data from DFT calculations that support the findings of this study and the modified source code for our charge density implementation are available from Rodrick Kuate Defo (email: rkuatedefo@princeton.edu) upon reasonable request.

\section*{APPENDIX: COMPUTATIONAL DETAILS}

We provide in this Appendix details of the computations. The atomic positions were relaxed until the magnitude of the Hellmann-Feynman forces was smaller than $10^{-4}$~eV$\cdot$\AA$^{-1}$ on each atom without spin-polarization and subsequently until the magnitude of the Hellmann-Feynman forces was smaller than $10^{-2}$~eV$\cdot$\AA$^{-1}$ on each atom with spin polarization to obtain defect levels and, for the stoichiometric conventional unit cell, the lattice parameters were concurrently relaxed. The wavefunctions were expanded in a plane wave basis with a cutoff energy of 500~eV for all systems and a Monkhorst-Pack grid of $6\times6\times2$ k-points was used for integrations in reciprocal space for SiS$_2$ and a Gamma centered grid of $4\times4\times2$ k-points was used for integrations in reciprocal space of $4H$-SiC. The relaxed lattice parameters of the stoichiometric unit cell were then used for the supercell structures. Formation energies and defect levels were calculated using a supercell of 108 atoms for SiS$_2$ ($3\times3\times1$ multiple of the stoichiometric unit cell) with appropriately scaled k-point grids. For $4H$-SiC, a supercell of 576 atoms ($6\times6\times2$ multiple of the stoichiometric unit cell) was used.

\bibliography{refs_BN}

\begin{thebibliography}{83}%
\makeatletter
\providecommand \@ifxundefined [1]{%
 \@ifx{#1\undefined}
}%
\providecommand \@ifnum [1]{%
 \ifnum #1\expandafter \@firstoftwo
 \else \expandafter \@secondoftwo
 \fi
}%
\providecommand \@ifx [1]{%
 \ifx #1\expandafter \@firstoftwo
 \else \expandafter \@secondoftwo
 \fi
}%
\providecommand \natexlab [1]{#1}%
\providecommand \enquote  [1]{``#1''}%
\providecommand \bibnamefont  [1]{#1}%
\providecommand \bibfnamefont [1]{#1}%
\providecommand \citenamefont [1]{#1}%
\providecommand \href@noop [0]{\@secondoftwo}%
\providecommand \href [0]{\begingroup \@sanitize@url \@href}%
\providecommand \@href[1]{\@@startlink{#1}\@@href}%
\providecommand \@@href[1]{\endgroup#1\@@endlink}%
\providecommand \@sanitize@url [0]{\catcode `\\12\catcode `\$12\catcode
  `\&12\catcode `\#12\catcode `\^12\catcode `\_12\catcode `\%12\relax}%
\providecommand \@@startlink[1]{}%
\providecommand \@@endlink[0]{}%
\providecommand \url  [0]{\begingroup\@sanitize@url \@url }%
\providecommand \@url [1]{\endgroup\@href {#1}{\urlprefix }}%
\providecommand \urlprefix  [0]{URL }%
\providecommand \Eprint [0]{\href }%
\providecommand \doibase [0]{http://dx.doi.org/}%
\providecommand \selectlanguage [0]{\@gobble}%
\providecommand \bibinfo  [0]{\@secondoftwo}%
\providecommand \bibfield  [0]{\@secondoftwo}%
\providecommand \translation [1]{[#1]}%
\providecommand \BibitemOpen [0]{}%
\providecommand \bibitemStop [0]{}%
\providecommand \bibitemNoStop [0]{.\EOS\space}%
\providecommand \EOS [0]{\spacefactor3000\relax}%
\providecommand \BibitemShut  [1]{\csname bibitem#1\endcsname}%
\let\auto@bib@innerbib\@empty
%</preamble>
\bibitem [{\citenamefont {Geim}\ and\ \citenamefont
  {Grigorieva}(2013)}]{Geim2013}%
  \BibitemOpen
  \bibfield  {author} {\bibinfo {author} {\bibfnamefont {A.~K.}\ \bibnamefont
  {Geim}}\ and\ \bibinfo {author} {\bibfnamefont {I.~V.}\ \bibnamefont
  {Grigorieva}},\ }\href {\doibase 10.1038/nature12385} {\bibfield  {journal}
  {\bibinfo  {journal} {Nature}\ }\textbf {\bibinfo {volume} {499}},\ \bibinfo
  {pages} {419} (\bibinfo {year} {2013})}\BibitemShut {NoStop}%
\bibitem [{\citenamefont {Toth}\ and\ \citenamefont
  {Aharonovich}(2019)}]{Toth2019}%
  \BibitemOpen
  \bibfield  {author} {\bibinfo {author} {\bibfnamefont {M.}~\bibnamefont
  {Toth}}\ and\ \bibinfo {author} {\bibfnamefont {I.}~\bibnamefont
  {Aharonovich}},\ }\href {\doibase 10.1146/annurev-physchem-042018-052628}
  {\bibfield  {journal} {\bibinfo  {journal} {Annual Review of Physical
  Chemistry}\ }\textbf {\bibinfo {volume} {70}},\ \bibinfo {pages} {123}
  (\bibinfo {year} {2019})}\BibitemShut {NoStop}%
\bibitem [{\citenamefont {Wu}\ \emph {et~al.}(2018)\citenamefont {Wu},
  \citenamefont {Fatemi}, \citenamefont {Gibson}, \citenamefont {Watanabe},
  \citenamefont {Taniguchi}, \citenamefont {Cava},\ and\ \citenamefont
  {Jarillo-Herrero}}]{Wu76}%
  \BibitemOpen
  \bibfield  {author} {\bibinfo {author} {\bibfnamefont {S.}~\bibnamefont
  {Wu}}, \bibinfo {author} {\bibfnamefont {V.}~\bibnamefont {Fatemi}}, \bibinfo
  {author} {\bibfnamefont {Q.~D.}\ \bibnamefont {Gibson}}, \bibinfo {author}
  {\bibfnamefont {K.}~\bibnamefont {Watanabe}}, \bibinfo {author}
  {\bibfnamefont {T.}~\bibnamefont {Taniguchi}}, \bibinfo {author}
  {\bibfnamefont {R.~J.}\ \bibnamefont {Cava}}, \ and\ \bibinfo {author}
  {\bibfnamefont {P.}~\bibnamefont {Jarillo-Herrero}},\ }\href {\doibase
  10.1126/science.aan6003} {\bibfield  {journal} {\bibinfo  {journal}
  {Science}\ }\textbf {\bibinfo {volume} {359}},\ \bibinfo {pages} {76}
  (\bibinfo {year} {2018})}\BibitemShut {NoStop}%
\bibitem [{\citenamefont {Cao}\ \emph {et~al.}(2018{\natexlab{a}})\citenamefont
  {Cao}, \citenamefont {Fatemi}, \citenamefont {Demir}, \citenamefont {Fang},
  \citenamefont {Tomarken}, \citenamefont {Luo}, \citenamefont
  {Sanchez-Yamagishi}, \citenamefont {Watanabe}, \citenamefont {Taniguchi},
  \citenamefont {Kaxiras}, \citenamefont {Ashoori},\ and\ \citenamefont
  {Jarillo-Herrero}}]{Cao2018corr}%
  \BibitemOpen
  \bibfield  {author} {\bibinfo {author} {\bibfnamefont {Y.}~\bibnamefont
  {Cao}}, \bibinfo {author} {\bibfnamefont {V.}~\bibnamefont {Fatemi}},
  \bibinfo {author} {\bibfnamefont {A.}~\bibnamefont {Demir}}, \bibinfo
  {author} {\bibfnamefont {S.}~\bibnamefont {Fang}}, \bibinfo {author}
  {\bibfnamefont {S.~L.}\ \bibnamefont {Tomarken}}, \bibinfo {author}
  {\bibfnamefont {J.~Y.}\ \bibnamefont {Luo}}, \bibinfo {author} {\bibfnamefont
  {J.~D.}\ \bibnamefont {Sanchez-Yamagishi}}, \bibinfo {author} {\bibfnamefont
  {K.}~\bibnamefont {Watanabe}}, \bibinfo {author} {\bibfnamefont
  {T.}~\bibnamefont {Taniguchi}}, \bibinfo {author} {\bibfnamefont
  {E.}~\bibnamefont {Kaxiras}}, \bibinfo {author} {\bibfnamefont {R.~C.}\
  \bibnamefont {Ashoori}}, \ and\ \bibinfo {author} {\bibfnamefont
  {P.}~\bibnamefont {Jarillo-Herrero}},\ }\href {\doibase 10.1038/nature26154}
  {\bibfield  {journal} {\bibinfo  {journal} {Nature}\ }\textbf {\bibinfo
  {volume} {556}},\ \bibinfo {pages} {80} (\bibinfo {year}
  {2018}{\natexlab{a}})}\BibitemShut {NoStop}%
\bibitem [{\citenamefont {Cao}\ \emph {et~al.}(2018{\natexlab{b}})\citenamefont
  {Cao}, \citenamefont {Fatemi}, \citenamefont {Fang}, \citenamefont
  {Watanabe}, \citenamefont {Taniguchi}, \citenamefont {Kaxiras},\ and\
  \citenamefont {Jarillo-Herrero}}]{Cao2018unc}%
  \BibitemOpen
  \bibfield  {author} {\bibinfo {author} {\bibfnamefont {Y.}~\bibnamefont
  {Cao}}, \bibinfo {author} {\bibfnamefont {V.}~\bibnamefont {Fatemi}},
  \bibinfo {author} {\bibfnamefont {S.}~\bibnamefont {Fang}}, \bibinfo {author}
  {\bibfnamefont {K.}~\bibnamefont {Watanabe}}, \bibinfo {author}
  {\bibfnamefont {T.}~\bibnamefont {Taniguchi}}, \bibinfo {author}
  {\bibfnamefont {E.}~\bibnamefont {Kaxiras}}, \ and\ \bibinfo {author}
  {\bibfnamefont {P.}~\bibnamefont {Jarillo-Herrero}},\ }\href {\doibase
  10.1038/nature26160} {\bibfield  {journal} {\bibinfo  {journal} {Nature}\
  }\textbf {\bibinfo {volume} {556}},\ \bibinfo {pages} {43} (\bibinfo {year}
  {2018}{\natexlab{b}})}\BibitemShut {NoStop}%
\bibitem [{\citenamefont {Gruber}\ \emph {et~al.}(1997)\citenamefont {Gruber},
  \citenamefont {Dr{\"a}benstedt}, \citenamefont {Tietz}, \citenamefont
  {Fleury}, \citenamefont {Wrachtrup},\ and\ \citenamefont
  {Borczyskowski}}]{Gruber2012}%
  \BibitemOpen
  \bibfield  {author} {\bibinfo {author} {\bibfnamefont {A.}~\bibnamefont
  {Gruber}}, \bibinfo {author} {\bibfnamefont {A.}~\bibnamefont
  {Dr{\"a}benstedt}}, \bibinfo {author} {\bibfnamefont {C.}~\bibnamefont
  {Tietz}}, \bibinfo {author} {\bibfnamefont {L.}~\bibnamefont {Fleury}},
  \bibinfo {author} {\bibfnamefont {J.}~\bibnamefont {Wrachtrup}}, \ and\
  \bibinfo {author} {\bibfnamefont {C.~v.}\ \bibnamefont {Borczyskowski}},\
  }\href {\doibase 10.1126/science.276.5321.2012} {\bibfield  {journal}
  {\bibinfo  {journal} {Science}\ }\textbf {\bibinfo {volume} {276}},\ \bibinfo
  {pages} {2012} (\bibinfo {year} {1997})}\BibitemShut {NoStop}%
\bibitem [{\citenamefont {Bar-Gill}\ \emph {et~al.}(2013)\citenamefont
  {Bar-Gill}, \citenamefont {Pham}, \citenamefont {Jarmola}, \citenamefont
  {Budker},\ and\ \citenamefont {Walsworth}}]{Gill2013}%
  \BibitemOpen
  \bibfield  {author} {\bibinfo {author} {\bibfnamefont {N.}~\bibnamefont
  {Bar-Gill}}, \bibinfo {author} {\bibfnamefont {L.~M.}\ \bibnamefont {Pham}},
  \bibinfo {author} {\bibfnamefont {A.}~\bibnamefont {Jarmola}}, \bibinfo
  {author} {\bibfnamefont {D.}~\bibnamefont {Budker}}, \ and\ \bibinfo {author}
  {\bibfnamefont {R.~L.}\ \bibnamefont {Walsworth}},\ }\href
  {https://doi.org/10.1038/ncomms2771} {\bibfield  {journal} {\bibinfo
  {journal} {Nat. Commun.}\ }\textbf {\bibinfo {volume} {4}},\ \bibinfo {pages}
  {1743} (\bibinfo {year} {2013})}\BibitemShut {NoStop}%
\bibitem [{\citenamefont {Hensen}\ \emph {et~al.}(2015)\citenamefont {Hensen},
  \citenamefont {Bernien}, \citenamefont {Dr{\'e}au}, \citenamefont {Reiserer},
  \citenamefont {Kalb}, \citenamefont {Blok}, \citenamefont {Ruitenberg},
  \citenamefont {Vermeulen}, \citenamefont {Schouten}, \citenamefont
  {Abell{\'a}n}, \citenamefont {Amaya}, \citenamefont {Pruneri}, \citenamefont
  {Mitchell}, \citenamefont {Markham}, \citenamefont {Twitchen}, \citenamefont
  {Elkouss}, \citenamefont {Wehner}, \citenamefont {Taminiau},\ and\
  \citenamefont {Hanson}}]{Hensen2015}%
  \BibitemOpen
  \bibfield  {author} {\bibinfo {author} {\bibfnamefont {B.}~\bibnamefont
  {Hensen}}, \bibinfo {author} {\bibfnamefont {H.}~\bibnamefont {Bernien}},
  \bibinfo {author} {\bibfnamefont {A.~E.}\ \bibnamefont {Dr{\'e}au}}, \bibinfo
  {author} {\bibfnamefont {A.}~\bibnamefont {Reiserer}}, \bibinfo {author}
  {\bibfnamefont {N.}~\bibnamefont {Kalb}}, \bibinfo {author} {\bibfnamefont
  {M.~S.}\ \bibnamefont {Blok}}, \bibinfo {author} {\bibfnamefont
  {J.}~\bibnamefont {Ruitenberg}}, \bibinfo {author} {\bibfnamefont {R.~F.~L.}\
  \bibnamefont {Vermeulen}}, \bibinfo {author} {\bibfnamefont {R.~N.}\
  \bibnamefont {Schouten}}, \bibinfo {author} {\bibfnamefont {C.}~\bibnamefont
  {Abell{\'a}n}}, \bibinfo {author} {\bibfnamefont {W.}~\bibnamefont {Amaya}},
  \bibinfo {author} {\bibfnamefont {V.}~\bibnamefont {Pruneri}}, \bibinfo
  {author} {\bibfnamefont {M.~W.}\ \bibnamefont {Mitchell}}, \bibinfo {author}
  {\bibfnamefont {M.}~\bibnamefont {Markham}}, \bibinfo {author} {\bibfnamefont
  {D.~J.}\ \bibnamefont {Twitchen}}, \bibinfo {author} {\bibfnamefont
  {D.}~\bibnamefont {Elkouss}}, \bibinfo {author} {\bibfnamefont
  {S.}~\bibnamefont {Wehner}}, \bibinfo {author} {\bibfnamefont {T.~H.}\
  \bibnamefont {Taminiau}}, \ and\ \bibinfo {author} {\bibfnamefont
  {R.}~\bibnamefont {Hanson}},\ }\href {\doibase 10.1038/nature15759}
  {\bibfield  {journal} {\bibinfo  {journal} {Nature}\ }\textbf {\bibinfo
  {volume} {526}},\ \bibinfo {pages} {682} (\bibinfo {year}
  {2015})}\BibitemShut {NoStop}%
\bibitem [{\citenamefont {Siyushev}\ \emph {et~al.}(2017)\citenamefont
  {Siyushev}, \citenamefont {Metsch}, \citenamefont {Ijaz}, \citenamefont
  {Binder}, \citenamefont {Bhaskar}, \citenamefont {Sukachev}, \citenamefont
  {Sipahigil}, \citenamefont {Evans}, \citenamefont {Nguyen}, \citenamefont
  {Lukin}, \citenamefont {Hemmer}, \citenamefont {Palyanov}, \citenamefont
  {Kupriyanov}, \citenamefont {Borzdov}, \citenamefont {Rogers},\ and\
  \citenamefont {Jelezko}}]{Siyushev2017}%
  \BibitemOpen
  \bibfield  {author} {\bibinfo {author} {\bibfnamefont {P.}~\bibnamefont
  {Siyushev}}, \bibinfo {author} {\bibfnamefont {M.~H.}\ \bibnamefont
  {Metsch}}, \bibinfo {author} {\bibfnamefont {A.}~\bibnamefont {Ijaz}},
  \bibinfo {author} {\bibfnamefont {J.~M.}\ \bibnamefont {Binder}}, \bibinfo
  {author} {\bibfnamefont {M.~K.}\ \bibnamefont {Bhaskar}}, \bibinfo {author}
  {\bibfnamefont {D.~D.}\ \bibnamefont {Sukachev}}, \bibinfo {author}
  {\bibfnamefont {A.}~\bibnamefont {Sipahigil}}, \bibinfo {author}
  {\bibfnamefont {R.~E.}\ \bibnamefont {Evans}}, \bibinfo {author}
  {\bibfnamefont {C.~T.}\ \bibnamefont {Nguyen}}, \bibinfo {author}
  {\bibfnamefont {M.~D.}\ \bibnamefont {Lukin}}, \bibinfo {author}
  {\bibfnamefont {P.~R.}\ \bibnamefont {Hemmer}}, \bibinfo {author}
  {\bibfnamefont {Y.~N.}\ \bibnamefont {Palyanov}}, \bibinfo {author}
  {\bibfnamefont {I.~N.}\ \bibnamefont {Kupriyanov}}, \bibinfo {author}
  {\bibfnamefont {Y.~M.}\ \bibnamefont {Borzdov}}, \bibinfo {author}
  {\bibfnamefont {L.~J.}\ \bibnamefont {Rogers}}, \ and\ \bibinfo {author}
  {\bibfnamefont {F.}~\bibnamefont {Jelezko}},\ }\href {\doibase
  10.1103/PhysRevB.96.081201} {\bibfield  {journal} {\bibinfo  {journal} {Phys.
  Rev. B}\ }\textbf {\bibinfo {volume} {96}},\ \bibinfo {pages} {081201(R)}
  (\bibinfo {year} {2017})}\BibitemShut {NoStop}%
\bibitem [{\citenamefont {Becker}\ \emph {et~al.}(2018)\citenamefont {Becker},
  \citenamefont {Pingault}, \citenamefont {Gro\ss{}}, \citenamefont
  {G\"undo\u{g}an}, \citenamefont {Kukharchyk}, \citenamefont {Markham},
  \citenamefont {Edmonds}, \citenamefont {Atat\"ure}, \citenamefont {Bushev},\
  and\ \citenamefont {Becher}}]{Becker2018}%
  \BibitemOpen
  \bibfield  {author} {\bibinfo {author} {\bibfnamefont {J.~N.}\ \bibnamefont
  {Becker}}, \bibinfo {author} {\bibfnamefont {B.}~\bibnamefont {Pingault}},
  \bibinfo {author} {\bibfnamefont {D.}~\bibnamefont {Gro\ss{}}}, \bibinfo
  {author} {\bibfnamefont {M.}~\bibnamefont {G\"undo\u{g}an}}, \bibinfo
  {author} {\bibfnamefont {N.}~\bibnamefont {Kukharchyk}}, \bibinfo {author}
  {\bibfnamefont {M.}~\bibnamefont {Markham}}, \bibinfo {author} {\bibfnamefont
  {A.}~\bibnamefont {Edmonds}}, \bibinfo {author} {\bibfnamefont
  {M.}~\bibnamefont {Atat\"ure}}, \bibinfo {author} {\bibfnamefont
  {P.}~\bibnamefont {Bushev}}, \ and\ \bibinfo {author} {\bibfnamefont
  {C.}~\bibnamefont {Becher}},\ }\href {\doibase
  10.1103/PhysRevLett.120.053603} {\bibfield  {journal} {\bibinfo  {journal}
  {Phys. Rev. Lett.}\ }\textbf {\bibinfo {volume} {120}},\ \bibinfo {pages}
  {053603} (\bibinfo {year} {2018})}\BibitemShut {NoStop}%
\bibitem [{\citenamefont {Elias}\ \emph {et~al.}(2019)\citenamefont {Elias},
  \citenamefont {Valvin}, \citenamefont {Pelini}, \citenamefont {Summerfield},
  \citenamefont {Mellor}, \citenamefont {Cheng}, \citenamefont {Eaves},
  \citenamefont {Foxon}, \citenamefont {Beton}, \citenamefont {Novikov},
  \citenamefont {Gil},\ and\ \citenamefont {Cassabois}}]{Elias2019}%
  \BibitemOpen
  \bibfield  {author} {\bibinfo {author} {\bibfnamefont {C.}~\bibnamefont
  {Elias}}, \bibinfo {author} {\bibfnamefont {P.}~\bibnamefont {Valvin}},
  \bibinfo {author} {\bibfnamefont {T.}~\bibnamefont {Pelini}}, \bibinfo
  {author} {\bibfnamefont {A.}~\bibnamefont {Summerfield}}, \bibinfo {author}
  {\bibfnamefont {C.~J.}\ \bibnamefont {Mellor}}, \bibinfo {author}
  {\bibfnamefont {T.~S.}\ \bibnamefont {Cheng}}, \bibinfo {author}
  {\bibfnamefont {L.}~\bibnamefont {Eaves}}, \bibinfo {author} {\bibfnamefont
  {C.~T.}\ \bibnamefont {Foxon}}, \bibinfo {author} {\bibfnamefont {P.~H.}\
  \bibnamefont {Beton}}, \bibinfo {author} {\bibfnamefont {S.~V.}\ \bibnamefont
  {Novikov}}, \bibinfo {author} {\bibfnamefont {B.}~\bibnamefont {Gil}}, \ and\
  \bibinfo {author} {\bibfnamefont {G.}~\bibnamefont {Cassabois}},\ }\href
  {\doibase 10.1038/s41467-019-10610-5} {\bibfield  {journal} {\bibinfo
  {journal} {Nature Communications}\ }\textbf {\bibinfo {volume} {10}},\
  \bibinfo {pages} {2639} (\bibinfo {year} {2019})}\BibitemShut {NoStop}%
\bibitem [{\citenamefont {Tran}\ \emph {et~al.}(2015)\citenamefont {Tran},
  \citenamefont {Bray}, \citenamefont {Ford}, \citenamefont {Toth},\ and\
  \citenamefont {Aharonovich}}]{Tran2}%
  \BibitemOpen
  \bibfield  {author} {\bibinfo {author} {\bibfnamefont {T.~T.}\ \bibnamefont
  {Tran}}, \bibinfo {author} {\bibfnamefont {K.}~\bibnamefont {Bray}}, \bibinfo
  {author} {\bibfnamefont {M.~J.}\ \bibnamefont {Ford}}, \bibinfo {author}
  {\bibfnamefont {M.}~\bibnamefont {Toth}}, \ and\ \bibinfo {author}
  {\bibfnamefont {I.}~\bibnamefont {Aharonovich}},\ }\href
  {https://doi.org/10.1038/nnano.2015.242} {\bibfield  {journal} {\bibinfo
  {journal} {Nature Nanotechnology}\ }\textbf {\bibinfo {volume} {11}},\
  \bibinfo {pages} {37} (\bibinfo {year} {2015})}\BibitemShut {NoStop}%
\bibitem [{\citenamefont {Iv{\'a}dy}\ \emph {et~al.}(2020)\citenamefont
  {Iv{\'a}dy}, \citenamefont {Barcza}, \citenamefont {Thiering}, \citenamefont
  {Li}, \citenamefont {Hamdi}, \citenamefont {Chou}, \citenamefont {Legeza},\
  and\ \citenamefont {Gali}}]{IvadyhBN}%
  \BibitemOpen
  \bibfield  {author} {\bibinfo {author} {\bibfnamefont {V.}~\bibnamefont
  {Iv{\'a}dy}}, \bibinfo {author} {\bibfnamefont {G.}~\bibnamefont {Barcza}},
  \bibinfo {author} {\bibfnamefont {G.}~\bibnamefont {Thiering}}, \bibinfo
  {author} {\bibfnamefont {S.}~\bibnamefont {Li}}, \bibinfo {author}
  {\bibfnamefont {H.}~\bibnamefont {Hamdi}}, \bibinfo {author} {\bibfnamefont
  {J.-P.}\ \bibnamefont {Chou}}, \bibinfo {author} {\bibfnamefont
  {{\"O}.}~\bibnamefont {Legeza}}, \ and\ \bibinfo {author} {\bibfnamefont
  {A.}~\bibnamefont {Gali}},\ }\href {\doibase 10.1038/s41524-020-0305-x}
  {\bibfield  {journal} {\bibinfo  {journal} {npj Computational Materials}\
  }\textbf {\bibinfo {volume} {6}},\ \bibinfo {pages} {41} (\bibinfo {year}
  {2020})}\BibitemShut {NoStop}%
\bibitem [{\citenamefont {Abdi}\ \emph {et~al.}(2018)\citenamefont {Abdi},
  \citenamefont {Chou}, \citenamefont {Gali},\ and\ \citenamefont
  {Plenio}}]{Mehdi2018}%
  \BibitemOpen
  \bibfield  {author} {\bibinfo {author} {\bibfnamefont {M.}~\bibnamefont
  {Abdi}}, \bibinfo {author} {\bibfnamefont {J.-P.}\ \bibnamefont {Chou}},
  \bibinfo {author} {\bibfnamefont {A.}~\bibnamefont {Gali}}, \ and\ \bibinfo
  {author} {\bibfnamefont {M.~B.}\ \bibnamefont {Plenio}},\ }\href {\doibase
  10.1021/acsphotonics.7b01442} {\bibfield  {journal} {\bibinfo  {journal} {ACS
  Photonics}\ }\textbf {\bibinfo {volume} {5}},\ \bibinfo {pages} {1967}
  (\bibinfo {year} {2018})}\BibitemShut {NoStop}%
\bibitem [{\citenamefont {Gottscholl}\ \emph {et~al.}(2020)\citenamefont
  {Gottscholl}, \citenamefont {Kianinia}, \citenamefont {Soltamov},
  \citenamefont {Orlinskii}, \citenamefont {Mamin}, \citenamefont {Bradac},
  \citenamefont {Kasper}, \citenamefont {Krambrock}, \citenamefont {Sperlich},
  \citenamefont {Toth}, \citenamefont {Aharonovich},\ and\ \citenamefont
  {Dyakonov}}]{Gottscholl2020}%
  \BibitemOpen
  \bibfield  {author} {\bibinfo {author} {\bibfnamefont {A.}~\bibnamefont
  {Gottscholl}}, \bibinfo {author} {\bibfnamefont {M.}~\bibnamefont
  {Kianinia}}, \bibinfo {author} {\bibfnamefont {V.}~\bibnamefont {Soltamov}},
  \bibinfo {author} {\bibfnamefont {S.}~\bibnamefont {Orlinskii}}, \bibinfo
  {author} {\bibfnamefont {G.}~\bibnamefont {Mamin}}, \bibinfo {author}
  {\bibfnamefont {C.}~\bibnamefont {Bradac}}, \bibinfo {author} {\bibfnamefont
  {C.}~\bibnamefont {Kasper}}, \bibinfo {author} {\bibfnamefont
  {K.}~\bibnamefont {Krambrock}}, \bibinfo {author} {\bibfnamefont
  {A.}~\bibnamefont {Sperlich}}, \bibinfo {author} {\bibfnamefont
  {M.}~\bibnamefont {Toth}}, \bibinfo {author} {\bibfnamefont {I.}~\bibnamefont
  {Aharonovich}}, \ and\ \bibinfo {author} {\bibfnamefont {V.}~\bibnamefont
  {Dyakonov}},\ }\href {\doibase 10.1038/s41563-020-0619-6} {\bibfield
  {journal} {\bibinfo  {journal} {Nature Materials}\ }\textbf {\bibinfo
  {volume} {19}},\ \bibinfo {pages} {540} (\bibinfo {year} {2020})}\BibitemShut
  {NoStop}%
\bibitem [{\citenamefont {Pla{\v s}ienka}\ \emph {et~al.}(2016)\citenamefont
  {Pla{\v s}ienka}, \citenamefont {Marto{\v n}{\'a}k},\ and\ \citenamefont
  {Tosatti}}]{Plasienka2016}%
  \BibitemOpen
  \bibfield  {author} {\bibinfo {author} {\bibfnamefont {D.}~\bibnamefont
  {Pla{\v s}ienka}}, \bibinfo {author} {\bibfnamefont {R.}~\bibnamefont
  {Marto{\v n}{\'a}k}}, \ and\ \bibinfo {author} {\bibfnamefont
  {E.}~\bibnamefont {Tosatti}},\ }\href {\doibase 10.1038/srep37694} {\bibfield
   {journal} {\bibinfo  {journal} {Scientific Reports}\ }\textbf {\bibinfo
  {volume} {6}},\ \bibinfo {pages} {37694} (\bibinfo {year}
  {2016})}\BibitemShut {NoStop}%
\bibitem [{\citenamefont {Kuate~Defo}\ \emph
  {et~al.}(2019{\natexlab{a}})\citenamefont {Kuate~Defo}, \citenamefont
  {Kaxiras},\ and\ \citenamefont {Richardson}}]{Kuate8}%
  \BibitemOpen
  \bibfield  {author} {\bibinfo {author} {\bibfnamefont {R.}~\bibnamefont
  {Kuate~Defo}}, \bibinfo {author} {\bibfnamefont {E.}~\bibnamefont {Kaxiras}},
  \ and\ \bibinfo {author} {\bibfnamefont {S.~L.}\ \bibnamefont {Richardson}},\
  }\href {\doibase 10.1063/1.5123227} {\bibfield  {journal} {\bibinfo
  {journal} {Journal of Applied Physics}\ }\textbf {\bibinfo {volume} {126}},\
  \bibinfo {pages} {195103} (\bibinfo {year} {2019}{\natexlab{a}})}\BibitemShut
  {NoStop}%
\bibitem [{\citenamefont {Doherty}\ \emph {et~al.}(2013)\citenamefont
  {Doherty}, \citenamefont {Manson}, \citenamefont {Delaney}, \citenamefont
  {Jelezko}, \citenamefont {Wrachtrup},\ and\ \citenamefont
  {Hollenberg}}]{DOHERTY20131}%
  \BibitemOpen
  \bibfield  {author} {\bibinfo {author} {\bibfnamefont {M.~W.}\ \bibnamefont
  {Doherty}}, \bibinfo {author} {\bibfnamefont {N.~B.}\ \bibnamefont {Manson}},
  \bibinfo {author} {\bibfnamefont {P.}~\bibnamefont {Delaney}}, \bibinfo
  {author} {\bibfnamefont {F.}~\bibnamefont {Jelezko}}, \bibinfo {author}
  {\bibfnamefont {J.}~\bibnamefont {Wrachtrup}}, \ and\ \bibinfo {author}
  {\bibfnamefont {L.~C.~L.}\ \bibnamefont {Hollenberg}},\ }\href {\doibase
  https://doi.org/10.1016/j.physrep.2013.02.001} {\bibfield  {journal}
  {\bibinfo  {journal} {Physics Reports}\ }\textbf {\bibinfo {volume} {528}},\
  \bibinfo {pages} {1} (\bibinfo {year} {2013})},\ \bibinfo {note} {the
  nitrogen-vacancy colour centre in diamond}\BibitemShut {NoStop}%
\bibitem [{\citenamefont {Gali}\ \emph {et~al.}(2010)\citenamefont {Gali},
  \citenamefont {G{\"a}llstr{\"o}m}, \citenamefont {Son},\ and\ \citenamefont
  {Janz{\'e}n}}]{Gali2010}%
  \BibitemOpen
  \bibfield  {author} {\bibinfo {author} {\bibfnamefont {A.}~\bibnamefont
  {Gali}}, \bibinfo {author} {\bibfnamefont {A.}~\bibnamefont
  {G{\"a}llstr{\"o}m}}, \bibinfo {author} {\bibfnamefont {N.}~\bibnamefont
  {Son}}, \ and\ \bibinfo {author} {\bibfnamefont {E.}~\bibnamefont
  {Janz{\'e}n}},\ }in\ \href {\doibase
  10.4028/www.scientific.net/MSF.645-648.395} {\emph {\bibinfo {booktitle}
  {Silicon Carbide and Related Materials 2009}}},\ \bibinfo {series and number}
  {Materials Science Forum}\ (\bibinfo  {publisher} {Trans Tech Publications
  Ltd},\ \bibinfo {year} {2010})\ p.\ \bibinfo {pages} {395}\BibitemShut
  {NoStop}%
\bibitem [{\citenamefont {Lee}\ \emph {et~al.}(2013)\citenamefont {Lee},
  \citenamefont {Widmann}, \citenamefont {Rendler}, \citenamefont {Doherty},
  \citenamefont {Babinec}, \citenamefont {Yang}, \citenamefont {Eyer},
  \citenamefont {Siyushev}, \citenamefont {Hausmann}, \citenamefont {Loncar},
  \citenamefont {Bodrog}, \citenamefont {Gali}, \citenamefont {Manson},
  \citenamefont {Fedder},\ and\ \citenamefont {Wrachtrup}}]{SangYun2013}%
  \BibitemOpen
  \bibfield  {author} {\bibinfo {author} {\bibfnamefont {S.-Y.}\ \bibnamefont
  {Lee}}, \bibinfo {author} {\bibfnamefont {M.}~\bibnamefont {Widmann}},
  \bibinfo {author} {\bibfnamefont {T.}~\bibnamefont {Rendler}}, \bibinfo
  {author} {\bibfnamefont {M.~W.}\ \bibnamefont {Doherty}}, \bibinfo {author}
  {\bibfnamefont {T.~M.}\ \bibnamefont {Babinec}}, \bibinfo {author}
  {\bibfnamefont {S.}~\bibnamefont {Yang}}, \bibinfo {author} {\bibfnamefont
  {M.}~\bibnamefont {Eyer}}, \bibinfo {author} {\bibfnamefont {P.}~\bibnamefont
  {Siyushev}}, \bibinfo {author} {\bibfnamefont {B.~J.~M.}\ \bibnamefont
  {Hausmann}}, \bibinfo {author} {\bibfnamefont {M.}~\bibnamefont {Loncar}},
  \bibinfo {author} {\bibfnamefont {Z.}~\bibnamefont {Bodrog}}, \bibinfo
  {author} {\bibfnamefont {A.}~\bibnamefont {Gali}}, \bibinfo {author}
  {\bibfnamefont {N.~B.}\ \bibnamefont {Manson}}, \bibinfo {author}
  {\bibfnamefont {H.}~\bibnamefont {Fedder}}, \ and\ \bibinfo {author}
  {\bibfnamefont {J.}~\bibnamefont {Wrachtrup}},\ }\href {\doibase
  10.1038/nnano.2013.104} {\bibfield  {journal} {\bibinfo  {journal} {Nature
  Nanotechnology}\ }\textbf {\bibinfo {volume} {8}},\ \bibinfo {pages} {487}
  (\bibinfo {year} {2013})}\BibitemShut {NoStop}%
\bibitem [{\citenamefont {Weber}\ \emph {et~al.}(2010)\citenamefont {Weber},
  \citenamefont {Koehl}, \citenamefont {Varley}, \citenamefont {Janotti},
  \citenamefont {Buckley}, \citenamefont {Van~de Walle},\ and\ \citenamefont
  {Awschalom}}]{Weber8513}%
  \BibitemOpen
  \bibfield  {author} {\bibinfo {author} {\bibfnamefont {J.~R.}\ \bibnamefont
  {Weber}}, \bibinfo {author} {\bibfnamefont {W.~F.}\ \bibnamefont {Koehl}},
  \bibinfo {author} {\bibfnamefont {J.~B.}\ \bibnamefont {Varley}}, \bibinfo
  {author} {\bibfnamefont {A.}~\bibnamefont {Janotti}}, \bibinfo {author}
  {\bibfnamefont {B.~B.}\ \bibnamefont {Buckley}}, \bibinfo {author}
  {\bibfnamefont {C.~G.}\ \bibnamefont {Van~de Walle}}, \ and\ \bibinfo
  {author} {\bibfnamefont {D.~D.}\ \bibnamefont {Awschalom}},\ }\href {\doibase
  10.1073/pnas.1003052107} {\bibfield  {journal} {\bibinfo  {journal}
  {Proceedings of the National Academy of Sciences}\ }\textbf {\bibinfo
  {volume} {107}},\ \bibinfo {pages} {8513} (\bibinfo {year}
  {2010})}\BibitemShut {NoStop}%
\bibitem [{\citenamefont {Soltamov}\ \emph {et~al.}(2012)\citenamefont
  {Soltamov}, \citenamefont {Soltamova}, \citenamefont {Baranov},\ and\
  \citenamefont {Proskuryakov}}]{Soltamov2012}%
  \BibitemOpen
  \bibfield  {author} {\bibinfo {author} {\bibfnamefont {V.~A.}\ \bibnamefont
  {Soltamov}}, \bibinfo {author} {\bibfnamefont {A.~A.}\ \bibnamefont
  {Soltamova}}, \bibinfo {author} {\bibfnamefont {P.~G.}\ \bibnamefont
  {Baranov}}, \ and\ \bibinfo {author} {\bibfnamefont {I.~I.}\ \bibnamefont
  {Proskuryakov}},\ }\href {\doibase 10.1103/PhysRevLett.108.226402} {\bibfield
   {journal} {\bibinfo  {journal} {Phys. Rev. Lett.}\ }\textbf {\bibinfo
  {volume} {108}},\ \bibinfo {pages} {226402} (\bibinfo {year}
  {2012})}\BibitemShut {NoStop}%
\bibitem [{\citenamefont {Koehl}\ \emph {et~al.}(2011)\citenamefont {Koehl},
  \citenamefont {Buckley}, \citenamefont {Heremans}, \citenamefont {Calusine},\
  and\ \citenamefont {Awschalom}}]{Koehl2011}%
  \BibitemOpen
  \bibfield  {author} {\bibinfo {author} {\bibfnamefont {W.~F.}\ \bibnamefont
  {Koehl}}, \bibinfo {author} {\bibfnamefont {B.~B.}\ \bibnamefont {Buckley}},
  \bibinfo {author} {\bibfnamefont {F.~J.}\ \bibnamefont {Heremans}}, \bibinfo
  {author} {\bibfnamefont {G.}~\bibnamefont {Calusine}}, \ and\ \bibinfo
  {author} {\bibfnamefont {D.~D.}\ \bibnamefont {Awschalom}},\ }\href {\doibase
  10.1038/nature10562} {\bibfield  {journal} {\bibinfo  {journal} {Nature}\
  }\textbf {\bibinfo {volume} {479}},\ \bibinfo {pages} {84} (\bibinfo {year}
  {2011})}\BibitemShut {NoStop}%
\bibitem [{\citenamefont {Kraus}\ \emph {et~al.}(2014)\citenamefont {Kraus},
  \citenamefont {Soltamov}, \citenamefont {Riedel}, \citenamefont {V{\"a}th},
  \citenamefont {Fuchs}, \citenamefont {Sperlich}, \citenamefont {Baranov},
  \citenamefont {Dyakonov},\ and\ \citenamefont {Astakhov}}]{Kraus2014}%
  \BibitemOpen
  \bibfield  {author} {\bibinfo {author} {\bibfnamefont {H.}~\bibnamefont
  {Kraus}}, \bibinfo {author} {\bibfnamefont {V.~A.}\ \bibnamefont {Soltamov}},
  \bibinfo {author} {\bibfnamefont {D.}~\bibnamefont {Riedel}}, \bibinfo
  {author} {\bibfnamefont {S.}~\bibnamefont {V{\"a}th}}, \bibinfo {author}
  {\bibfnamefont {F.}~\bibnamefont {Fuchs}}, \bibinfo {author} {\bibfnamefont
  {A.}~\bibnamefont {Sperlich}}, \bibinfo {author} {\bibfnamefont {P.~G.}\
  \bibnamefont {Baranov}}, \bibinfo {author} {\bibfnamefont {V.}~\bibnamefont
  {Dyakonov}}, \ and\ \bibinfo {author} {\bibfnamefont {G.~V.}\ \bibnamefont
  {Astakhov}},\ }\href {\doibase 10.1038/nphys2826} {\bibfield  {journal}
  {\bibinfo  {journal} {Nature Physics}\ }\textbf {\bibinfo {volume} {10}},\
  \bibinfo {pages} {157} (\bibinfo {year} {2014})}\BibitemShut {NoStop}%
\bibitem [{\citenamefont {Wahl}\ \emph {et~al.}(2020)\citenamefont {Wahl},
  \citenamefont {Correia}, \citenamefont {Villarreal}, \citenamefont
  {Bourgeois}, \citenamefont {Gulka}, \citenamefont {Nesl\'adek}, \citenamefont
  {Vantomme},\ and\ \citenamefont {Pereira}}]{Wahl}%
  \BibitemOpen
  \bibfield  {author} {\bibinfo {author} {\bibfnamefont {U.}~\bibnamefont
  {Wahl}}, \bibinfo {author} {\bibfnamefont {J.~G.}\ \bibnamefont {Correia}},
  \bibinfo {author} {\bibfnamefont {R.}~\bibnamefont {Villarreal}}, \bibinfo
  {author} {\bibfnamefont {E.}~\bibnamefont {Bourgeois}}, \bibinfo {author}
  {\bibfnamefont {M.}~\bibnamefont {Gulka}}, \bibinfo {author} {\bibfnamefont
  {M.}~\bibnamefont {Nesl\'adek}}, \bibinfo {author} {\bibfnamefont
  {A.}~\bibnamefont {Vantomme}}, \ and\ \bibinfo {author} {\bibfnamefont
  {L.~M.~C.}\ \bibnamefont {Pereira}},\ }\href {\doibase
  10.1103/PhysRevLett.125.045301} {\bibfield  {journal} {\bibinfo  {journal}
  {Phys. Rev. Lett.}\ }\textbf {\bibinfo {volume} {125}},\ \bibinfo {pages}
  {045301} (\bibinfo {year} {2020})}\BibitemShut {NoStop}%
\bibitem [{\citenamefont {Wang}\ \emph {et~al.}(2019)\citenamefont {Wang},
  \citenamefont {Li}, \citenamefont {Yan}, \citenamefont {Liu}, \citenamefont
  {Guo}, \citenamefont {Zhang}, \citenamefont {Zhou}, \citenamefont {Guo},
  \citenamefont {Lin}, \citenamefont {Cui}, \citenamefont {Xu}, \citenamefont
  {Xu}, \citenamefont {Li},\ and\ \citenamefont {Guo}}]{Wang_2019}%
  \BibitemOpen
  \bibfield  {author} {\bibinfo {author} {\bibfnamefont {J.-F.}\ \bibnamefont
  {Wang}}, \bibinfo {author} {\bibfnamefont {Q.}~\bibnamefont {Li}}, \bibinfo
  {author} {\bibfnamefont {F.-F.}\ \bibnamefont {Yan}}, \bibinfo {author}
  {\bibfnamefont {H.}~\bibnamefont {Liu}}, \bibinfo {author} {\bibfnamefont
  {G.-P.}\ \bibnamefont {Guo}}, \bibinfo {author} {\bibfnamefont {W.-P.}\
  \bibnamefont {Zhang}}, \bibinfo {author} {\bibfnamefont {X.}~\bibnamefont
  {Zhou}}, \bibinfo {author} {\bibfnamefont {L.-P.}\ \bibnamefont {Guo}},
  \bibinfo {author} {\bibfnamefont {Z.-H.}\ \bibnamefont {Lin}}, \bibinfo
  {author} {\bibfnamefont {J.-M.}\ \bibnamefont {Cui}}, \bibinfo {author}
  {\bibfnamefont {X.-Y.}\ \bibnamefont {Xu}}, \bibinfo {author} {\bibfnamefont
  {J.-S.}\ \bibnamefont {Xu}}, \bibinfo {author} {\bibfnamefont {C.-F.}\
  \bibnamefont {Li}}, \ and\ \bibinfo {author} {\bibfnamefont {G.-C.}\
  \bibnamefont {Guo}},\ }\href {\doibase 10.1021/acsphotonics.9b00451}
  {\bibfield  {journal} {\bibinfo  {journal} {ACS Photonics}\ }\textbf
  {\bibinfo {volume} {6}},\ \bibinfo {pages} {1736} (\bibinfo {year}
  {2019})}\BibitemShut {NoStop}%
\bibitem [{\citenamefont {Dong}\ \emph {et~al.}(2019)\citenamefont {Dong},
  \citenamefont {Doherty},\ and\ \citenamefont {Economou}}]{Dong}%
  \BibitemOpen
  \bibfield  {author} {\bibinfo {author} {\bibfnamefont {W.}~\bibnamefont
  {Dong}}, \bibinfo {author} {\bibfnamefont {M.~W.}\ \bibnamefont {Doherty}}, \
  and\ \bibinfo {author} {\bibfnamefont {S.~E.}\ \bibnamefont {Economou}},\
  }\href {\doibase 10.1103/PhysRevB.99.184102} {\bibfield  {journal} {\bibinfo
  {journal} {Phys. Rev. B}\ }\textbf {\bibinfo {volume} {99}},\ \bibinfo
  {pages} {184102} (\bibinfo {year} {2019})}\BibitemShut {NoStop}%
\bibitem [{\citenamefont {Green}\ \emph {et~al.}(2017)\citenamefont {Green},
  \citenamefont {Mottishaw}, \citenamefont {Breeze}, \citenamefont {Edmonds},
  \citenamefont {D'Haenens-Johansson}, \citenamefont {Doherty}, \citenamefont
  {Williams}, \citenamefont {Twitchen},\ and\ \citenamefont
  {Newton}}]{Green_2017}%
  \BibitemOpen
  \bibfield  {author} {\bibinfo {author} {\bibfnamefont {B.~L.}\ \bibnamefont
  {Green}}, \bibinfo {author} {\bibfnamefont {S.}~\bibnamefont {Mottishaw}},
  \bibinfo {author} {\bibfnamefont {B.~G.}\ \bibnamefont {Breeze}}, \bibinfo
  {author} {\bibfnamefont {A.~M.}\ \bibnamefont {Edmonds}}, \bibinfo {author}
  {\bibfnamefont {U.~F.~S.}\ \bibnamefont {D'Haenens-Johansson}}, \bibinfo
  {author} {\bibfnamefont {M.~W.}\ \bibnamefont {Doherty}}, \bibinfo {author}
  {\bibfnamefont {S.~D.}\ \bibnamefont {Williams}}, \bibinfo {author}
  {\bibfnamefont {D.~J.}\ \bibnamefont {Twitchen}}, \ and\ \bibinfo {author}
  {\bibfnamefont {M.~E.}\ \bibnamefont {Newton}},\ }\href {\doibase
  10.1103/PhysRevLett.119.096402} {\bibfield  {journal} {\bibinfo  {journal}
  {Phys. Rev. Lett.}\ }\textbf {\bibinfo {volume} {119}},\ \bibinfo {pages}
  {096402} (\bibinfo {year} {2017})}\BibitemShut {NoStop}%
\bibitem [{\citenamefont {Green}\ \emph {et~al.}(2019)\citenamefont {Green},
  \citenamefont {Doherty}, \citenamefont {Nako}, \citenamefont {Manson},
  \citenamefont {D'Haenens-Johansson}, \citenamefont {Williams}, \citenamefont
  {Twitchen},\ and\ \citenamefont {Newton}}]{Green_2019}%
  \BibitemOpen
  \bibfield  {author} {\bibinfo {author} {\bibfnamefont {B.~L.}\ \bibnamefont
  {Green}}, \bibinfo {author} {\bibfnamefont {M.~W.}\ \bibnamefont {Doherty}},
  \bibinfo {author} {\bibfnamefont {E.}~\bibnamefont {Nako}}, \bibinfo {author}
  {\bibfnamefont {N.~B.}\ \bibnamefont {Manson}}, \bibinfo {author}
  {\bibfnamefont {U.~F.~S.}\ \bibnamefont {D'Haenens-Johansson}}, \bibinfo
  {author} {\bibfnamefont {S.~D.}\ \bibnamefont {Williams}}, \bibinfo {author}
  {\bibfnamefont {D.~J.}\ \bibnamefont {Twitchen}}, \ and\ \bibinfo {author}
  {\bibfnamefont {M.~E.}\ \bibnamefont {Newton}},\ }\href {\doibase
  10.1103/PhysRevB.99.161112} {\bibfield  {journal} {\bibinfo  {journal} {Phys.
  Rev. B}\ }\textbf {\bibinfo {volume} {99}},\ \bibinfo {pages} {161112(R)}
  (\bibinfo {year} {2019})}\BibitemShut {NoStop}%
\bibitem [{\citenamefont {Kuate~Defo}\ \emph {et~al.}(2018)\citenamefont
  {Kuate~Defo}, \citenamefont {Zhang}, \citenamefont {Bracher}, \citenamefont
  {Kim}, \citenamefont {Hu},\ and\ \citenamefont {Kaxiras}}]{Kuate}%
  \BibitemOpen
  \bibfield  {author} {\bibinfo {author} {\bibfnamefont {R.}~\bibnamefont
  {Kuate~Defo}}, \bibinfo {author} {\bibfnamefont {X.}~\bibnamefont {Zhang}},
  \bibinfo {author} {\bibfnamefont {D.}~\bibnamefont {Bracher}}, \bibinfo
  {author} {\bibfnamefont {G.}~\bibnamefont {Kim}}, \bibinfo {author}
  {\bibfnamefont {E.}~\bibnamefont {Hu}}, \ and\ \bibinfo {author}
  {\bibfnamefont {E.}~\bibnamefont {Kaxiras}},\ }\href {\doibase
  10.1103/PhysRevB.98.104103} {\bibfield  {journal} {\bibinfo  {journal} {Phys.
  Rev. B}\ }\textbf {\bibinfo {volume} {98}},\ \bibinfo {pages} {104103}
  (\bibinfo {year} {2018})}\BibitemShut {NoStop}%
\bibitem [{\citenamefont {Kuate~Defo}\ \emph
  {et~al.}(2019{\natexlab{b}})\citenamefont {Kuate~Defo}, \citenamefont
  {Wang},\ and\ \citenamefont {Manjunathaiah}}]{DEFO2019}%
  \BibitemOpen
  \bibfield  {author} {\bibinfo {author} {\bibfnamefont {R.}~\bibnamefont
  {Kuate~Defo}}, \bibinfo {author} {\bibfnamefont {R.}~\bibnamefont {Wang}}, \
  and\ \bibinfo {author} {\bibfnamefont {M.}~\bibnamefont {Manjunathaiah}},\
  }\href {http://www.sciencedirect.com/science/article/pii/S1877750318304320}
  {\bibfield  {journal} {\bibinfo  {journal} {Journal of Computational
  Science}\ } (\bibinfo {year} {2019}{\natexlab{b}})}\BibitemShut {NoStop}%
\bibitem [{\citenamefont {Sohn}\ \emph {et~al.}(2018)\citenamefont {Sohn},
  \citenamefont {Meesala}, \citenamefont {Pingault}, \citenamefont {Atikian},
  \citenamefont {Holzgrafe}, \citenamefont {G{\"u}ndo{\u g}an}, \citenamefont
  {Stavrakas}, \citenamefont {Stanley}, \citenamefont {Sipahigil},
  \citenamefont {Choi}, \citenamefont {Zhang}, \citenamefont {Pacheco},
  \citenamefont {Abraham}, \citenamefont {Bielejec}, \citenamefont {Lukin},
  \citenamefont {Atat{\"u}re},\ and\ \citenamefont {Lon{\v
  c}ar}}]{sohn2018cont}%
  \BibitemOpen
  \bibfield  {author} {\bibinfo {author} {\bibfnamefont {Y.-I.}\ \bibnamefont
  {Sohn}}, \bibinfo {author} {\bibfnamefont {S.}~\bibnamefont {Meesala}},
  \bibinfo {author} {\bibfnamefont {B.}~\bibnamefont {Pingault}}, \bibinfo
  {author} {\bibfnamefont {H.~A.}\ \bibnamefont {Atikian}}, \bibinfo {author}
  {\bibfnamefont {J.}~\bibnamefont {Holzgrafe}}, \bibinfo {author}
  {\bibfnamefont {M.}~\bibnamefont {G{\"u}ndo{\u g}an}}, \bibinfo {author}
  {\bibfnamefont {C.}~\bibnamefont {Stavrakas}}, \bibinfo {author}
  {\bibfnamefont {M.~J.}\ \bibnamefont {Stanley}}, \bibinfo {author}
  {\bibfnamefont {A.}~\bibnamefont {Sipahigil}}, \bibinfo {author}
  {\bibfnamefont {J.}~\bibnamefont {Choi}}, \bibinfo {author} {\bibfnamefont
  {M.}~\bibnamefont {Zhang}}, \bibinfo {author} {\bibfnamefont {J.~L.}\
  \bibnamefont {Pacheco}}, \bibinfo {author} {\bibfnamefont {J.}~\bibnamefont
  {Abraham}}, \bibinfo {author} {\bibfnamefont {E.}~\bibnamefont {Bielejec}},
  \bibinfo {author} {\bibfnamefont {M.~D.}\ \bibnamefont {Lukin}}, \bibinfo
  {author} {\bibfnamefont {M.}~\bibnamefont {Atat{\"u}re}}, \ and\ \bibinfo
  {author} {\bibfnamefont {M.}~\bibnamefont {Lon{\v c}ar}},\ }\href {\doibase
  10.1038/s41467-018-04340-3} {\bibfield  {journal} {\bibinfo  {journal} {Nat.
  Commun.}\ }\textbf {\bibinfo {volume} {9}},\ \bibinfo {pages} {2012}
  (\bibinfo {year} {2018})}\BibitemShut {NoStop}%
\bibitem [{\citenamefont {Davidsson}\ \emph {et~al.}(2021)\citenamefont
  {Davidsson}, \citenamefont {Iv\'ady}, \citenamefont {Armiento},\ and\
  \citenamefont {Abrikosov}}]{davidsson2021adaq}%
  \BibitemOpen
  \bibfield  {author} {\bibinfo {author} {\bibfnamefont {J.}~\bibnamefont
  {Davidsson}}, \bibinfo {author} {\bibfnamefont {V.}~\bibnamefont {Iv\'ady}},
  \bibinfo {author} {\bibfnamefont {R.}~\bibnamefont {Armiento}}, \ and\
  \bibinfo {author} {\bibfnamefont {I.~A.}\ \bibnamefont {Abrikosov}},\
  }\href@noop {} {\enquote {\bibinfo {title} {Adaq: Automatic workflows for
  magneto-optical properties of point defects in semiconductors},}\ } (\bibinfo
  {year} {2021}),\ \Eprint {http://arxiv.org/abs/2008.12539} {arXiv:2008.12539
  [cond-mat.mtrl-sci]} \BibitemShut {NoStop}%
\bibitem [{\citenamefont {Broberg}\ \emph {et~al.}(2018)\citenamefont
  {Broberg}, \citenamefont {Medasani}, \citenamefont {Zimmermann},
  \citenamefont {Yu}, \citenamefont {Canning}, \citenamefont {Haranczyk},
  \citenamefont {Asta},\ and\ \citenamefont {Hautier}}]{BROBERG2018165}%
  \BibitemOpen
  \bibfield  {author} {\bibinfo {author} {\bibfnamefont {D.}~\bibnamefont
  {Broberg}}, \bibinfo {author} {\bibfnamefont {B.}~\bibnamefont {Medasani}},
  \bibinfo {author} {\bibfnamefont {N.~E.}\ \bibnamefont {Zimmermann}},
  \bibinfo {author} {\bibfnamefont {G.}~\bibnamefont {Yu}}, \bibinfo {author}
  {\bibfnamefont {A.}~\bibnamefont {Canning}}, \bibinfo {author} {\bibfnamefont
  {M.}~\bibnamefont {Haranczyk}}, \bibinfo {author} {\bibfnamefont
  {M.}~\bibnamefont {Asta}}, \ and\ \bibinfo {author} {\bibfnamefont
  {G.}~\bibnamefont {Hautier}},\ }\href {\doibase
  https://doi.org/10.1016/j.cpc.2018.01.004} {\bibfield  {journal} {\bibinfo
  {journal} {Computer Physics Communications}\ }\textbf {\bibinfo {volume}
  {226}},\ \bibinfo {pages} {165} (\bibinfo {year} {2018})}\BibitemShut
  {NoStop}%
\bibitem [{\citenamefont {Janak}(1978)}]{Janak1978}%
  \BibitemOpen
  \bibfield  {author} {\bibinfo {author} {\bibfnamefont {J.~F.}\ \bibnamefont
  {Janak}},\ }\href {\doibase 10.1103/PhysRevB.18.7165} {\bibfield  {journal}
  {\bibinfo  {journal} {Phys. Rev. B}\ }\textbf {\bibinfo {volume} {18}},\
  \bibinfo {pages} {7165} (\bibinfo {year} {1978})}\BibitemShut {NoStop}%
\bibitem [{\citenamefont {Koopmans}(1934)}]{KOOPMANS1934}%
  \BibitemOpen
  \bibfield  {author} {\bibinfo {author} {\bibfnamefont {T.}~\bibnamefont
  {Koopmans}},\ }\href {\doibase https://doi.org/10.1016/S0031-8914(34)90011-2}
  {\bibfield  {journal} {\bibinfo  {journal} {Physica}\ }\textbf {\bibinfo
  {volume} {1}},\ \bibinfo {pages} {104 } (\bibinfo {year} {1934})}\BibitemShut
  {NoStop}%
\bibitem [{\citenamefont {Slater}\ and\ \citenamefont
  {Wood}(1970)}]{Slater1970}%
  \BibitemOpen
  \bibfield  {author} {\bibinfo {author} {\bibfnamefont {J.~C.}\ \bibnamefont
  {Slater}}\ and\ \bibinfo {author} {\bibfnamefont {J.~H.}\ \bibnamefont
  {Wood}},\ }\href {\doibase https://doi.org/10.1002/qua.560050703} {\bibfield
  {journal} {\bibinfo  {journal} {International Journal of Quantum Chemistry}\
  }\textbf {\bibinfo {volume} {5}},\ \bibinfo {pages} {3} (\bibinfo {year}
  {1970})}\BibitemShut {NoStop}%
\bibitem [{\citenamefont {Kaxiras}(2003)}]{Kaxiras}%
  \BibitemOpen
  \bibfield  {author} {\bibinfo {author} {\bibfnamefont {E.}~\bibnamefont
  {Kaxiras}},\ }\href@noop {} {\emph {\bibinfo {title} {{Atomic and Electronic
  Structure of Solids }}}}\ (\bibinfo  {publisher} {Cambridge University
  Press},\ \bibinfo {address} {New York},\ \bibinfo {year} {2003})\BibitemShut
  {NoStop}%
\bibitem [{\citenamefont {Johnson}(1988)}]{Johnson1988}%
  \BibitemOpen
  \bibfield  {author} {\bibinfo {author} {\bibfnamefont {D.~D.}\ \bibnamefont
  {Johnson}},\ }\href {\doibase 10.1103/PhysRevB.38.12807} {\bibfield
  {journal} {\bibinfo  {journal} {Phys. Rev. B}\ }\textbf {\bibinfo {volume}
  {38}},\ \bibinfo {pages} {12807} (\bibinfo {year} {1988})}\BibitemShut
  {NoStop}%
\bibitem [{\citenamefont {Slater}\ \emph {et~al.}(1969)\citenamefont {Slater},
  \citenamefont {Mann}, \citenamefont {Wilson},\ and\ \citenamefont
  {Wood}}]{PhysRev.184.672}%
  \BibitemOpen
  \bibfield  {author} {\bibinfo {author} {\bibfnamefont {J.~C.}\ \bibnamefont
  {Slater}}, \bibinfo {author} {\bibfnamefont {J.~B.}\ \bibnamefont {Mann}},
  \bibinfo {author} {\bibfnamefont {T.~M.}\ \bibnamefont {Wilson}}, \ and\
  \bibinfo {author} {\bibfnamefont {J.~H.}\ \bibnamefont {Wood}},\ }\href
  {\doibase 10.1103/PhysRev.184.672} {\bibfield  {journal} {\bibinfo  {journal}
  {Phys. Rev.}\ }\textbf {\bibinfo {volume} {184}},\ \bibinfo {pages} {672}
  (\bibinfo {year} {1969})}\BibitemShut {NoStop}%
\bibitem [{\citenamefont {Hadjisavvas}\ and\ \citenamefont
  {Theophilou}(1985)}]{PhysRevA.32.720}%
  \BibitemOpen
  \bibfield  {author} {\bibinfo {author} {\bibfnamefont {N.}~\bibnamefont
  {Hadjisavvas}}\ and\ \bibinfo {author} {\bibfnamefont {A.}~\bibnamefont
  {Theophilou}},\ }\href {\doibase 10.1103/PhysRevA.32.720} {\bibfield
  {journal} {\bibinfo  {journal} {Phys. Rev. A}\ }\textbf {\bibinfo {volume}
  {32}},\ \bibinfo {pages} {720} (\bibinfo {year} {1985})}\BibitemShut
  {NoStop}%
\bibitem [{\citenamefont {Kohn}\ and\ \citenamefont {Sham}(1965)}]{Kohn1965}%
  \BibitemOpen
  \bibfield  {author} {\bibinfo {author} {\bibfnamefont {W.}~\bibnamefont
  {Kohn}}\ and\ \bibinfo {author} {\bibfnamefont {L.~J.}\ \bibnamefont
  {Sham}},\ }\href {\doibase 10.1103/PhysRev.140.A1133} {\bibfield  {journal}
  {\bibinfo  {journal} {Phys. Rev.}\ }\textbf {\bibinfo {volume} {140}},\
  \bibinfo {pages} {A1133} (\bibinfo {year} {1965})}\BibitemShut {NoStop}%
\bibitem [{\citenamefont {Kresse}\ and\ \citenamefont
  {Hafner}(1993)}]{Kresse1}%
  \BibitemOpen
  \bibfield  {author} {\bibinfo {author} {\bibfnamefont {G.}~\bibnamefont
  {Kresse}}\ and\ \bibinfo {author} {\bibfnamefont {J.}~\bibnamefont
  {Hafner}},\ }\href {\doibase 10.1103/PhysRevB.47.558} {\bibfield  {journal}
  {\bibinfo  {journal} {Phys. Rev. B}\ }\textbf {\bibinfo {volume} {47}},\
  \bibinfo {pages} {558} (\bibinfo {year} {1993})}\BibitemShut {NoStop}%
\bibitem [{\citenamefont {Kresse}\ and\ \citenamefont
  {Furthm\"uller}(1996)}]{Kresse2}%
  \BibitemOpen
  \bibfield  {author} {\bibinfo {author} {\bibfnamefont {G.}~\bibnamefont
  {Kresse}}\ and\ \bibinfo {author} {\bibfnamefont {J.}~\bibnamefont
  {Furthm\"uller}},\ }\href {\doibase 10.1103/PhysRevB.54.11169} {\bibfield
  {journal} {\bibinfo  {journal} {Phys. Rev. B}\ }\textbf {\bibinfo {volume}
  {54}},\ \bibinfo {pages} {11169} (\bibinfo {year} {1996})}\BibitemShut
  {NoStop}%
\bibitem [{\citenamefont {Kresse}\ and\ \citenamefont
  {Joubert}(1999)}]{Kresse3}%
  \BibitemOpen
  \bibfield  {author} {\bibinfo {author} {\bibfnamefont {G.}~\bibnamefont
  {Kresse}}\ and\ \bibinfo {author} {\bibfnamefont {D.}~\bibnamefont
  {Joubert}},\ }\href {\doibase 10.1103/PhysRevB.59.1758} {\bibfield  {journal}
  {\bibinfo  {journal} {Phys. Rev. B}\ }\textbf {\bibinfo {volume} {59}},\
  \bibinfo {pages} {1758} (\bibinfo {year} {1999})}\BibitemShut {NoStop}%
\bibitem [{\citenamefont {Giannozzi}\ \emph {et~al.}(2009)\citenamefont
  {Giannozzi}, \citenamefont {Baroni}, \citenamefont {Bonini}, \citenamefont
  {Calandra}, \citenamefont {Car}, \citenamefont {Cavazzoni}, \citenamefont
  {Ceresoli}, \citenamefont {Chiarotti}, \citenamefont {Cococcioni},
  \citenamefont {Dabo}, \citenamefont {Corso}, \citenamefont {de~Gironcoli},
  \citenamefont {Fabris}, \citenamefont {Fratesi}, \citenamefont {Gebauer},
  \citenamefont {Gerstmann}, \citenamefont {Gougoussis}, \citenamefont
  {Kokalj}, \citenamefont {Lazzeri}, \citenamefont {Martin-Samos},
  \citenamefont {Marzari}, \citenamefont {Mauri}, \citenamefont {Mazzarello},
  \citenamefont {Paolini}, \citenamefont {Pasquarello}, \citenamefont
  {Paulatto}, \citenamefont {Sbraccia}, \citenamefont {Scandolo}, \citenamefont
  {Sclauzero}, \citenamefont {Seitsonen}, \citenamefont {Smogunov},
  \citenamefont {Umari},\ and\ \citenamefont {Wentzcovitch}}]{Giannozzi_2009}%
  \BibitemOpen
  \bibfield  {author} {\bibinfo {author} {\bibfnamefont {P.}~\bibnamefont
  {Giannozzi}}, \bibinfo {author} {\bibfnamefont {S.}~\bibnamefont {Baroni}},
  \bibinfo {author} {\bibfnamefont {N.}~\bibnamefont {Bonini}}, \bibinfo
  {author} {\bibfnamefont {M.}~\bibnamefont {Calandra}}, \bibinfo {author}
  {\bibfnamefont {R.}~\bibnamefont {Car}}, \bibinfo {author} {\bibfnamefont
  {C.}~\bibnamefont {Cavazzoni}}, \bibinfo {author} {\bibfnamefont
  {D.}~\bibnamefont {Ceresoli}}, \bibinfo {author} {\bibfnamefont {G.~L.}\
  \bibnamefont {Chiarotti}}, \bibinfo {author} {\bibfnamefont {M.}~\bibnamefont
  {Cococcioni}}, \bibinfo {author} {\bibfnamefont {I.}~\bibnamefont {Dabo}},
  \bibinfo {author} {\bibfnamefont {A.~D.}\ \bibnamefont {Corso}}, \bibinfo
  {author} {\bibfnamefont {S.}~\bibnamefont {de~Gironcoli}}, \bibinfo {author}
  {\bibfnamefont {S.}~\bibnamefont {Fabris}}, \bibinfo {author} {\bibfnamefont
  {G.}~\bibnamefont {Fratesi}}, \bibinfo {author} {\bibfnamefont
  {R.}~\bibnamefont {Gebauer}}, \bibinfo {author} {\bibfnamefont
  {U.}~\bibnamefont {Gerstmann}}, \bibinfo {author} {\bibfnamefont
  {C.}~\bibnamefont {Gougoussis}}, \bibinfo {author} {\bibfnamefont
  {A.}~\bibnamefont {Kokalj}}, \bibinfo {author} {\bibfnamefont
  {M.}~\bibnamefont {Lazzeri}}, \bibinfo {author} {\bibfnamefont
  {L.}~\bibnamefont {Martin-Samos}}, \bibinfo {author} {\bibfnamefont
  {N.}~\bibnamefont {Marzari}}, \bibinfo {author} {\bibfnamefont
  {F.}~\bibnamefont {Mauri}}, \bibinfo {author} {\bibfnamefont
  {R.}~\bibnamefont {Mazzarello}}, \bibinfo {author} {\bibfnamefont
  {S.}~\bibnamefont {Paolini}}, \bibinfo {author} {\bibfnamefont
  {A.}~\bibnamefont {Pasquarello}}, \bibinfo {author} {\bibfnamefont
  {L.}~\bibnamefont {Paulatto}}, \bibinfo {author} {\bibfnamefont
  {C.}~\bibnamefont {Sbraccia}}, \bibinfo {author} {\bibfnamefont
  {S.}~\bibnamefont {Scandolo}}, \bibinfo {author} {\bibfnamefont
  {G.}~\bibnamefont {Sclauzero}}, \bibinfo {author} {\bibfnamefont {A.~P.}\
  \bibnamefont {Seitsonen}}, \bibinfo {author} {\bibfnamefont {A.}~\bibnamefont
  {Smogunov}}, \bibinfo {author} {\bibfnamefont {P.}~\bibnamefont {Umari}}, \
  and\ \bibinfo {author} {\bibfnamefont {R.~M.}\ \bibnamefont {Wentzcovitch}},\
  }\href {\doibase 10.1088/0953-8984/21/39/395502} {\bibfield  {journal}
  {\bibinfo  {journal} {Journal of Physics: Condensed Matter}\ }\textbf
  {\bibinfo {volume} {21}},\ \bibinfo {pages} {395502} (\bibinfo {year}
  {2009})}\BibitemShut {NoStop}%
\bibitem [{\citenamefont {Giannozzi}\ \emph {et~al.}(2017)\citenamefont
  {Giannozzi}, \citenamefont {Andreussi}, \citenamefont {Brumme}, \citenamefont
  {Bunau}, \citenamefont {Nardelli}, \citenamefont {Calandra}, \citenamefont
  {Car}, \citenamefont {Cavazzoni}, \citenamefont {Ceresoli}, \citenamefont
  {Cococcioni}, \citenamefont {Colonna}, \citenamefont {Carnimeo},
  \citenamefont {Corso}, \citenamefont {de~Gironcoli}, \citenamefont {Delugas},
  \citenamefont {DiStasio}, \citenamefont {Ferretti}, \citenamefont {Floris},
  \citenamefont {Fratesi}, \citenamefont {Fugallo}, \citenamefont {Gebauer},
  \citenamefont {Gerstmann}, \citenamefont {Giustino}, \citenamefont {Gorni},
  \citenamefont {Jia}, \citenamefont {Kawamura}, \citenamefont {Ko},
  \citenamefont {Kokalj}, \citenamefont {KÃŒ{\c{c}}ÃŒkbenli}, \citenamefont
  {Lazzeri}, \citenamefont {Marsili}, \citenamefont {Marzari}, \citenamefont
  {Mauri}, \citenamefont {Nguyen}, \citenamefont {Nguyen}, \citenamefont {de-la
  Roza}, \citenamefont {Paulatto}, \citenamefont {Ponc{\'{e}}}, \citenamefont
  {Rocca}, \citenamefont {Sabatini}, \citenamefont {Santra}, \citenamefont
  {Schlipf}, \citenamefont {Seitsonen}, \citenamefont {Smogunov}, \citenamefont
  {Timrov}, \citenamefont {Thonhauser}, \citenamefont {Umari}, \citenamefont
  {Vast}, \citenamefont {Wu},\ and\ \citenamefont {Baroni}}]{Giannozzi_2017}%
  \BibitemOpen
  \bibfield  {author} {\bibinfo {author} {\bibfnamefont {P.}~\bibnamefont
  {Giannozzi}}, \bibinfo {author} {\bibfnamefont {O.}~\bibnamefont
  {Andreussi}}, \bibinfo {author} {\bibfnamefont {T.}~\bibnamefont {Brumme}},
  \bibinfo {author} {\bibfnamefont {O.}~\bibnamefont {Bunau}}, \bibinfo
  {author} {\bibfnamefont {M.~B.}\ \bibnamefont {Nardelli}}, \bibinfo {author}
  {\bibfnamefont {M.}~\bibnamefont {Calandra}}, \bibinfo {author}
  {\bibfnamefont {R.}~\bibnamefont {Car}}, \bibinfo {author} {\bibfnamefont
  {C.}~\bibnamefont {Cavazzoni}}, \bibinfo {author} {\bibfnamefont
  {D.}~\bibnamefont {Ceresoli}}, \bibinfo {author} {\bibfnamefont
  {M.}~\bibnamefont {Cococcioni}}, \bibinfo {author} {\bibfnamefont
  {N.}~\bibnamefont {Colonna}}, \bibinfo {author} {\bibfnamefont
  {I.}~\bibnamefont {Carnimeo}}, \bibinfo {author} {\bibfnamefont {A.~D.}\
  \bibnamefont {Corso}}, \bibinfo {author} {\bibfnamefont {S.}~\bibnamefont
  {de~Gironcoli}}, \bibinfo {author} {\bibfnamefont {P.}~\bibnamefont
  {Delugas}}, \bibinfo {author} {\bibfnamefont {R.~A.}\ \bibnamefont
  {DiStasio}}, \bibinfo {author} {\bibfnamefont {A.}~\bibnamefont {Ferretti}},
  \bibinfo {author} {\bibfnamefont {A.}~\bibnamefont {Floris}}, \bibinfo
  {author} {\bibfnamefont {G.}~\bibnamefont {Fratesi}}, \bibinfo {author}
  {\bibfnamefont {G.}~\bibnamefont {Fugallo}}, \bibinfo {author} {\bibfnamefont
  {R.}~\bibnamefont {Gebauer}}, \bibinfo {author} {\bibfnamefont
  {U.}~\bibnamefont {Gerstmann}}, \bibinfo {author} {\bibfnamefont
  {F.}~\bibnamefont {Giustino}}, \bibinfo {author} {\bibfnamefont
  {T.}~\bibnamefont {Gorni}}, \bibinfo {author} {\bibfnamefont
  {J.}~\bibnamefont {Jia}}, \bibinfo {author} {\bibfnamefont {M.}~\bibnamefont
  {Kawamura}}, \bibinfo {author} {\bibfnamefont {H.-Y.}\ \bibnamefont {Ko}},
  \bibinfo {author} {\bibfnamefont {A.}~\bibnamefont {Kokalj}}, \bibinfo
  {author} {\bibfnamefont {E.}~\bibnamefont {KÃŒ{\c{c}}ÃŒkbenli}}, \bibinfo
  {author} {\bibfnamefont {M.}~\bibnamefont {Lazzeri}}, \bibinfo {author}
  {\bibfnamefont {M.}~\bibnamefont {Marsili}}, \bibinfo {author} {\bibfnamefont
  {N.}~\bibnamefont {Marzari}}, \bibinfo {author} {\bibfnamefont
  {F.}~\bibnamefont {Mauri}}, \bibinfo {author} {\bibfnamefont {N.~L.}\
  \bibnamefont {Nguyen}}, \bibinfo {author} {\bibfnamefont {H.-V.}\
  \bibnamefont {Nguyen}}, \bibinfo {author} {\bibfnamefont {A.~O.}\
  \bibnamefont {de-la Roza}}, \bibinfo {author} {\bibfnamefont
  {L.}~\bibnamefont {Paulatto}}, \bibinfo {author} {\bibfnamefont
  {S.}~\bibnamefont {Ponc{\'{e}}}}, \bibinfo {author} {\bibfnamefont
  {D.}~\bibnamefont {Rocca}}, \bibinfo {author} {\bibfnamefont
  {R.}~\bibnamefont {Sabatini}}, \bibinfo {author} {\bibfnamefont
  {B.}~\bibnamefont {Santra}}, \bibinfo {author} {\bibfnamefont
  {M.}~\bibnamefont {Schlipf}}, \bibinfo {author} {\bibfnamefont {A.~P.}\
  \bibnamefont {Seitsonen}}, \bibinfo {author} {\bibfnamefont {A.}~\bibnamefont
  {Smogunov}}, \bibinfo {author} {\bibfnamefont {I.}~\bibnamefont {Timrov}},
  \bibinfo {author} {\bibfnamefont {T.}~\bibnamefont {Thonhauser}}, \bibinfo
  {author} {\bibfnamefont {P.}~\bibnamefont {Umari}}, \bibinfo {author}
  {\bibfnamefont {N.}~\bibnamefont {Vast}}, \bibinfo {author} {\bibfnamefont
  {X.}~\bibnamefont {Wu}}, \ and\ \bibinfo {author} {\bibfnamefont
  {S.}~\bibnamefont {Baroni}},\ }\href {\doibase 10.1088/1361-648x/aa8f79}
  {\bibfield  {journal} {\bibinfo  {journal} {Journal of Physics: Condensed
  Matter}\ }\textbf {\bibinfo {volume} {29}},\ \bibinfo {pages} {465901}
  (\bibinfo {year} {2017})}\BibitemShut {NoStop}%
\bibitem [{\citenamefont {Perdew}\ \emph {et~al.}(1996)\citenamefont {Perdew},
  \citenamefont {Burke},\ and\ \citenamefont {Ernzerhof}}]{Perdew2}%
  \BibitemOpen
  \bibfield  {author} {\bibinfo {author} {\bibfnamefont {J.~P.}\ \bibnamefont
  {Perdew}}, \bibinfo {author} {\bibfnamefont {K.}~\bibnamefont {Burke}}, \
  and\ \bibinfo {author} {\bibfnamefont {M.}~\bibnamefont {Ernzerhof}},\ }\href
  {\doibase 10.1103/PhysRevLett.77.3865} {\bibfield  {journal} {\bibinfo
  {journal} {Phys. Rev. Lett.}\ }\textbf {\bibinfo {volume} {77}},\ \bibinfo
  {pages} {3865} (\bibinfo {year} {1996})}\BibitemShut {NoStop}%
\bibitem [{\citenamefont {Heyd}\ \emph {et~al.}(2003)\citenamefont {Heyd},
  \citenamefont {Scuseria},\ and\ \citenamefont {Ernzerhof}}]{Heyd}%
  \BibitemOpen
  \bibfield  {author} {\bibinfo {author} {\bibfnamefont {J.}~\bibnamefont
  {Heyd}}, \bibinfo {author} {\bibfnamefont {G.~E.}\ \bibnamefont {Scuseria}},
  \ and\ \bibinfo {author} {\bibfnamefont {M.}~\bibnamefont {Ernzerhof}},\
  }\href {\doibase 10.1063/1.1564060} {\bibfield  {journal} {\bibinfo
  {journal} {The Journal of Chemical Physics}\ }\textbf {\bibinfo {volume}
  {118}},\ \bibinfo {pages} {8207} (\bibinfo {year} {2003})}\BibitemShut
  {NoStop}%
\bibitem [{\citenamefont {Krukau}\ \emph {et~al.}(2006)\citenamefont {Krukau},
  \citenamefont {Vydrov}, \citenamefont {Izmaylov},\ and\ \citenamefont
  {Scuseria}}]{Krukau}%
  \BibitemOpen
  \bibfield  {author} {\bibinfo {author} {\bibfnamefont {A.~V.}\ \bibnamefont
  {Krukau}}, \bibinfo {author} {\bibfnamefont {O.~A.}\ \bibnamefont {Vydrov}},
  \bibinfo {author} {\bibfnamefont {A.~F.}\ \bibnamefont {Izmaylov}}, \ and\
  \bibinfo {author} {\bibfnamefont {G.~E.}\ \bibnamefont {Scuseria}},\ }\href
  {\doibase 10.1063/1.2404663} {\bibfield  {journal} {\bibinfo  {journal} {The
  Journal of Chemical Physics}\ }\textbf {\bibinfo {volume} {125}},\ \bibinfo
  {pages} {224106} (\bibinfo {year} {2006})}\BibitemShut {NoStop}%
\bibitem [{\citenamefont {Feenstra}\ \emph {et~al.}(2013)\citenamefont
  {Feenstra}, \citenamefont {Srivastava}, \citenamefont {Gao}, \citenamefont
  {Widom}, \citenamefont {Diaconescu}, \citenamefont {Ohta}, \citenamefont
  {Kellogg}, \citenamefont {Robinson},\ and\ \citenamefont
  {Vlassiouk}}]{Feenstra2013}%
  \BibitemOpen
  \bibfield  {author} {\bibinfo {author} {\bibfnamefont {R.~M.}\ \bibnamefont
  {Feenstra}}, \bibinfo {author} {\bibfnamefont {N.}~\bibnamefont
  {Srivastava}}, \bibinfo {author} {\bibfnamefont {Q.}~\bibnamefont {Gao}},
  \bibinfo {author} {\bibfnamefont {M.}~\bibnamefont {Widom}}, \bibinfo
  {author} {\bibfnamefont {B.}~\bibnamefont {Diaconescu}}, \bibinfo {author}
  {\bibfnamefont {T.}~\bibnamefont {Ohta}}, \bibinfo {author} {\bibfnamefont
  {G.~L.}\ \bibnamefont {Kellogg}}, \bibinfo {author} {\bibfnamefont {J.~T.}\
  \bibnamefont {Robinson}}, \ and\ \bibinfo {author} {\bibfnamefont {I.~V.}\
  \bibnamefont {Vlassiouk}},\ }\href {\doibase 10.1103/PhysRevB.87.041406}
  {\bibfield  {journal} {\bibinfo  {journal} {Phys. Rev. B}\ }\textbf {\bibinfo
  {volume} {87}},\ \bibinfo {pages} {041406(R)} (\bibinfo {year}
  {2013})}\BibitemShut {NoStop}%
\bibitem [{\citenamefont {Togo}\ and\ \citenamefont {Tanaka}(2015)}]{Togo}%
  \BibitemOpen
  \bibfield  {author} {\bibinfo {author} {\bibfnamefont {A.}~\bibnamefont
  {Togo}}\ and\ \bibinfo {author} {\bibfnamefont {I.}~\bibnamefont {Tanaka}},\
  }\href {\doibase https://doi.org/10.1016/j.scriptamat.2015.07.021} {\bibfield
   {journal} {\bibinfo  {journal} {Scr. Mater.}\ }\textbf {\bibinfo {volume}
  {108}},\ \bibinfo {pages} {1} (\bibinfo {year} {2015})}\BibitemShut {NoStop}%
\bibitem [{\citenamefont {Zhang}\ and\ \citenamefont
  {Northrup}(1991)}]{zhang1991chemical}%
  \BibitemOpen
  \bibfield  {author} {\bibinfo {author} {\bibfnamefont {S.~B.}\ \bibnamefont
  {Zhang}}\ and\ \bibinfo {author} {\bibfnamefont {J.~E.}\ \bibnamefont
  {Northrup}},\ }\href {\doibase 10.1103/PhysRevLett.67.2339} {\bibfield
  {journal} {\bibinfo  {journal} {Phys. Rev. Lett.}\ }\textbf {\bibinfo
  {volume} {67}},\ \bibinfo {pages} {2339} (\bibinfo {year}
  {1991})}\BibitemShut {NoStop}%
\bibitem [{\citenamefont {Freysoldt}\ \emph {et~al.}(2014)\citenamefont
  {Freysoldt}, \citenamefont {Grabowski}, \citenamefont {Hickel}, \citenamefont
  {Neugebauer}, \citenamefont {Kresse}, \citenamefont {Janotti},\ and\
  \citenamefont {Van~de Walle}}]{RevModPhys.86.253}%
  \BibitemOpen
  \bibfield  {author} {\bibinfo {author} {\bibfnamefont {C.}~\bibnamefont
  {Freysoldt}}, \bibinfo {author} {\bibfnamefont {B.}~\bibnamefont
  {Grabowski}}, \bibinfo {author} {\bibfnamefont {T.}~\bibnamefont {Hickel}},
  \bibinfo {author} {\bibfnamefont {J.}~\bibnamefont {Neugebauer}}, \bibinfo
  {author} {\bibfnamefont {G.}~\bibnamefont {Kresse}}, \bibinfo {author}
  {\bibfnamefont {A.}~\bibnamefont {Janotti}}, \ and\ \bibinfo {author}
  {\bibfnamefont {C.~G.}\ \bibnamefont {Van~de Walle}},\ }\href {\doibase
  10.1103/RevModPhys.86.253} {\bibfield  {journal} {\bibinfo  {journal} {Rev.
  Mod. Phys.}\ }\textbf {\bibinfo {volume} {86}},\ \bibinfo {pages} {253}
  (\bibinfo {year} {2014})}\BibitemShut {NoStop}%
\bibitem [{\citenamefont {Vinichenko}\ \emph {et~al.}(2017)\citenamefont
  {Vinichenko}, \citenamefont {Sensoy}, \citenamefont {Friend},\ and\
  \citenamefont {Kaxiras}}]{Vinichenko}%
  \BibitemOpen
  \bibfield  {author} {\bibinfo {author} {\bibfnamefont {D.}~\bibnamefont
  {Vinichenko}}, \bibinfo {author} {\bibfnamefont {M.~G.}\ \bibnamefont
  {Sensoy}}, \bibinfo {author} {\bibfnamefont {C.~M.}\ \bibnamefont {Friend}},
  \ and\ \bibinfo {author} {\bibfnamefont {E.}~\bibnamefont {Kaxiras}},\ }\href
  {\doibase 10.1103/PhysRevB.95.235310} {\bibfield  {journal} {\bibinfo
  {journal} {Phys. Rev. B}\ }\textbf {\bibinfo {volume} {95}},\ \bibinfo
  {pages} {235310} (\bibinfo {year} {2017})}\BibitemShut {NoStop}%
\bibitem [{\citenamefont {Kuate~Defo}\ \emph {et~al.}(2016)\citenamefont
  {Kuate~Defo}, \citenamefont {Fang}, \citenamefont {Shirodkar}, \citenamefont
  {Tritsaris}, \citenamefont {Dimoulas},\ and\ \citenamefont
  {Kaxiras}}]{Kuate2016}%
  \BibitemOpen
  \bibfield  {author} {\bibinfo {author} {\bibfnamefont {R.}~\bibnamefont
  {Kuate~Defo}}, \bibinfo {author} {\bibfnamefont {S.}~\bibnamefont {Fang}},
  \bibinfo {author} {\bibfnamefont {S.~N.}\ \bibnamefont {Shirodkar}}, \bibinfo
  {author} {\bibfnamefont {G.~A.}\ \bibnamefont {Tritsaris}}, \bibinfo {author}
  {\bibfnamefont {A.}~\bibnamefont {Dimoulas}}, \ and\ \bibinfo {author}
  {\bibfnamefont {E.}~\bibnamefont {Kaxiras}},\ }\href {\doibase
  10.1103/PhysRevB.94.155310} {\bibfield  {journal} {\bibinfo  {journal} {Phys.
  Rev. B}\ }\textbf {\bibinfo {volume} {94}},\ \bibinfo {pages} {155310}
  (\bibinfo {year} {2016})}\BibitemShut {NoStop}%
\bibitem [{\citenamefont {Steudel}(2003)}]{Steudel}%
  \BibitemOpen
  \bibfield  {author} {\bibinfo {author} {\bibfnamefont {R.}~\bibnamefont
  {Steudel}},\ }\href {https://books.google.com/books?id=R-zg\_aev9NAC} {\emph
  {\bibinfo {title} {{Elemental Sulfur and Sulfur-Rich Compounds I }}}}\
  (\bibinfo  {publisher} {Springer},\ \bibinfo {address} {New York},\ \bibinfo
  {year} {2003})\BibitemShut {NoStop}%
\bibitem [{\citenamefont {Zhou}\ \emph {et~al.}(2019)\citenamefont {Zhou},
  \citenamefont {Shen}, \citenamefont {Costa}, \citenamefont {Persson},
  \citenamefont {Ong}, \citenamefont {Huck}, \citenamefont {Lu}, \citenamefont
  {Ma}, \citenamefont {Chen}, \citenamefont {Tang},\ and\ \citenamefont
  {Feng}}]{2Dmatpedia}%
  \BibitemOpen
  \bibfield  {author} {\bibinfo {author} {\bibfnamefont {J.}~\bibnamefont
  {Zhou}}, \bibinfo {author} {\bibfnamefont {L.}~\bibnamefont {Shen}}, \bibinfo
  {author} {\bibfnamefont {M.~D.}\ \bibnamefont {Costa}}, \bibinfo {author}
  {\bibfnamefont {K.~A.}\ \bibnamefont {Persson}}, \bibinfo {author}
  {\bibfnamefont {S.~P.}\ \bibnamefont {Ong}}, \bibinfo {author} {\bibfnamefont
  {P.}~\bibnamefont {Huck}}, \bibinfo {author} {\bibfnamefont {Y.}~\bibnamefont
  {Lu}}, \bibinfo {author} {\bibfnamefont {X.}~\bibnamefont {Ma}}, \bibinfo
  {author} {\bibfnamefont {Y.}~\bibnamefont {Chen}}, \bibinfo {author}
  {\bibfnamefont {H.}~\bibnamefont {Tang}}, \ and\ \bibinfo {author}
  {\bibfnamefont {Y.~P.}\ \bibnamefont {Feng}},\ }\href {\doibase
  10.1038/s41597-019-0097-3} {\bibfield  {journal} {\bibinfo  {journal}
  {Scientific Data}\ }\textbf {\bibinfo {volume} {6}},\ \bibinfo {pages} {86}
  (\bibinfo {year} {2019})},\ \Eprint {http://arxiv.org/abs/www.2DMatPedia.org}
  {www.2DMatPedia.org} \BibitemShut {NoStop}%
\bibitem [{\citenamefont {Haastrup}\ \emph {et~al.}(2018)\citenamefont
  {Haastrup}, \citenamefont {Strange}, \citenamefont {Pandey}, \citenamefont
  {Deilmann}, \citenamefont {Schmidt}, \citenamefont {Hinsche}, \citenamefont
  {Gjerding}, \citenamefont {Torelli}, \citenamefont {Larsen}, \citenamefont
  {Riis-Jensen}, \citenamefont {Gath}, \citenamefont {Jacobsen}, \citenamefont
  {Mortensen}, \citenamefont {Olsen},\ and\ \citenamefont
  {Thygesen}}]{Haastrup2018}%
  \BibitemOpen
  \bibfield  {author} {\bibinfo {author} {\bibfnamefont {S.}~\bibnamefont
  {Haastrup}}, \bibinfo {author} {\bibfnamefont {M.}~\bibnamefont {Strange}},
  \bibinfo {author} {\bibfnamefont {M.}~\bibnamefont {Pandey}}, \bibinfo
  {author} {\bibfnamefont {T.}~\bibnamefont {Deilmann}}, \bibinfo {author}
  {\bibfnamefont {P.~S.}\ \bibnamefont {Schmidt}}, \bibinfo {author}
  {\bibfnamefont {N.~F.}\ \bibnamefont {Hinsche}}, \bibinfo {author}
  {\bibfnamefont {M.~N.}\ \bibnamefont {Gjerding}}, \bibinfo {author}
  {\bibfnamefont {D.}~\bibnamefont {Torelli}}, \bibinfo {author} {\bibfnamefont
  {P.~M.}\ \bibnamefont {Larsen}}, \bibinfo {author} {\bibfnamefont {A.~C.}\
  \bibnamefont {Riis-Jensen}}, \bibinfo {author} {\bibfnamefont
  {J.}~\bibnamefont {Gath}}, \bibinfo {author} {\bibfnamefont {K.~W.}\
  \bibnamefont {Jacobsen}}, \bibinfo {author} {\bibfnamefont {J.~J.}\
  \bibnamefont {Mortensen}}, \bibinfo {author} {\bibfnamefont {T.}~\bibnamefont
  {Olsen}}, \ and\ \bibinfo {author} {\bibfnamefont {K.~S.}\ \bibnamefont
  {Thygesen}},\ }\href {\doibase 10.1088/2053-1583/aacfc1} {\bibfield
  {journal} {\bibinfo  {journal} {2D Materials}\ }\textbf {\bibinfo {volume}
  {5}},\ \bibinfo {pages} {042002} (\bibinfo {year} {2018})}\BibitemShut
  {NoStop}%
\bibitem [{\citenamefont {Gjerding}\ \emph {et~al.}(2021)\citenamefont
  {Gjerding}, \citenamefont {Taghizadeh}, \citenamefont {Rasmussen},
  \citenamefont {Ali}, \citenamefont {Bertoldo}, \citenamefont {Deilmann},
  \citenamefont {Holguin}, \citenamefont {Kn{\o}sgaard}, \citenamefont {Kruse},
  \citenamefont {Manti}, \citenamefont {Pedersen}, \citenamefont {Skovhus},
  \citenamefont {Svendsen}, \citenamefont {Mortensen}, \citenamefont {Olsen},\
  and\ \citenamefont {Thygesen}}]{gjerding2021recent}%
  \BibitemOpen
  \bibfield  {author} {\bibinfo {author} {\bibfnamefont {M.~N.}\ \bibnamefont
  {Gjerding}}, \bibinfo {author} {\bibfnamefont {A.}~\bibnamefont
  {Taghizadeh}}, \bibinfo {author} {\bibfnamefont {A.}~\bibnamefont
  {Rasmussen}}, \bibinfo {author} {\bibfnamefont {S.}~\bibnamefont {Ali}},
  \bibinfo {author} {\bibfnamefont {F.}~\bibnamefont {Bertoldo}}, \bibinfo
  {author} {\bibfnamefont {T.}~\bibnamefont {Deilmann}}, \bibinfo {author}
  {\bibfnamefont {U.~P.}\ \bibnamefont {Holguin}}, \bibinfo {author}
  {\bibfnamefont {N.~R.}\ \bibnamefont {Kn{\o}sgaard}}, \bibinfo {author}
  {\bibfnamefont {M.}~\bibnamefont {Kruse}}, \bibinfo {author} {\bibfnamefont
  {S.}~\bibnamefont {Manti}}, \bibinfo {author} {\bibfnamefont {T.~G.}\
  \bibnamefont {Pedersen}}, \bibinfo {author} {\bibfnamefont {T.}~\bibnamefont
  {Skovhus}}, \bibinfo {author} {\bibfnamefont {M.~K.}\ \bibnamefont
  {Svendsen}}, \bibinfo {author} {\bibfnamefont {J.~J.}\ \bibnamefont
  {Mortensen}}, \bibinfo {author} {\bibfnamefont {T.}~\bibnamefont {Olsen}}, \
  and\ \bibinfo {author} {\bibfnamefont {K.~S.}\ \bibnamefont {Thygesen}},\
  }\href@noop {} {\enquote {\bibinfo {title} {Recent progress of the
  computational 2d materials database (c2db)},}\ } (\bibinfo {year} {2021}),\
  \Eprint {http://arxiv.org/abs/2102.03029} {arXiv:2102.03029
  [cond-mat.mtrl-sci]} \BibitemShut {NoStop}%
\bibitem [{\citenamefont {Ferrenti}\ \emph {et~al.}(2020)\citenamefont
  {Ferrenti}, \citenamefont {de~Leon}, \citenamefont {Thompson},\ and\
  \citenamefont {Cava}}]{Ferrenti2020}%
  \BibitemOpen
  \bibfield  {author} {\bibinfo {author} {\bibfnamefont {A.~M.}\ \bibnamefont
  {Ferrenti}}, \bibinfo {author} {\bibfnamefont {N.~P.}\ \bibnamefont
  {de~Leon}}, \bibinfo {author} {\bibfnamefont {J.~D.}\ \bibnamefont
  {Thompson}}, \ and\ \bibinfo {author} {\bibfnamefont {R.~J.}\ \bibnamefont
  {Cava}},\ }\href {\doibase 10.1038/s41524-020-00391-7} {\bibfield  {journal}
  {\bibinfo  {journal} {npj Computational Materials}\ }\textbf {\bibinfo
  {volume} {6}},\ \bibinfo {pages} {126} (\bibinfo {year} {2020})}\BibitemShut
  {NoStop}%
\bibitem [{\citenamefont {Gali}\ \emph {et~al.}(2008)\citenamefont {Gali},
  \citenamefont {Fyta},\ and\ \citenamefont {Kaxiras}}]{Gali}%
  \BibitemOpen
  \bibfield  {author} {\bibinfo {author} {\bibfnamefont {A.}~\bibnamefont
  {Gali}}, \bibinfo {author} {\bibfnamefont {M.}~\bibnamefont {Fyta}}, \ and\
  \bibinfo {author} {\bibfnamefont {E.}~\bibnamefont {Kaxiras}},\ }\href
  {\doibase 10.1103/PhysRevB.77.155206} {\bibfield  {journal} {\bibinfo
  {journal} {Phys. Rev. B}\ }\textbf {\bibinfo {volume} {77}},\ \bibinfo
  {pages} {155206} (\bibinfo {year} {2008})}\BibitemShut {NoStop}%
\bibitem [{\citenamefont {Lannoo}(1992)}]{Lannoo1992}%
  \BibitemOpen
  \bibfield  {author} {\bibinfo {author} {\bibfnamefont {M.}~\bibnamefont
  {Lannoo}},\ }in\ \href {\doibase
  https://doi.org/10.1016/B978-0-444-88855-6.50012-7} {\emph {\bibinfo
  {booktitle} {Basic Properties of Semiconductors}}},\ \bibinfo {series and
  number} {Handbook on Semiconductors},\ \bibinfo {editor} {edited by\ \bibinfo
  {editor} {\bibfnamefont {P.~T.}\ \bibnamefont {Landsberg}}}\ (\bibinfo
  {publisher} {Elsevier},\ \bibinfo {address} {Amsterdam},\ \bibinfo {year}
  {1992})\ pp.\ \bibinfo {pages} {113--160}\BibitemShut {NoStop}%
\bibitem [{\citenamefont {Levinshtein}\ \emph {et~al.}(2001)\citenamefont
  {Levinshtein}, \citenamefont {Rumyantsev},\ and\ \citenamefont
  {Shur}}]{Levinshtein}%
  \BibitemOpen
  \bibfield  {author} {\bibinfo {author} {\bibfnamefont {M.~E.}\ \bibnamefont
  {Levinshtein}}, \bibinfo {author} {\bibfnamefont {S.~L.}\ \bibnamefont
  {Rumyantsev}}, \ and\ \bibinfo {author} {\bibfnamefont {M.~S.}\ \bibnamefont
  {Shur}},\ }\href@noop {} {\emph {\bibinfo {title} {Properties of Advanced
  Semiconductor Materials: GaN, AlN, InN, BN, SiC, SiGe}}}\ (\bibinfo
  {publisher} {John Wiley \& Sons, Inc.},\ \bibinfo {address} {New York},\
  \bibinfo {year} {2001})\BibitemShut {NoStop}%
\bibitem [{\citenamefont {Yan}\ \emph {et~al.}(2020)\citenamefont {Yan},
  \citenamefont {Li}, \citenamefont {Kang}, \citenamefont {Wei},\ and\
  \citenamefont {Huang}}]{Yan2020}%
  \BibitemOpen
  \bibfield  {author} {\bibinfo {author} {\bibfnamefont {X.}~\bibnamefont
  {Yan}}, \bibinfo {author} {\bibfnamefont {P.}~\bibnamefont {Li}}, \bibinfo
  {author} {\bibfnamefont {L.}~\bibnamefont {Kang}}, \bibinfo {author}
  {\bibfnamefont {S.-H.}\ \bibnamefont {Wei}}, \ and\ \bibinfo {author}
  {\bibfnamefont {B.}~\bibnamefont {Huang}},\ }\href {\doibase
  10.1063/1.5140692} {\bibfield  {journal} {\bibinfo  {journal} {Journal of
  Applied Physics}\ }\textbf {\bibinfo {volume} {127}},\ \bibinfo {pages}
  {085702} (\bibinfo {year} {2020})}\BibitemShut {NoStop}%
\bibitem [{\citenamefont {Soykal}\ \emph {et~al.}(2016)\citenamefont {Soykal},
  \citenamefont {Dev},\ and\ \citenamefont {Economou}}]{Soykal}%
  \BibitemOpen
  \bibfield  {author} {\bibinfo {author} {\bibfnamefont {O.~O.}\ \bibnamefont
  {Soykal}}, \bibinfo {author} {\bibfnamefont {P.}~\bibnamefont {Dev}}, \ and\
  \bibinfo {author} {\bibfnamefont {S.~E.}\ \bibnamefont {Economou}},\ }\href
  {\doibase 10.1103/PhysRevB.93.081207} {\bibfield  {journal} {\bibinfo
  {journal} {Phys. Rev. B}\ }\textbf {\bibinfo {volume} {93}},\ \bibinfo
  {pages} {081207(R)} (\bibinfo {year} {2016})}\BibitemShut {NoStop}%
\bibitem [{\citenamefont {Bracher}\ \emph {et~al.}(2017)\citenamefont
  {Bracher}, \citenamefont {Zhang},\ and\ \citenamefont {Hu}}]{Bracher}%
  \BibitemOpen
  \bibfield  {author} {\bibinfo {author} {\bibfnamefont {D.~O.}\ \bibnamefont
  {Bracher}}, \bibinfo {author} {\bibfnamefont {X.}~\bibnamefont {Zhang}}, \
  and\ \bibinfo {author} {\bibfnamefont {E.~L.}\ \bibnamefont {Hu}},\ }\href
  {\doibase 10.1073/pnas.1704219114} {\bibfield  {journal} {\bibinfo  {journal}
  {Proceedings of the National Academy of Sciences}\ }\textbf {\bibinfo
  {volume} {114}},\ \bibinfo {pages} {4060} (\bibinfo {year}
  {2017})}\BibitemShut {NoStop}%
\bibitem [{\citenamefont {Paufler}(1987)}]{Paufler}%
  \BibitemOpen
  \bibfield  {author} {\bibinfo {author} {\bibfnamefont {P.}~\bibnamefont
  {Paufler}},\ }\href {\doibase 10.1002/crat.2170220911} {\bibfield  {journal}
  {\bibinfo  {journal} {Crystal Research and Technology}\ }\textbf {\bibinfo
  {volume} {22}},\ \bibinfo {pages} {1158} (\bibinfo {year}
  {1987})}\BibitemShut {NoStop}%
\bibitem [{\citenamefont {Iv\'ady}\ \emph {et~al.}(2017)\citenamefont
  {Iv\'ady}, \citenamefont {Davidsson}, \citenamefont {Son}, \citenamefont
  {Ohshima}, \citenamefont {Abrikosov},\ and\ \citenamefont
  {Gali}}]{Ivady2017}%
  \BibitemOpen
  \bibfield  {author} {\bibinfo {author} {\bibfnamefont {V.}~\bibnamefont
  {Iv\'ady}}, \bibinfo {author} {\bibfnamefont {J.}~\bibnamefont {Davidsson}},
  \bibinfo {author} {\bibfnamefont {N.~T.}\ \bibnamefont {Son}}, \bibinfo
  {author} {\bibfnamefont {T.}~\bibnamefont {Ohshima}}, \bibinfo {author}
  {\bibfnamefont {I.~A.}\ \bibnamefont {Abrikosov}}, \ and\ \bibinfo {author}
  {\bibfnamefont {A.}~\bibnamefont {Gali}},\ }\href {\doibase
  10.1103/PhysRevB.96.161114} {\bibfield  {journal} {\bibinfo  {journal} {Phys.
  Rev. B}\ }\textbf {\bibinfo {volume} {96}},\ \bibinfo {pages} {161114(R)}
  (\bibinfo {year} {2017})}\BibitemShut {NoStop}%
\bibitem [{\citenamefont {Udvarhelyi}\ \emph {et~al.}(2020)\citenamefont
  {Udvarhelyi}, \citenamefont {Thiering}, \citenamefont {Morioka},
  \citenamefont {Babin}, \citenamefont {Kaiser}, \citenamefont {Lukin},
  \citenamefont {Ohshima}, \citenamefont {Ul-Hassan}, \citenamefont {Son},
  \citenamefont {Vu\ifmmode \check{c}\else
  \v{c}\fi{}kovi\ifmmode~\acute{c}\else \'{c}\fi{}}, \citenamefont
  {Wrachtrup},\ and\ \citenamefont {Gali}}]{Udvarhelyi2020}%
  \BibitemOpen
  \bibfield  {author} {\bibinfo {author} {\bibfnamefont {P.}~\bibnamefont
  {Udvarhelyi}}, \bibinfo {author} {\bibfnamefont {G.~m.~H.}\ \bibnamefont
  {Thiering}}, \bibinfo {author} {\bibfnamefont {N.}~\bibnamefont {Morioka}},
  \bibinfo {author} {\bibfnamefont {C.}~\bibnamefont {Babin}}, \bibinfo
  {author} {\bibfnamefont {F.}~\bibnamefont {Kaiser}}, \bibinfo {author}
  {\bibfnamefont {D.}~\bibnamefont {Lukin}}, \bibinfo {author} {\bibfnamefont
  {T.}~\bibnamefont {Ohshima}}, \bibinfo {author} {\bibfnamefont
  {J.}~\bibnamefont {Ul-Hassan}}, \bibinfo {author} {\bibfnamefont {N.~T.}\
  \bibnamefont {Son}}, \bibinfo {author} {\bibfnamefont {J.}~\bibnamefont
  {Vu\ifmmode \check{c}\else \v{c}\fi{}kovi\ifmmode~\acute{c}\else
  \'{c}\fi{}}}, \bibinfo {author} {\bibfnamefont {J.}~\bibnamefont
  {Wrachtrup}}, \ and\ \bibinfo {author} {\bibfnamefont {A.}~\bibnamefont
  {Gali}},\ }\href {\doibase 10.1103/PhysRevApplied.13.054017} {\bibfield
  {journal} {\bibinfo  {journal} {Phys. Rev. Applied}\ }\textbf {\bibinfo
  {volume} {13}},\ \bibinfo {pages} {054017} (\bibinfo {year}
  {2020})}\BibitemShut {NoStop}%
\bibitem [{\citenamefont {de~Jong}\ \emph {et~al.}(2016)\citenamefont
  {de~Jong}, \citenamefont {Chen}, \citenamefont {Notestine}, \citenamefont
  {Persson}, \citenamefont {Ceder}, \citenamefont {Jain}, \citenamefont
  {Asta},\ and\ \citenamefont {Gamst}}]{deJong2016}%
  \BibitemOpen
  \bibfield  {author} {\bibinfo {author} {\bibfnamefont {M.}~\bibnamefont
  {de~Jong}}, \bibinfo {author} {\bibfnamefont {W.}~\bibnamefont {Chen}},
  \bibinfo {author} {\bibfnamefont {R.}~\bibnamefont {Notestine}}, \bibinfo
  {author} {\bibfnamefont {K.}~\bibnamefont {Persson}}, \bibinfo {author}
  {\bibfnamefont {G.}~\bibnamefont {Ceder}}, \bibinfo {author} {\bibfnamefont
  {A.}~\bibnamefont {Jain}}, \bibinfo {author} {\bibfnamefont {M.}~\bibnamefont
  {Asta}}, \ and\ \bibinfo {author} {\bibfnamefont {A.}~\bibnamefont {Gamst}},\
  }\href {\doibase 10.1038/srep34256} {\bibfield  {journal} {\bibinfo
  {journal} {Scientific Reports}\ }\textbf {\bibinfo {volume} {6}},\ \bibinfo
  {pages} {34256} (\bibinfo {year} {2016})}\BibitemShut {NoStop}%
\bibitem [{\citenamefont {Evers}\ \emph {et~al.}(2015)\citenamefont {Evers},
  \citenamefont {Mayer}, \citenamefont {Moeckl}, \citenamefont {Oehlinger},
  \citenamefont {Koeppe},\ and\ \citenamefont {Schnoeckel}}]{Evers2015}%
  \BibitemOpen
  \bibfield  {author} {\bibinfo {author} {\bibfnamefont {J.}~\bibnamefont
  {Evers}}, \bibinfo {author} {\bibfnamefont {P.}~\bibnamefont {Mayer}},
  \bibinfo {author} {\bibfnamefont {L.}~\bibnamefont {Moeckl}}, \bibinfo
  {author} {\bibfnamefont {G.}~\bibnamefont {Oehlinger}}, \bibinfo {author}
  {\bibfnamefont {R.}~\bibnamefont {Koeppe}}, \ and\ \bibinfo {author}
  {\bibfnamefont {H.}~\bibnamefont {Schnoeckel}},\ }\href@noop {} {\bibfield
  {journal} {\bibinfo  {journal} {Inorganic Chemistry}\ }\textbf {\bibinfo
  {volume} {54}},\ \bibinfo {pages} {1240} (\bibinfo {year}
  {2015})}\BibitemShut {NoStop}%
\bibitem [{\citenamefont {Smart}\ \emph {et~al.}(2020)\citenamefont {Smart},
  \citenamefont {Li}, \citenamefont {Xu},\ and\ \citenamefont
  {Ping}}]{smart2020intersystem}%
  \BibitemOpen
  \bibfield  {author} {\bibinfo {author} {\bibfnamefont {T.~J.}\ \bibnamefont
  {Smart}}, \bibinfo {author} {\bibfnamefont {K.}~\bibnamefont {Li}}, \bibinfo
  {author} {\bibfnamefont {J.}~\bibnamefont {Xu}}, \ and\ \bibinfo {author}
  {\bibfnamefont {Y.}~\bibnamefont {Ping}},\ }\href@noop {} {\enquote {\bibinfo
  {title} {Intersystem crossing and exciton-defect coupling of spin defects in
  hexagonal boron nitride},}\ } (\bibinfo {year} {2020}),\ \Eprint
  {http://arxiv.org/abs/2009.02830} {arXiv:2009.02830 [cond-mat.mtrl-sci]}
  \BibitemShut {NoStop}%
\bibitem [{\citenamefont {Johari}\ and\ \citenamefont
  {Shenoy}(2012)}]{Johari2012}%
  \BibitemOpen
  \bibfield  {author} {\bibinfo {author} {\bibfnamefont {P.}~\bibnamefont
  {Johari}}\ and\ \bibinfo {author} {\bibfnamefont {V.~B.}\ \bibnamefont
  {Shenoy}},\ }\href {\doibase 10.1021/nn301320r} {\bibfield  {journal}
  {\bibinfo  {journal} {ACS Nano}\ }\textbf {\bibinfo {volume} {6}},\ \bibinfo
  {pages} {5449} (\bibinfo {year} {2012})},\ \bibinfo {note} {pMID:
  22591011}\BibitemShut {NoStop}%
\bibitem [{\citenamefont {Kuc}\ \emph {et~al.}(2011)\citenamefont {Kuc},
  \citenamefont {Zibouche},\ and\ \citenamefont {Heine}}]{Kuc2011}%
  \BibitemOpen
  \bibfield  {author} {\bibinfo {author} {\bibfnamefont {A.}~\bibnamefont
  {Kuc}}, \bibinfo {author} {\bibfnamefont {N.}~\bibnamefont {Zibouche}}, \
  and\ \bibinfo {author} {\bibfnamefont {T.}~\bibnamefont {Heine}},\ }\href
  {\doibase 10.1103/PhysRevB.83.245213} {\bibfield  {journal} {\bibinfo
  {journal} {Phys. Rev. B}\ }\textbf {\bibinfo {volume} {83}},\ \bibinfo
  {pages} {245213} (\bibinfo {year} {2011})}\BibitemShut {NoStop}%
\bibitem [{\citenamefont {Ataca}\ and\ \citenamefont
  {Ciraci}(2011)}]{Ataca2011}%
  \BibitemOpen
  \bibfield  {author} {\bibinfo {author} {\bibfnamefont {C.}~\bibnamefont
  {Ataca}}\ and\ \bibinfo {author} {\bibfnamefont {S.}~\bibnamefont {Ciraci}},\
  }\href {\doibase 10.1021/jp2000442} {\bibfield  {journal} {\bibinfo
  {journal} {The Journal of Physical Chemistry C}\ }\textbf {\bibinfo {volume}
  {115}},\ \bibinfo {pages} {13303} (\bibinfo {year} {2011})}\BibitemShut
  {NoStop}%
\bibitem [{\citenamefont {Pfender}\ \emph {et~al.}(2017)\citenamefont
  {Pfender}, \citenamefont {Aslam}, \citenamefont {Sumiya}, \citenamefont
  {Onoda}, \citenamefont {Neumann}, \citenamefont {Isoya}, \citenamefont
  {Meriles},\ and\ \citenamefont {Wrachtrup}}]{Pfender2017}%
  \BibitemOpen
  \bibfield  {author} {\bibinfo {author} {\bibfnamefont {M.}~\bibnamefont
  {Pfender}}, \bibinfo {author} {\bibfnamefont {N.}~\bibnamefont {Aslam}},
  \bibinfo {author} {\bibfnamefont {H.}~\bibnamefont {Sumiya}}, \bibinfo
  {author} {\bibfnamefont {S.}~\bibnamefont {Onoda}}, \bibinfo {author}
  {\bibfnamefont {P.}~\bibnamefont {Neumann}}, \bibinfo {author} {\bibfnamefont
  {J.}~\bibnamefont {Isoya}}, \bibinfo {author} {\bibfnamefont {C.~A.}\
  \bibnamefont {Meriles}}, \ and\ \bibinfo {author} {\bibfnamefont
  {J.}~\bibnamefont {Wrachtrup}},\ }\href {\doibase 10.1038/s41467-017-00964-z}
  {\bibfield  {journal} {\bibinfo  {journal} {Nature Communications}\ }\textbf
  {\bibinfo {volume} {8}},\ \bibinfo {pages} {834} (\bibinfo {year}
  {2017})}\BibitemShut {NoStop}%
\bibitem [{\citenamefont {Sato}\ \emph {et~al.}(2020)\citenamefont {Sato},
  \citenamefont {Yoshida}, \citenamefont {Zen}, \citenamefont {Hachiya},
  \citenamefont {Goto}, \citenamefont {Sagawa},\ and\ \citenamefont
  {Ohgaki}}]{Sato2020}%
  \BibitemOpen
  \bibfield  {author} {\bibinfo {author} {\bibfnamefont {O.}~\bibnamefont
  {Sato}}, \bibinfo {author} {\bibfnamefont {K.}~\bibnamefont {Yoshida}},
  \bibinfo {author} {\bibfnamefont {H.}~\bibnamefont {Zen}}, \bibinfo {author}
  {\bibfnamefont {K.}~\bibnamefont {Hachiya}}, \bibinfo {author} {\bibfnamefont
  {T.}~\bibnamefont {Goto}}, \bibinfo {author} {\bibfnamefont {T.}~\bibnamefont
  {Sagawa}}, \ and\ \bibinfo {author} {\bibfnamefont {H.}~\bibnamefont
  {Ohgaki}},\ }\href {\doibase https://doi.org/10.1016/j.physleta.2019.126223}
  {\bibfield  {journal} {\bibinfo  {journal} {Physics Letters A}\ }\textbf
  {\bibinfo {volume} {384}},\ \bibinfo {pages} {126223} (\bibinfo {year}
  {2020})}\BibitemShut {NoStop}%
\bibitem [{\citenamefont {Snider}\ \emph {et~al.}(2020)\citenamefont {Snider},
  \citenamefont {Dasenbrock-Gammon}, \citenamefont {McBride}, \citenamefont
  {Debessai}, \citenamefont {Vindana}, \citenamefont {Vencatasamy},
  \citenamefont {Lawler}, \citenamefont {Salamat},\ and\ \citenamefont
  {Dias}}]{Snider2020}%
  \BibitemOpen
  \bibfield  {author} {\bibinfo {author} {\bibfnamefont {E.}~\bibnamefont
  {Snider}}, \bibinfo {author} {\bibfnamefont {N.}~\bibnamefont
  {Dasenbrock-Gammon}}, \bibinfo {author} {\bibfnamefont {R.}~\bibnamefont
  {McBride}}, \bibinfo {author} {\bibfnamefont {M.}~\bibnamefont {Debessai}},
  \bibinfo {author} {\bibfnamefont {H.}~\bibnamefont {Vindana}}, \bibinfo
  {author} {\bibfnamefont {K.}~\bibnamefont {Vencatasamy}}, \bibinfo {author}
  {\bibfnamefont {K.~V.}\ \bibnamefont {Lawler}}, \bibinfo {author}
  {\bibfnamefont {A.}~\bibnamefont {Salamat}}, \ and\ \bibinfo {author}
  {\bibfnamefont {R.~P.}\ \bibnamefont {Dias}},\ }\href {\doibase
  10.1038/s41586-020-2801-z} {\bibfield  {journal} {\bibinfo  {journal}
  {Nature}\ }\textbf {\bibinfo {volume} {586}},\ \bibinfo {pages} {373}
  (\bibinfo {year} {2020})}\BibitemShut {NoStop}%
\bibitem [{\citenamefont {Lesik}\ \emph {et~al.}(2019)\citenamefont {Lesik},
  \citenamefont {Plisson}, \citenamefont {Toraille}, \citenamefont {Renaud},
  \citenamefont {Occelli}, \citenamefont {Schmidt}, \citenamefont {Salord},
  \citenamefont {Delobbe}, \citenamefont {Debuisschert}, \citenamefont
  {Rondin}, \citenamefont {Loubeyre},\ and\ \citenamefont {Roch}}]{Lesik2019}%
  \BibitemOpen
  \bibfield  {author} {\bibinfo {author} {\bibfnamefont {M.}~\bibnamefont
  {Lesik}}, \bibinfo {author} {\bibfnamefont {T.}~\bibnamefont {Plisson}},
  \bibinfo {author} {\bibfnamefont {L.}~\bibnamefont {Toraille}}, \bibinfo
  {author} {\bibfnamefont {J.}~\bibnamefont {Renaud}}, \bibinfo {author}
  {\bibfnamefont {F.}~\bibnamefont {Occelli}}, \bibinfo {author} {\bibfnamefont
  {M.}~\bibnamefont {Schmidt}}, \bibinfo {author} {\bibfnamefont
  {O.}~\bibnamefont {Salord}}, \bibinfo {author} {\bibfnamefont
  {A.}~\bibnamefont {Delobbe}}, \bibinfo {author} {\bibfnamefont
  {T.}~\bibnamefont {Debuisschert}}, \bibinfo {author} {\bibfnamefont
  {L.}~\bibnamefont {Rondin}}, \bibinfo {author} {\bibfnamefont
  {P.}~\bibnamefont {Loubeyre}}, \ and\ \bibinfo {author} {\bibfnamefont
  {J.-F.}\ \bibnamefont {Roch}},\ }\href {\doibase 10.1126/science.aaw4329}
  {\bibfield  {journal} {\bibinfo  {journal} {Science}\ }\textbf {\bibinfo
  {volume} {366}},\ \bibinfo {pages} {1359} (\bibinfo {year}
  {2019})}\BibitemShut {NoStop}%
\bibitem [{\citenamefont {Yip}\ \emph {et~al.}(2019)\citenamefont {Yip},
  \citenamefont {Ho}, \citenamefont {Yu}, \citenamefont {Chen}, \citenamefont
  {Zhang}, \citenamefont {Kasahara}, \citenamefont {Mizukami}, \citenamefont
  {Shibauchi}, \citenamefont {Matsuda}, \citenamefont {Goh},\ and\
  \citenamefont {Yang}}]{Yip2019}%
  \BibitemOpen
  \bibfield  {author} {\bibinfo {author} {\bibfnamefont {K.~Y.}\ \bibnamefont
  {Yip}}, \bibinfo {author} {\bibfnamefont {K.~O.}\ \bibnamefont {Ho}},
  \bibinfo {author} {\bibfnamefont {K.~Y.}\ \bibnamefont {Yu}}, \bibinfo
  {author} {\bibfnamefont {Y.}~\bibnamefont {Chen}}, \bibinfo {author}
  {\bibfnamefont {W.}~\bibnamefont {Zhang}}, \bibinfo {author} {\bibfnamefont
  {S.}~\bibnamefont {Kasahara}}, \bibinfo {author} {\bibfnamefont
  {Y.}~\bibnamefont {Mizukami}}, \bibinfo {author} {\bibfnamefont
  {T.}~\bibnamefont {Shibauchi}}, \bibinfo {author} {\bibfnamefont
  {Y.}~\bibnamefont {Matsuda}}, \bibinfo {author} {\bibfnamefont {S.~K.}\
  \bibnamefont {Goh}}, \ and\ \bibinfo {author} {\bibfnamefont
  {S.}~\bibnamefont {Yang}},\ }\href {\doibase 10.1126/science.aaw4278}
  {\bibfield  {journal} {\bibinfo  {journal} {Science}\ }\textbf {\bibinfo
  {volume} {366}},\ \bibinfo {pages} {1355} (\bibinfo {year}
  {2019})}\BibitemShut {NoStop}%
\bibitem [{\citenamefont {Hsieh}\ \emph {et~al.}(2019)\citenamefont {Hsieh},
  \citenamefont {Bhattacharyya}, \citenamefont {Zu}, \citenamefont {Mittiga},
  \citenamefont {Smart}, \citenamefont {Machado}, \citenamefont {Kobrin},
  \citenamefont {H{\"o}hn}, \citenamefont {Rui}, \citenamefont {Kamrani},
  \citenamefont {Chatterjee}, \citenamefont {Choi}, \citenamefont {Zaletel},
  \citenamefont {Struzhkin}, \citenamefont {Moore}, \citenamefont {Levitas},
  \citenamefont {Jeanloz},\ and\ \citenamefont {Yao}}]{Hsieh2019}%
  \BibitemOpen
  \bibfield  {author} {\bibinfo {author} {\bibfnamefont {S.}~\bibnamefont
  {Hsieh}}, \bibinfo {author} {\bibfnamefont {P.}~\bibnamefont
  {Bhattacharyya}}, \bibinfo {author} {\bibfnamefont {C.}~\bibnamefont {Zu}},
  \bibinfo {author} {\bibfnamefont {T.}~\bibnamefont {Mittiga}}, \bibinfo
  {author} {\bibfnamefont {T.~J.}\ \bibnamefont {Smart}}, \bibinfo {author}
  {\bibfnamefont {F.}~\bibnamefont {Machado}}, \bibinfo {author} {\bibfnamefont
  {B.}~\bibnamefont {Kobrin}}, \bibinfo {author} {\bibfnamefont {T.~O.}\
  \bibnamefont {H{\"o}hn}}, \bibinfo {author} {\bibfnamefont {N.~Z.}\
  \bibnamefont {Rui}}, \bibinfo {author} {\bibfnamefont {M.}~\bibnamefont
  {Kamrani}}, \bibinfo {author} {\bibfnamefont {S.}~\bibnamefont {Chatterjee}},
  \bibinfo {author} {\bibfnamefont {S.}~\bibnamefont {Choi}}, \bibinfo {author}
  {\bibfnamefont {M.}~\bibnamefont {Zaletel}}, \bibinfo {author} {\bibfnamefont
  {V.~V.}\ \bibnamefont {Struzhkin}}, \bibinfo {author} {\bibfnamefont {J.~E.}\
  \bibnamefont {Moore}}, \bibinfo {author} {\bibfnamefont {V.~I.}\ \bibnamefont
  {Levitas}}, \bibinfo {author} {\bibfnamefont {R.}~\bibnamefont {Jeanloz}}, \
  and\ \bibinfo {author} {\bibfnamefont {N.~Y.}\ \bibnamefont {Yao}},\ }\href
  {\doibase 10.1126/science.aaw4352} {\bibfield  {journal} {\bibinfo  {journal}
  {Science}\ }\textbf {\bibinfo {volume} {366}},\ \bibinfo {pages} {1349}
  (\bibinfo {year} {2019})}\BibitemShut {NoStop}%
\bibitem [{\citenamefont {Towns}\ \emph {et~al.}(2014)\citenamefont {Towns},
  \citenamefont {Cockerill}, \citenamefont {Dahan}, \citenamefont {Foster},
  \citenamefont {Gaither}, \citenamefont {Grimshaw}, \citenamefont {Hazlewood},
  \citenamefont {Lathrop}, \citenamefont {Lifka}, \citenamefont {Peterson},
  \citenamefont {Roskies}, \citenamefont {Scott},\ and\ \citenamefont
  {Wilkins-Diehr}}]{Towns}%
  \BibitemOpen
  \bibfield  {author} {\bibinfo {author} {\bibfnamefont {J.}~\bibnamefont
  {Towns}}, \bibinfo {author} {\bibfnamefont {T.}~\bibnamefont {Cockerill}},
  \bibinfo {author} {\bibfnamefont {M.}~\bibnamefont {Dahan}}, \bibinfo
  {author} {\bibfnamefont {I.}~\bibnamefont {Foster}}, \bibinfo {author}
  {\bibfnamefont {K.}~\bibnamefont {Gaither}}, \bibinfo {author} {\bibfnamefont
  {A.}~\bibnamefont {Grimshaw}}, \bibinfo {author} {\bibfnamefont
  {V.}~\bibnamefont {Hazlewood}}, \bibinfo {author} {\bibfnamefont
  {S.}~\bibnamefont {Lathrop}}, \bibinfo {author} {\bibfnamefont
  {D.}~\bibnamefont {Lifka}}, \bibinfo {author} {\bibfnamefont {G.~D.}\
  \bibnamefont {Peterson}}, \bibinfo {author} {\bibfnamefont {R.}~\bibnamefont
  {Roskies}}, \bibinfo {author} {\bibfnamefont {J.~R.}\ \bibnamefont {Scott}},
  \ and\ \bibinfo {author} {\bibfnamefont {N.}~\bibnamefont {Wilkins-Diehr}},\
  }\href {\doibase 10.1109/MCSE.2014.80} {\bibfield  {journal} {\bibinfo
  {journal} {Computing in Science \& Engineering}\ }\textbf {\bibinfo {volume}
  {16}},\ \bibinfo {pages} {62} (\bibinfo {year} {2014})}\BibitemShut {NoStop}%
\end{thebibliography}%
\end{document}